\documentstyle[11pt,epsf]{article}

\def\singleandabitspaced{\baselineskip=\normalbaselineskip\multiply
    \baselineskip by 110\divide\baselineskip by 100}

\newcommand{ \NPB    }[3]{Nucl. Phys. {\bf B#1}, #3 (#2)}
\newcommand{ \PLB    }[3]{Phys. Lett. B {\bf #1}, #3 (#2)}
\newcommand{ \PLBold }[3]{Phys. Lett. {\bf #1B}, #3 (#2)}
\newcommand{ \PRD    }[3]{Phys. Rev. {\bf D#1}, #3 (#2)}

\newcommand{ \PRL    }[3]{Phys. Rev. Lett. {\bf #1}, #3 (#2)}
\newcommand{ \PREP   }[3]{Phys. Rep. {\bf #1}, #3 (#2)}

\newcommand{ \centeron }[2]{{\setbox0=\hbox{#1}\setbox1=\hbox{#2}\ifdim
                             \wd1>\wd0\kern.5\wd1\kern-.5\wd0\fi \copy0
                             \kern-.5\wd0\kern-.5\wd1\copy1\ifdim\wd0>\wd1
                             \kern.5\wd0\kern-.5\wd1\fi}}
\newcommand{ \ltap }{\>\centeron{\raise.35ex\hbox{$<$}}
                     {\lower.65ex\hbox{$\sim$}}\>}
\newcommand{ \gtap }{\>\centeron{\raise.35ex\hbox{$>$}}
                     {\lower.65ex\hbox{$\sim$}}\>}
\newcommand{ \gsim }{\mathrel{\gtap}}
\newcommand{ \lsim }{\mathrel{\ltap}}

\newcommand{ \slashchar }[1]{\setbox0=\hbox{$#1$}   
   \dimen0=\wd0                                     
   \setbox1=\hbox{/} \dimen1=\wd1                   
   \ifdim\dimen0>\dimen1                            
      \rlap{\hbox to \dimen0{\hfil/\hfil}}          
      #1                                            
   \else                                            
      \rlap{\hbox to \dimen1{\hfil$#1$\hfil}}       
      /                                             
   \fi}                                             %

\newcommand{ \N          }{ {\tilde N} }
\newcommand{ \Ni         }{ {\tilde N}_i }
\newcommand{ \Nj         }{ {\tilde N}_j }
\newcommand{ \NI         }{ {\tilde N}_1 }
\newcommand{ \NII        }{ {\tilde N}_2 }
\newcommand{ \NIII       }{ {\tilde N}_3 }
\newcommand{ \NIIII      }{ {\tilde N}_4 }
\newcommand{ \C          }{ {\tilde C} }
\newcommand{ \Ci         }{ {\tilde C}^{\pm}_i }
\newcommand{ \Cj         }{ {\tilde C}^{\pm}_j }

\newcommand{ \Cjmp       }{ {\tilde C}^{\mp}_j }

\newcommand{ \CI         }{ {\tilde C}_1 }
\newcommand{ \CII        }{ {\tilde C}_2 }
\newcommand{ \Cpm        }{ {\tilde C}^{\pm} }
\newcommand{ \CIpm       }{ {\tilde C}^{\pm}_1 }
\newcommand{ \CIIpm      }{ {\tilde C}^{\pm}_2 }

\newcommand{ \CIImp      }{ {\tilde C}^{\mp}_2 }
\newcommand{ \CIp        }{ {\tilde C}^{+}_1 }
\newcommand{ \CIm        }{ {\tilde C}^{-}_1 }
\newcommand{ \CIIp       }{ {\tilde C}^{+}_2 }
\newcommand{ \CIIm       }{ {\tilde C}^{-}_2 }
\newcommand{ \G          }{ {\tilde G} }
\newcommand{ \eL         }{ {\tilde e}_L }
\newcommand{ \eR         }{ {\tilde e}_R }
\newcommand{ \veL        }{ {\tilde \nu}_e }
\newcommand{ \Ha         }{ {\tilde H}_a }
\newcommand{ \Hb         }{ {\tilde H}_b }
\newcommand{ \tbeta      }{ {\tan \beta} }
\newcommand{ \dL         }{ \tilde d_L }
\newcommand{ \dR         }{ \tilde d_R }
\newcommand{ \uL         }{ \tilde u_L }
\newcommand{ \uR         }{ \tilde u_R }
\newcommand{ \tI         }{ {\tilde t}_1 }
\newcommand{ \tII        }{ {\tilde t}_2 }
\newcommand{ \tR         }{ {\tilde t}_R }
\newcommand{ \tL         }{ {\tilde t}_L }
\newcommand{ \ra         }{ \rightarrow }
\newcommand{ \ph         }{ \gamma }
\newcommand{ \lL         }{ { \tilde l}_L }
\newcommand{ \lR         }{ { \tilde l}_R }
\newcommand{ \vL         }{ { \tilde \nu} }

\newcommand{ \E          }{ { \slashchar{E} } }
\newcommand{ \Et         }{ { \slashchar{E}_T } }

\newcommand{ \BR         }{ {\cal B} }
\newcommand{ \eegg       }{ {ee\gamma\gamma} }

\newcommand{ \phino      }{ \tilde{\gamma} }
\newcommand{ \Zino       }{ \tilde{Z} }

\newcommand{ \sthw       }{ \sin\theta_{\scriptscriptstyle W} }
\newcommand{ \cthw       }{ \cos\theta_{\scriptscriptstyle W} }
\newcommand{ \sthwq      }{ \sin^2\!\theta_{\scriptscriptstyle W} }
\newcommand{ \cthwq      }{ \cos^2\!\theta_{\scriptscriptstyle W} }

\newcommand{ \sindb      }{ \sin 2\beta }
\newcommand{ \cosdb      }{ \cos 2\beta } 

\begin{document}

\begin{titlepage}
\begin{flushright}
{\large
 hep-ph/9607414 \\
 (accepted Phys.\ Rev.\ D) \\
 July 1996 \\
}
\end{flushright}

\vskip 1.0cm

\begin{center}
{\LARGE\bf Low energy supersymmetry with a neutralino LSP }

{\LARGE\bf and the CDF $\eegg + \Et$ event }

\vskip 1cm

{\large
 S. Ambrosanio$^{*,}$\footnote{{\tt ambros@umich.edu}},
 G. L. Kane\footnote{{\tt gkane@umich.edu}},
 Graham D. Kribs\footnote{{\tt kribs@umich.edu}},
 Stephen P. Martin\footnote{{\tt spmartin@umich.edu}}
} \\
\vskip 4pt
{\it Randall Physics Laboratory, University of Michigan,\\
     Ann Arbor, MI 48109--1120 } \\
\vskip 10pt
{\large
S. Mrenna\footnote{{\tt mrenna@hep.anl.gov}}} \\
\vskip 4pt
{\it High Energy Physics Division, Argonne National Laboratory, \\
     Argonne, IL 60439 } \\

\vskip 2.0cm

\begin{abstract}
\singleandabitspaced

We present a refined and expanded analysis of the CDF $\eegg + \Et$
event as superpartner production, assuming the lightest neutralino 
is the lightest supersymmetric particle.  A general low-energy
Lagrangian is constrained by a minimum cross section
times branching ratio into two electrons and two photons,
kinematics consistent with the event, and LEP1-LEP130 data.  
We examine how the
supersymmetric parameters depend on the kinematics, branching ratios 
and experimental predictions with a selectron interpretation of 
the event, and discuss to what extent these are modified by other 
interpretations.  Predictions for imminent CERN LEP upgrades
and the present and future Fermilab Tevatron are presented.  
Finally, we briefly discuss 
the possible connection to other phenomena including a light stop, 
the neutralino relic density, the shift in $R_b$ and the 
associated shift in $\alpha_s$, 
and implications for the form of the theory.
\end{abstract}

\end{center}

\vskip 1.0cm

\noindent
$^*$Supported mainly by a INFN postdoctoral fellowship, Italy.

\end{titlepage}
\setcounter{footnote}{0}
\setcounter{page}{2}
\setcounter{section}{0}
\setcounter{subsection}{0}
\setcounter{subsubsection}{0}

\singleandabitspaced
\section{Introduction}
\label{introduction-section}
\indent

Minimal low energy supersymmetry provides the most promising
framework to extend the Standard Model (SM).  Such extensions
take the form of complete models that encompass the gauge
group structure and particle content of the SM, along with the 
supersymmetrized interactions and superpartners.  General 
low energy theories of supersymmetry have over 100 parameters
in addition to the SM parameters; such parameters can
certainly be constrained by direct collider searches, but in 
general one needs more information or more assumptions 
to do calculations that examine many parts of the remaining parameter 
space.  In many cases only one or a few parameters enter the
calculation of a given observable, so useful predictions
can often be made from a small subset of the supersymmetric
parameters without loss of generality.  
The two obvious approaches to reduce the parameter 
space are to use theoretical assumptions, and 
(direct and indirect) experimental constraints.

In Ref.~\cite{PRL} we showed that the CDF $\eegg + \Et$ 
event~\cite{event} at the
Fermilab Tevatron could be interpreted in low energy supersymmetry 
with roughly the expected rate and kinematics.  If we assume
this interpretation is correct and
the event {\em is}\/ due to supersymmetry, then we can 
reduce the parameter space by searching for sets of parameters
that satisfy the event's constraints.  We use the term `model'
to describe a distinct set of parameters, but of course all of
our `models' parameterize only one basic supersymmetric low 
energy Lagrangian.  The primary difficulty in deriving precise
parameter constraints (hence predictions) 
is the somewhat arbitrary notion of interpreting one event in terms 
of a cross section times branching ratio.  Instead of advocating 
a particular lower (or upper) threshold value, we vary the 
value in a reasonable range and show the effect 
on parameter space and predictions.  In this way we attempt 
to give an appreciation for the robustness or confidence of 
particular constraints or predictions.

We work within a general low energy ($\equiv$ electroweak scale) 
supersymmetric theory without assuming common scalar or gaugino masses 
at the unification scale~\cite{unification}.  To determine branching
ratios and scalar interaction contributions to cross sections, 
we do assume squark mass degeneracy except possibly for the light stop 
$\tI$, and a mass degeneracy among sleptons with 
the same electroweak quantum numbers.  Such assumptions are
not crucial to our analysis, and could be removed if necessary.
We assume R-parity 
is exactly conserved, so the lightest supersymmetric
particle (LSP) is stable (consistent with the $\eegg + \Et$
event where the two LSPs escape the CDF detector).
Finally, throughout this paper 
we assume the LSP is the 
lightest neutralino $\NI$, and not the gravitino.  Analyses 
of the $\eegg + \Et$ event
assuming the LSP is a light gravitino have been presented 
by us~\cite{PRL,Grav} and in other 
Refs.~\cite{DimopoulosPRL,DimopoulosSecond}. 
One cannot distinguish these scenarios based solely on the 
$\eegg + \Et$ event, although it is likely that associated 
phenomenology can distinguish the scenarios.  In this paper
we assume that $\NI$ is the LSP, or is at least long-lived
enough to escape the detector.  If $\NI$ is identified as a
stable LSP, then it is a possible cold dark matter 
particle~\cite{HiggsinoLSP}.

In minimal low energy supersymmetry the possibility
of one-loop radiative decay of neutralinos~\cite{Komatsu,HaberWyler,%
AmbrosMele1,AmbrosMele2} leads to signals with hard
isolated photons plus missing energy in the final state,
a signal predicted many years prior to the $\eegg + \Et$ event.
This is by no means the only mechanism to produce photons
plus missing energy, but it does allow the 
interpretation of the $\eegg + \Et$ event as selectron production 
$p\overline{p} \ra {\tilde e}^+{\tilde e}^- (+ X)$, with the
selectron ${\tilde e}$ decaying mainly into the next-to-lightest
neutralino $\NII$ and an electron, followed by $\NII \ra \NI\ph$.
It is also possible to imagine other interpretations 
that involve the radiative decay of $\NII$, but for which the 
initial superpartner
production is different.  The two possibilities in this class
that we consider below are chargino pair production and neutralino
pair production.

The plan of the paper is as follows.  In Sec.~\ref{kinematics-sec}
we discuss the kinematics of the $\eegg + \Et$ event
in the selectron interpretation, the chargino interpretation,
the neutralino interpretation, and other interpretations.
Using superpartner mass constraints established from 
the $\eegg + \Et$ event kinematics, we discuss low energy 
supersymmetric model building in Sec.~\ref{model-building-section}.  
Here we present a discussion of the radiative neutralino branching ratio,
slepton decay and constraints from LEP.  
In Sec.~\ref{numerical-results-section} we discuss
the results obtained from a numerical scan of the parameter
space, using the structure built up from 
Sec.~\ref{model-building-section}.  The bulk of
our results are contained in Sec.~\ref{numerical-results-section},
where we discuss the model building results, the 
chargino/neutralino/slepton branching ratios, 
and predictions for LEP and Tevatron.  
In Sec.~\ref{light-stop-sec} we discuss the possibility
of explaining the $\eegg + \Et$ event with the
further assumption of a light stop $\tI$.  Finally, 
in Sec.~\ref{conclusions-sec}, we present our
concluding remarks, including a summary of such 
questions as distinguishing left- and right-selectrons,
and the main channels that can confirm the $\eegg + \Et$
event is due to supersymmetry with an LSP=$\NI$.
In Appendix~\ref{char-sec} we discuss the
viability of the chargino interpretation, 
and the results of attempts at model building. 
In Appendix~\ref{sample-models-sec} we give four sample models in
the selectron interpretation.

{\em Note added:}  As we were completing this paper, three
other papers appeared which discuss the CDF $\eegg + \Et$ 
event in various 
contexts~\cite{Mohapatra,LopezNanopoulos,Hisano}.

\section{Kinematics of the $\eegg + \Et$ event}
\label{kinematics-sec}
\indent

The kinematical requirements on the intermediate particles
involved in the $\eegg + \Et$ event are stringent, and for 
completeness we present a refined analysis based on the 
procedure outlined in Ref.~\cite{PRL}.  There are three basic 
possibilities for intermediate (s)particles; 
we will present these in
terms of LSP$=\NI$ interpretations, but the analysis 
is generic and could be applied to any set of intermediate
particles that satisfy the criteria below.  All decays
are assumed to occur close to the apparent vertex, 
which would be true of any LSP$=\NI$ interpretation.
The procedure we use to find kinematical constraints is
to begin with the information on the observed 
particles~\cite{event}, assume two- or three- body decays
as appropriate, randomly select unconstrained momentum components 
of the unobserved particles on both 
sides of the decay chain, and then reconstruct
the intermediate particle masses based on all possible
pairings of electrons and photons.  The masses of identical particles 
on both sides of the decay chain are required to be within 
2.5 GeV to `pass' the kinematic cut.  The net transverse
momentum in the event from adding both the observed particles and
the LSPs is assumed to be $|p_T| \lsim 20$ GeV\@.

\subsection{Selectron interpretation}
\indent

\begin{figure}
\vspace*{1cm}
\centering
\epsfxsize=4in
\hspace*{0in}
\epsffile{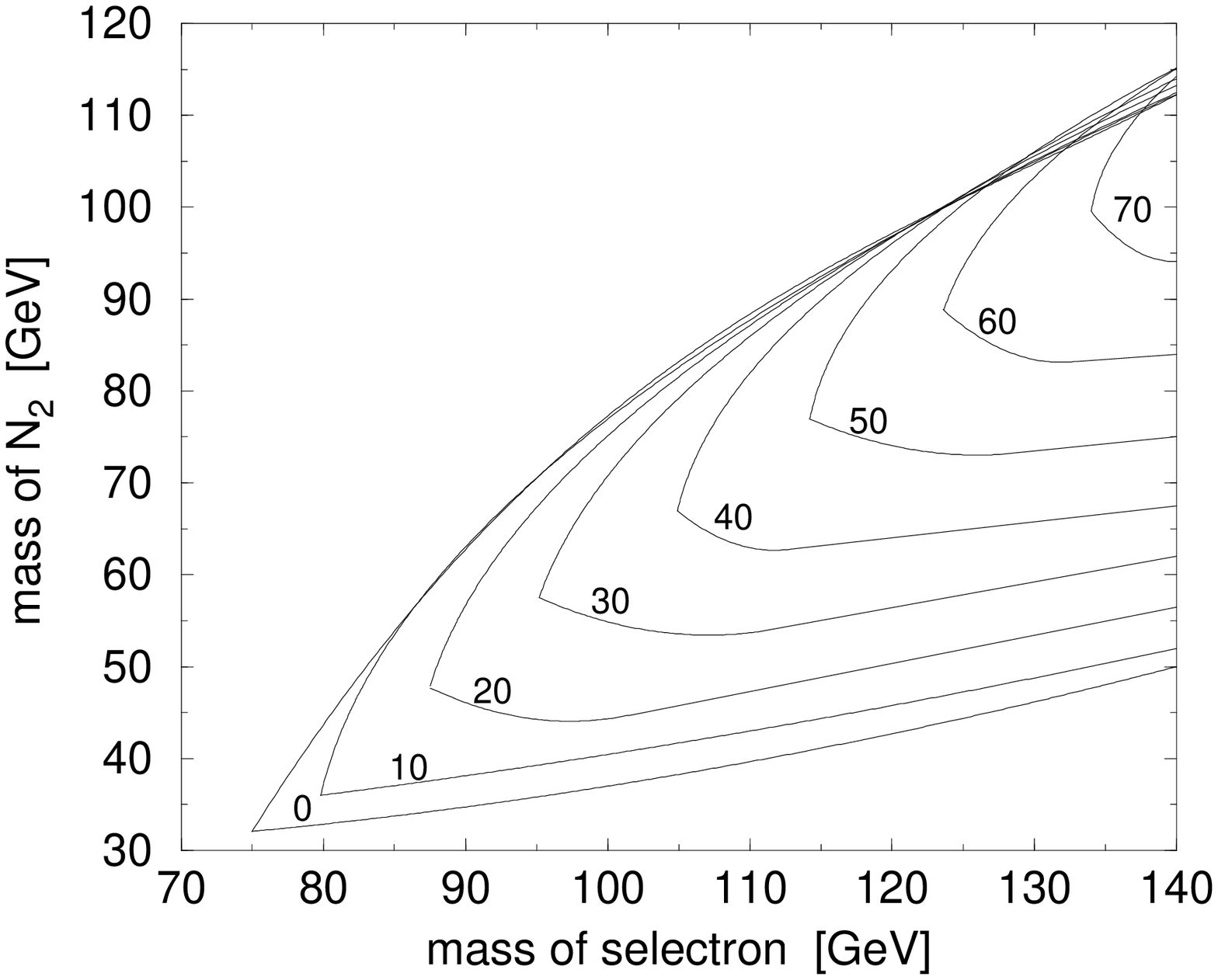}
\caption{The kinematically allowed region of the
$\eegg + \Et$ event in the $m_{\tilde e}$--$m_{\NII}$ plane
is shown for various values of $m_{\NI}$ in the 
selectron interpretation.  The allowed regions 
for $m_{\NI} = 0$, $10$, $20$, $30$, $40$, $50$, $60$, $70$~GeV
are to the inside and right of the indicated lines.
The allowed region for any given $m_{\NI}$ is roughly a subset of any
lower $m_{\NI}$, except for large values of $m_{\NII}$.
Since the lines are derived from the momenta of the $\eegg + \Et$ event,
they are only as precise as the associated measurement of momenta.
}
\label{kinematics-selectron-fig}
\end{figure}

The first possibility is selectron production 
$p\overline{p} \ra {\tilde e}^+{\tilde e}^- (+ X)$ and decay via
the 2-body mode  ${\tilde e} \ra e \NII$ followed by 
$\NII \ra \NI\ph$.  All sparticles are assumed to be on
mass shell.  The general result is summarized in
Fig.~\ref{kinematics-selectron-fig}, where the allowed regions 
in the $m_{\tilde e}$--$m_{\NII}$ plane are given for a series of 
maximum values of $m_{\NI}$.  The choice to cut off the graph at 
$m_{\tilde e} = 140$ GeV is motivated by a rough lower
limit on the selectron cross section, which will be
made precise in Sec.~\ref{model-building-results-subsec}.  
Since the electron and photon momenta have experimental 
uncertainties, the kinematic results that we derive from the 
event will have associated uncertainties.
Analytic forms of the constraints have been extracted and are 
presented in Table~\ref{kinematics-cuts-table}; a few observations 
are in order that will be useful in model building:

\begin{itemize}
\item[1.  ] $m_{\NI} \lsim (50,74)$ GeV, for $m_{\tilde e} < (115,137)$ GeV\@.
\item[2.  ] $m_{\NII} - m_{\NI} > 21$ GeV, this value increasing to $30$ GeV
      as $m_{\NI} \ra 0$ GeV\@.
\item[3.  ] $m_{\tilde e} - m_{\NII} \gsim 20$ GeV, this value increasing
      for decreasing $m_{\tilde e}$.
\item[4.  ] Given $m_{\NI} \gsim 33$ GeV, then $m_{\tilde e} \gsim 100$ GeV\@.
\item[5.  ] Only one pairing of electron and photon gives consistent
            kinematics for $m_{\tilde e} \lsim 125$ GeV\@.
\end{itemize}

The non-trivial mass differences that are required are 
not surprising, since all of the particles in the event
have large (transverse) energy.  We incorporate the
mass difference constraints as well as the
constraints on the ranges of $m_{\NI}$, $m_{\NII}$, 
and $m_{\tilde e}$ in our model building efforts.

\begin{table}
\renewcommand{\baselinestretch}{1.2}\small\normalsize
\begin{center}
\begin{tabular}{rcl} \hline
$m_{\tilde e}$ & $>$ & $75$ GeV \\
$m_{\NII}$     & $<$ & 
   $-0.00722 m_{\tilde e}^2 + 2.71 m_{\tilde e} - 122$ GeV
   \hspace{0.8cm} [$m_{\tilde e}$ in GeV] \\
$m_{\NII}$     & $>$ & $0.286 m_{\tilde e} + 10$ GeV \\
$m_{\NII}$     & $<$ & $0.167 m_{\NI} + 101$ GeV \\
$m_{\NII}$     & $>$ & $0.955 m_{\NI}      + 25$ GeV \\
$m_{\NI}$      & $<$ & $1.06  m_{\tilde e} - 71$ GeV \\ \hline
\end{tabular}
\end{center}
\caption{Kinematical constraints in the selectron interpretation.}
\label{kinematics-cuts-table}
\end{table}

\subsection{Chargino interpretation}
\label{char-kinematics-subsec}
\indent

The second possibility is chargino production 
$p\overline{p} \ra {\tilde C}_{i} {\tilde C}_{j}$ ($i,j = 1,2$), 
with three possible decay
chains:  3-body $\C \ra \NII e \nu_e$
(through an off-shell or possibly on-shell $W$),
2-body $\C \ra {\tilde e} \nu_e$ or $\C \ra {\tilde \nu_e} e$.
For either 2-body decay, the on-shell slepton proceeds
through another 2-body decay ${\tilde e}({\tilde \nu}_e) 
\ra e(\nu_e) \NII$, then the photons are obtained 
through $\NII \ra \NI\ph$.
Calculating consistent kinematics requires specifying the six
unknown momenta of the two neutrinos as well as the unknown
LSP momenta in the final state.
This is too complicated to delineate any rigorous exclusion 
regions using the randomized momenta procedure 
as in the selectron interpretation.
However, we have checked that it is possible to generate 
consistent kinematics for $m_{\C} > 95$ GeV, assuming
the 2-body decay $\C \ra {\tilde l} l'$ and that
all (s)particles are on-shell.
The rough regions where we were able to find
kinematical solutions have $m_{\tilde l} \gsim 60$, $75$
for slepton $=$ sneutrino, selectron.  
In addition, we found solutions with
$m_{\NII} \gsim 20$ GeV, $m_{\NII} - m_{\NI} \gsim 10$ GeV,
$m_{\C} \gsim {\rm max}[ 2.5 m_{\NII} - 95, 
  \> 1.5 m_{\NI} + 65, \> 95]$~GeV\@.
Thus a solution in the selectron interpretation
need not be a solution in the chargino interpretation, and
vice versa.

\subsection{Neutralino interpretation}
\indent

The third possibility is neutralino production, e.g.\
$p\overline{p} \ra \NII {\tilde N}_j$, where either of the
heavier neutralinos $j = 3$,$4$ decay 
as ${\tilde N}_j \ra l^+l^-\NII$, followed
by the usual $\NII \ra \NI\ph$.  This interpretation
contrasts with the first two by producing both leptons
from one side of the decay, however it is
calculable as in the selectron scenario (since the only
unknown final state momenta are the two neutralinos).
The invariant mass of the electron pair can be extracted
from the event $m_{e^+e^-} \sim 160$ GeV~\cite{event}, 
which implies the mass difference between $m_{{\tilde N}_j}$ 
and $m_{\NII}$ must also be greater than $160$ GeV\@.
This is almost certainly too high for a reasonable
Tevatron cross section while retaining a reasonable 
$m_{\NII}$ and proper neutralino mixing 
to have $\NII \ra \NI\ph$.  Further, in the particular
case where the branching ratio for the decay 
${\tilde N}_j \ra \NII Z$ is large, then a 
lepton pair from $Z \ra l^+l^-$ will always reconstruct
to to an invariant mass of about $m_Z$.  
Thus, a neutralino interpretation 
of the $\eegg + \Et$ event seems extremely unlikely, 
and we will not consider it further.

\subsection{Other interpretations}
\indent

Other supersymmetric interpretations with a neutralino LSP 
are in principle possible, and are based on variants 
of selectron production, chargino production
or neutralino production.  The differences lie in
the particular decay from which the electrons originate,
plus possibly other invisible phenomena (neutrinos).   
In all cases the photon is obtained from the 
decay $\NII \ra \NI \ph$, and as a consequence the photon
always appears in the last step of the decay chain.  
One example is stau production
$p\overline{p} \ra {\tilde \tau}^+ {\tilde \tau}^- (+ X)$ with
the decay ${\tilde \tau} \ra \tau \NII$, followed by 
$\tau \ra e \, (+ \> \nu_\tau \nu_e )$.  The total branching ratio 
is suppressed compared with selectron production by a 
factor $\BR(\tau \ra e \nu_\tau \nu_e)^2 \sim 0.03$, hence the
rate into $\eegg$ is much smaller than selectron production.
Another example is a variant of the selectron interpretation with
a chargino $\C$ that is lighter than the selectron, such that
the decay $\eL \ra \nu_e \C (\ra \NII e\nu_e)$ is dominant.
In this case it is probably not possible to have a large 
decay $\eL \ra \C \nu_e$, with 
both $\eL \ra {\tilde N}_{1,2} e$ suppressed.  Further,
$\C \ra \NI e \nu_e$ has to be suppressed with respect to
$\C \ra \NII e \nu_e$, which is difficult especially
in the presence of $\NII \ra \NI \ph$.
Finally, with four neutrinos carrying off invisible momentum 
it seems difficult to have a large probability for 
the high energy electrons required in the final state,
since the selectrons have to be light to have a large
$\eegg$ rate.

\section{Model building}
\label{model-building-section}
\indent

The kinematics of the event have illustrated two viable
sources of $\eegg + \Et$ events:  slepton production
or chargino production.  In either case, the essential
ingredient to getting photons is through the one-loop radiative
decay of neutralinos.  To proceed, we first
define the relevant parameters of the low-energy 
supersymmetric theory, including the chargino and neutralino
mass matrices.  This sets the stage for the discussion
of the radiative neutralino branching ratio.  We also discuss
the treatment of the squark, slepton and Higgs sectors
and the relevant mixings, as well as discussing the selectron
branching ratios.  Once the models have been 
constructed, we describe the constraints imposed on 
the parameters from experiment.

The main focus of this paper is on the selectron
interpretation and not the chargino interpretation, 
since it is made clear in Appendix~\ref{char-sec} that 
the chargino interpretation is difficult for many reasons.
However, in the following we have attempted to provide a 
general discussion of the model building, since 
radiative neutralino decay is required in both
interpretations.  

\subsection{Supersymmetric parameters}
\label{susy-parameters-subsection}
\indent

The chargino and neutralino tree-level masses and mixings are 
determined by specifying the gaugino soft masses $M_1$ and $M_2$, 
the ratio of the Higgs vacuum expectation values 
$\tbeta \equiv$~$\langle$${H}_2^0$$\rangle$/%
$\langle$${H}_1^0$$\rangle$
and the Higgs superfield mass parameter $\mu$.  The form
of the mass matrices is well known, but it will prove useful
in the discussion of the radiative branching ratio to have the 
expressions in the particular basis as follows.  Note that
we assume no relation between $M_1$ and $M_2$.

The chargino mass matrix 
in the ($-i {\tilde W}^\pm$, ${\tilde H}^\pm$) basis is
\begin{equation}
{\cal M}_{\Cpm} = \left( \begin{array}{cc} 
                  M_2                      &  \sqrt{2} M_W \sin \beta \\
                  \sqrt{2} M_W \cos \beta  &  \mu 
                  \end{array} \right), 
\label{char-mass-mat-eq}
\end{equation}
and can be diagonalized by a biunitary transformation 
$U^* {\cal M}_{\Cpm} V^{-1}$ 
to yield the masses and mixing matrices $U$, $V$
(as well as fixing the sign convention of $\mu$, 
consistent with Ref.~\cite{HaberKane}).  The chargino masses 
can be found from the analytic expression
\begin{eqnarray}
m_{{\tilde C}_{1,2}}^2 &=& 
    \frac{1}{2} \bigg\{ M_2^2 + \mu^2 + 2 M_W^2 \nonumber \\
 & & \qquad \mp \sqrt{ (M_2^2 - \mu^2)^2 + 4 M_W^4 \cos^2 2 \beta
            + 4 M_W^2 (M_2^2 + \mu^2 + 2 M_2 \mu \sin 2 \beta)
            } \bigg\}.
\label{char-mass-eq}
\end{eqnarray}
The neutralino mass matrix in the ($-i \phino$, $-i \Zino$, 
$\Ha$, $\Hb$) basis is
\begin{equation}
{\cal M}_{\tilde N} = \left( \begin{array}{cccc}
  M_1 \cthwq + M_2 \sthwq  & (M_2 - M_1) \sthw\cthw   & 0   & 0             \\ 
 (M_2 - M_1)\sthw\cthw     & M_1 \sthwq + M_2 \cthwq  & M_Z & 0             \\ 
    0                      &      M_Z      &  \mu \sindb    &  -\mu \cosdb  \\ 
    0                      &      0        &  -\mu \cosdb   &  -\mu \sindb  \\ 
\end{array} \right),
\label{neut-mass-mat-eq}
\end{equation}
and can be diagonalized by a unitary transformation 
$N^* {\cal M}_{\tilde N} N^{-1}$ to yield the four neutralino mass 
eigenvalues $\epsilon_i m_{\Ni}$ 
and the mixing matrix $N$ that we assume to be real and 
orthogonal (exact expressions for the mixings and masses
can be found in~\cite{NeutAnalytic,BBO2}).
The sign of the neutralino mass eigenvalue $\epsilon_i$ 
enters the supersymmetric Feynman rules, while the physical masses
$m_{\Ni}$ are always positive with the ordering
$0 \le m_{\NI} \le m_{\NII} \le m_{\NIII} \le m_{\NIIII}$.
The ($\phino$, $\Zino$) basis is related to the 
(${\tilde B}$, ${\tilde W}^3$) basis through
\begin{equation}
\left( \begin{array}{c} 
       \phino \\
       \Zino 
       \end{array} \right) = 
\left( \begin{array}{cc}
        \cthw & \sthw \\
        -\sthw & \cthw 
        \end{array} \right)
\left( \begin{array}{c}
       {\tilde B} \\
       {\tilde W}^3 
       \end{array} \right),
\end{equation}
and the ($\Ha$, $\Hb$) basis is related to 
the (${\tilde H}_1^0$, ${\tilde H}_2^0$) basis through
\begin{equation}
\left( \begin{array}{c} 
       \Ha \\
       \Hb 
       \end{array} \right) = 
\left( \begin{array}{cc}
        \cos \beta  & -\sin \beta \\
        \sin \beta & \cos \beta  
        \end{array} \right)
\left( \begin{array}{c}
       {\tilde H}_1^0 \\
       {\tilde H}_2^0 
       \end{array} \right).
\end{equation}
Our notation follows Refs.~\cite{HaberKane,BartlNeutMatrix}, with
${\tilde H}_1^0$ and ${\tilde H}_2^0$ coupling to the down- and up-type
fermions respectively.  The production cross sections for charginos 
and neutralinos at LEP and at the Tevatron involve graphs with 
$s$-channel gauge boson exchange and $t$-channel slepton or squark
exchange.  In Table~\ref{nn_cc_nc-table}, we itemize the dependence
of each chargino/neutralino cross section on the squark or slepton mass.

\begin{table}
\renewcommand{\baselinestretch}{1.2}\small\normalsize
\begin{center}
\begin{tabular}{ccc} \hline\hline
Process     &           LEP           &  Tevatron \\ \hline
$\Ni\Nj$    &  $m_{\eL}$, $m_{\eR}$   &  $m_{\uL}$, $m_{\dL}$, 
                                         $m_{\uR}$, $m_{\dR}$ \\
$\Ci\Cjmp$  &  $m_{\veL}$             &  $m_{\uL}$, $m_{\dL}$ \\
$\Ni\Cj$    &    -                    &  $m_{\uL}$, $m_{\dL}$ \\ \hline\hline
\end{tabular}
\end{center}
\caption{Chargino and neutralino cross sections at LEP and Tevatron
depend on $M_1$, $M_2$, $\tbeta$, $\mu$ and the particular 
superpartner masses as above.  (The Tevatron 
cross sections also depend on the second family masses, 
but these contributions are generally suppressed by Cabbibo 
mixing and a small parton distribution $f_{q|p}$ in the proton.)
}
\label{nn_cc_nc-table}
\end{table}

The gluino does not enter phenomenology directly associated
with the $\eegg + \Et$ event.  Its tree-level mass is given 
by the soft mass parameter $M_3$ that is unconstrained without 
gaugino mass unification.  
There need be no relation between $M_1$, $M_2$, and $M_3$, 
and we do not assume one.  However, one could imagine that
the non-Abelian masses $M_2$, $M_3$ are equal at the unification 
scale, with the U(1) mass $M_1$ related to them in a more subtle way.
Ref.~\cite{KaneMrennaGluino}
has suggested that the gluino may play a dramatic role at the 
Tevatron, if the lightest stop $\tI$ has a 
mass ${\cal O}(50)$ GeV\@.  However, for the primary purposes 
of this paper we can focus on phenomenology that is independent 
of the gluino.  In Sec.~\ref{light-stop-sec} we elaborate 
on the possibility of models that can generate 
an $\eegg + \Et$ event with the additional assumption 
of a light stop.

The slepton sector is defined by the masses $m_{\lL}$ and $m_{\lR}$,
with $m_{\vL}$ related by the SU(2)$_L$ sum rule
\begin{equation}
m_{\vL}^2 = m_{\lL}^2 - M_W^2 | \cos 2 \beta |,
\label{sum-rule-eq}
\end{equation}
for $\tbeta > 1$, and the couplings to gauge bosons and gauginos
fixed by the SM gauge group.  Slepton
production cross sections at the Tevatron are given 
in Refs.~\cite{DEQ,BaerSlepton,PRL}, and depend {\em only}\/ on
the mass of the slepton.  We assume slepton mass degeneracy 
(motivated by the absence of lepton flavor changing decays), 
although it is not required by the theory nor the $\eegg + \Et$ 
event.  Where necessary, we remark on the effect of removing
this assumption on associated phenomenology.  We also assume
$L$--$R$ mixing in the slepton sector can be neglected.

The squark sector in our model building is defined 
for simplicity by a common 
squark mass $m_{\tilde q}$, the stop masses $m_{\tI}$, 
$m_{{\tilde t}_2}$ and the stop mixing angle $\theta_{\tilde t}$.  
In this way we achieve a useful reduction of parameter 
space through $m_{\tilde q} = m_{\uL} = m_{\dL} = m_{\uR} 
= m_{\dR} = \ldots$, and we further assume for simplicity
$m_{{\tilde t}_2} = m_{\tilde q}$.  
These assumptions can be removed if data becomes
sensitive to them.
The stop mass eigenstates
are defined by
\begin{equation}
\left( \begin{array}{c} \tI \\ 
                        \tII \\
       \end{array} \right) = 
\left( \begin{array}{cc} \cos \theta_{\tilde t}  & \sin \theta_{\tilde t} \\
                         -\sin \theta_{\tilde t} & \cos \theta_{\tilde t} \\
       \end{array} \right) 
\left( \begin{array}{c} \tL \\
                       \tR \\
       \end{array} \right) 
\end{equation}
with the stop trilinear
coupling $A_t$ (and the soft masses $m_{\tilde Q}$, 
$m_{\tR}$)
uniquely determined by $m_{{\tilde t}_{1,2}}$
and the mixing angle $\theta_{\tilde t}$, for a given
$\mu$ and $\tbeta$.  We assume all other
$L$--$R$ squark mixing can be neglected.

The Higgs sector is determined from $\tbeta$, the neutral CP-odd Higgs
mass $m_A$, and higher order corrections~\cite{GoodHiggsRef,ERZ}.  
We include one-loop corrections from stops~\cite{ERZ}, 
and neglect all other contributions. 
In this framework we calculate the charged Higgs mass $m_{H^\pm}$,
the neutral CP-even Higgs masses $m_h$, $m_H$ and the mixing angle
$\alpha$ from the above parameters.  The Higgs sector enters 
the radiative neutralino decay through the charged Higgs boson, 
and the branching ratios for the heavier superpartners into 
one or more of $h$, $A$, $H$, or $H^\pm$ (neglecting off-shell
Higgs exchange in 3-body $\C,\N \ra \C,\N f\overline{f}$ decays).  

\subsection{Radiative decay of neutralinos}
\label{radiative-decay-subsection}
\indent

The radiative decay of neutralinos has been well 
studied~\cite{Komatsu,HaberWyler,AmbrosMele1,AmbrosMele2},
and it suffices to review the mechanism that enhances the radiative 
branching ratio with respect to the traditional 3-body 
$\NII \ra \NI f\overline{f}$ decays, as pertaining to 
the $\eegg + \Et$ event.  
We exclusively discuss $\NII \ra \NI\ph$, since heavier
neutralinos always have sizeable tree-level branching 
ratios into 2- or 3-body channels,
causing the radiative branching ratio to be
negligible.

There exists both a kinematical and a dynamical mechanism that can give
an enhancement of the radiative neutralino 
decay~\cite{HaberWyler,AmbrosMele2}.  
The kinematic enhancement can only occur when the
mass difference $m_{\NII}-m_{\NI}$ is small ${\cal O}(10)$~GeV, 
so that other decay modes are closed or suppressed.   
However, the kinematics in the selectron interpretation 
enforce $m_{\NII} - m_{\NI} > 21$ GeV
by Observation 2, and so a kinematic enhancement 
of the radiative branching ratio is not crucial for
our purposes (although see Sec.~\ref{model-building-results-subsec}
for exceptions).

The dynamic enhancement of the radiative decay occurs as follows.
First, examine the limit when $\tbeta \ra 1$ and 
$(M_1 - M_2) \ra 0$~\cite{BartlNeutMatrix}; the neutralino mass
matrix (already written in a suggestive form
in Eq.\ (\ref{neut-mass-mat-eq})) becomes particularly
simple,
\begin{equation}
{\cal M}_{\tilde N} = \left( \begin{array}{cccc}
  M_2  &    0   &   0   &    0   \\ 
   0   &   M_2  &  M_Z  &    0   \\ 
   0   &   M_Z  &  \mu  &    0   \\ 
   0   &    0   &   0   &  -\mu  \\ 
\end{array} \right) \quad {\rm for} \quad 
\left\{ \begin{array}{c}
        \tbeta = 1 \\
        M_1 = M_2
        \end{array} \right. .
\label{neut-mass-mat-special-eq}
\end{equation}
In this limit two neutralinos become pure photino ($\phino$) 
and Higgsino ($\Hb$) states, with masses $M_2$ and $|\mu|$ 
respectively.  The other two neutralinos are mixtures of 
$\Zino$--$\Ha$, with masses 
\begin{equation}
m_{\Zino-\Ha} = 
\frac{1}{2} \left| M_2 + \mu \pm \sqrt{ (M_2 - \mu)^2 + 4 M_Z^2 } \right|.
\label{heavy-neut-masses-limit-eq}
\end{equation}
For pure $\phino$ and $\Hb$ states, the
tree-level couplings $\phino \Hb Z$, 
$\phino \Hb h (A)$, and $\Hb {\tilde f} f$
(in the limit $m_f \ra 0$) go to zero, leaving
the one-loop `effective' coupling $\phino \Hb \ph$
dominant.  Thus, by associating ${\tilde N}_{1,2}$ with $\phino$,
$\Hb$, then the one loop decay $\NII \ra \NI\ph$ is dominant.
One consequence of requiring the two lightest neutralinos to 
be either of the states $\phino$ or $\Hb$ (hence the heavier 
two neutralino masses are given by 
Eq.~(\ref{heavy-neut-masses-limit-eq})) is that the required
mass ordering $m_{{\tilde N}_{1,2}} < m_{{\tilde N}_{3,4}}$ implies
\begin{equation}
M_1 (= M_2), \; |\mu| < 
\frac{1}{2} \left| M_2 + \mu \pm \sqrt{ (M_2 - \mu)^2 + 4 M_Z^2 } \right|.
\end{equation}
See Ref.~\cite{AmbrosMele2} for a more comprehensive treatment
of this issue.  What is not determined by requiring a 
large radiative branching ratio by this mechanism
is which one of the two lightest neutralinos 
is the photino or Higgsino.

\begin{figure}
\centerline{
\hfill 
\epsfxsize=0.55\textwidth
\epsffile{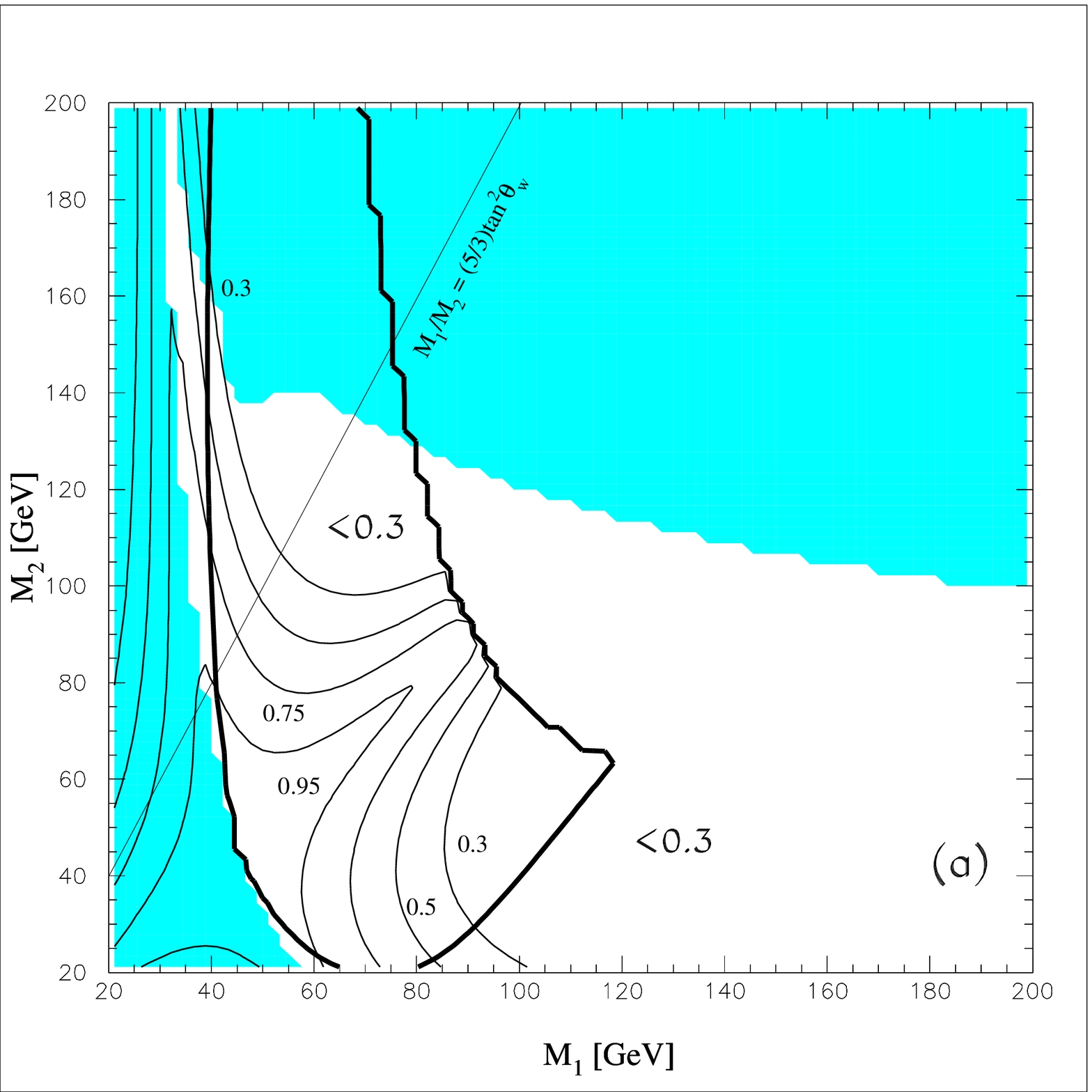}
\hfill 
\epsfxsize=0.55\textwidth 
\epsffile{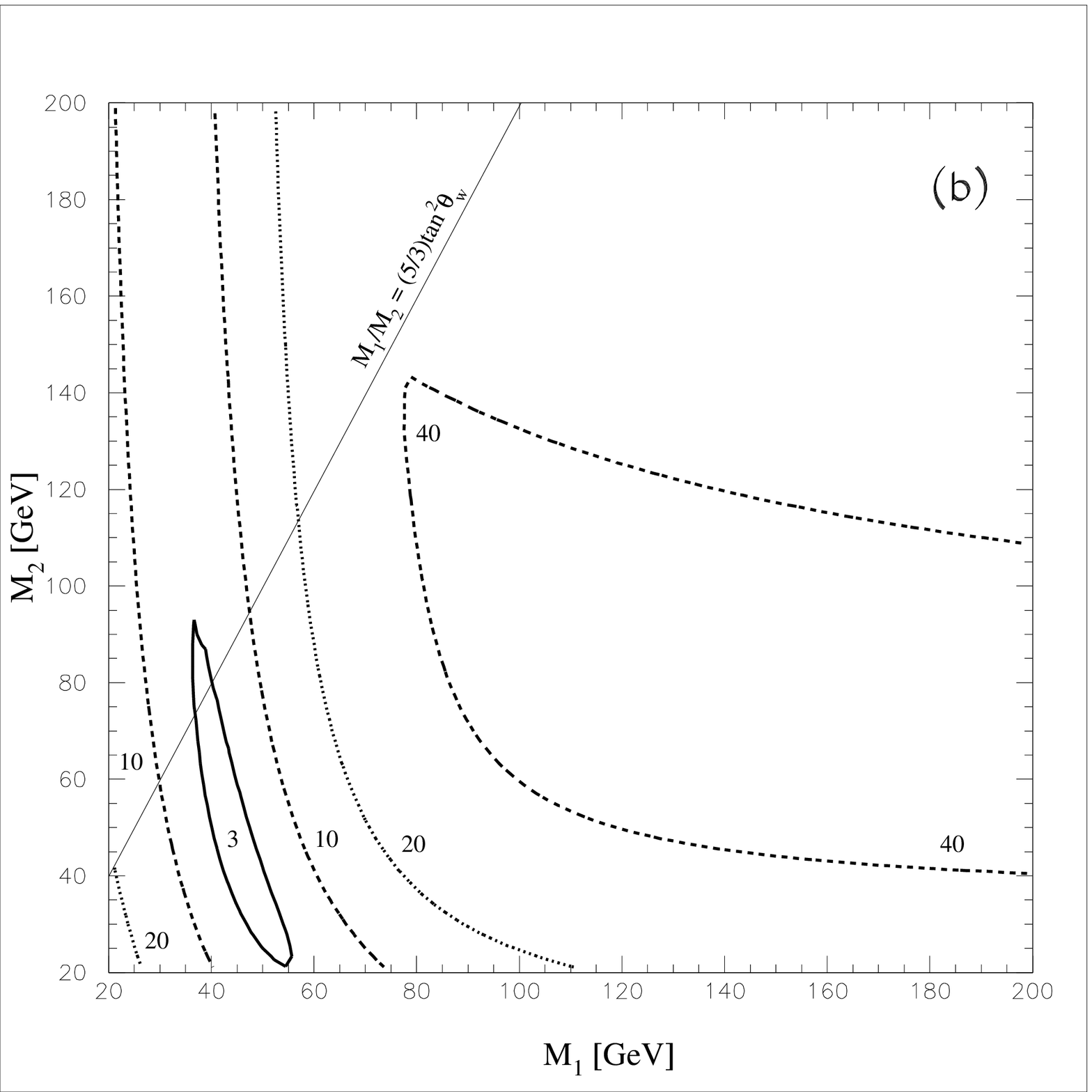}
\hfill } 
\caption{(a) Contour plot for the branching ratio 
of the radiative neutralino decay 
$\NII\ra\NI\gamma$ in the $M_1$--$M_2$ plane for the case 
$\tbeta=1.2$, $\mu = -45$ GeV, $m_{\tilde{e}_{L,R}} = 110$ GeV,
$m_{\tilde q} = m_{{\tilde t}_{1,2}} = 500$ GeV, 
and $m_A = 400$ GeV\@.  The $\BR( \NII \ra \NI\ph ) = 0.9$,
$0.75$, $0.5$, $0.3$ 
levels are shown and labeled.  The LEP excluded region is shaded.  
The solid thick line outlines the region where 
$\langle\NI|\tilde{H}^0_b\rangle^2\langle\NII | 
\tilde{\gamma}\rangle^2 > 0.7$.  
(b) Contour plot in the same plane with the parameters
above, showing the mass difference of the two lightest neutralinos 
in GeV\@.  This figure is a result of the general 
radiative neutralino decay analysis of 
Ref.~\protect\cite{AmbrosMele2}.}  
\label{radiative-M1-M2-fig}
\end{figure}

The extent to which a large radiative branching ratio is possible
in general (and in particular through the dynamical mechanism
without the exact relations above) can be evaluated
semi-analytically and numerically~\cite{AmbrosMele2}.  
As an example, Fig.~\ref{radiative-M1-M2-fig}(a) shows contours
of the branching ratio of $\NII \ra \NI\ph$ in the $M_1$--$M_2$ plane, 
for $\mu = -45$ GeV, $m_{\eL} = m_{\eR} = 110$ GeV, 
$m_A = 400$ GeV, $\tbeta = 1.2$, and 
all squarks heavy $m_{\tI} = m_{\tilde q} = 500$ GeV\@.
The thick solid line bounding the region defined by
$\langle\NI|\tilde{H}^0_b\rangle^2 
 \langle\NII | \tilde{\gamma}\rangle^2 > 0.7$ anticipates the 
constraint on selectron decay from the $\eegg + \Et$ event 
(see Sec.~\ref{slepton-decay-subsection} below).
Contours in the mass difference $m_{\NII}-m_{\NI} > 3$, 
$10$, $20$, $40$~GeV are shown in
Fig.~\ref{radiative-M1-M2-fig}(b).  
Since the selectron interpretation requires
a large mass difference $m_{\NII} - m_{\NI} > 21$~GeV, only a 
fairly small region of parameter space remains satisfying the 
constraint of a large radiative neutralino branching ratio.
For example, the region bounded by 
$\BR( \NII \ra \NI\ph ) > 0.5$, $m_{\NII} - m_{\NI} > 20$~GeV,
and the LEP exclusion region
is characterized by roughly $0.6 < M_2/M_1 < 1.5$ for
$60 < M_1 < 90$~GeV, $45 < M_2 < 90$~GeV, where the constraints
on $M_2/M_1$ are stronger for larger values of $M_1$, $M_2$.
Of course this example only applies to the choice of $\mu$,
$\tbeta$, $m_{\tilde e}$, $m_{\tilde q}$, $m_A$ values as 
above, but it gives a 
reasonable illustration of the constraints.  The
region with a large radiative neutralino decay centered 
on the line $M_1 = M_2$ persists as $|\mu|$ is increased
or decreased (the region shifts up or down the $M_1 = M_2$ line), 
but tends to shrink (and eventually disappear) as $\tbeta$ 
is increased or the squark or slepton masses are decreased.

\subsection{Slepton decay}
\label{slepton-decay-subsection}
\indent

In the selectron interpretation, the branching ratio of the
selectrons ${\tilde e} \ra e \NII$ is crucial to 
produce an $\eegg + \Et$ event.  In general,
sleptons couple to the gauginos through the usual supersymmetrized
gauge interactions, and also to the Higgsinos through the Yukawa
couplings.  The Yukawa couplings $\lambda_{\tilde l} \sim m_l/M_W$
are strongly suppressed by small lepton masses, and
for our purposes can be neglected.  Since the radiative branching
ratio $\NII \ra \NI\ph$ requires one of ${\tilde N}_{1,2}$ to
be mostly a photino and the other mostly a Higgsino, then
the requirement that the selectron decays as 
${\tilde e} \ra e \NII$ implies the photino-Higgsino 
content of the neutralinos is unique and determined
\begin{eqnarray}
\langle \NI  | \Hb \rangle^2 &\approx& 1 \nonumber \\
\langle \NII | \> \phino \> \rangle^2 &\approx& 1 .
\label{neutralino-content-eq}
\end{eqnarray}
Based on Sec.~\ref{radiative-decay-subsection}, 
this implies $|\mu| < M_1 \; (= M_2)$, 
in the limit of pure states.

If the $\eegg + \Et$ event is due to $\eL\eL$ production, 
one must also consider the branching fraction of $\eL$ 
to charginos if kinematically accessible.
In the kinematics of the selectron interpretation
no such decay was considered, and naively it would seem
possible to suppress this decay through a judicious choice of
chargino mixings.  However, it is also possible that
$\eL$ production occurs with the selectron decay 
$\eL \ra \CI \nu_e$, then the decay $\CI \ra e \NII$.  
In the $\tbeta = 1$ limit (with neutralinos pure states) 
the masses of the charginos simplifies considerably 
from Eq.~(\ref{char-mass-eq}) to
\begin{equation}
m_{{\tilde C}_{1,2}} =  \frac{1}{2} \left| M_2 + \mu
                        \pm \sqrt{ (M_2 - \mu)^2 + 4 M_W^2 } \right| .
\label{char-mass-tanbeta-limit-eq}
\end{equation}
This expression is the same as Eq.~(\ref{heavy-neut-masses-limit-eq})
with $M_Z \ra M_W$, and shows that the chargino masses are
directly correlated with the heavier two neutralino masses.  
It is a simple matter to show that 
$m_{\CI} > m_{\NII}$ is always true
(in the $\tbeta = 1$, $M_1 = M_2$ limit),
while the coupling of $\CI$ to $\NII$~$(= \phino)$
and $\NI$~$(= \Hb)$ is dependent 
on the gaugino-Higgsino mixings of the chargino.
The $\eL$--$\C$--$\nu_e$ couplings are also proportional to the 
gaugino component of $\C$
and so a full numerical calculation is necessary to determine
the relative size of the branching fractions.  This will
be presented in Sec.~\ref{chargino-BR-subsec}.

\subsection{Constraints from LEP}
\label{LEP-constraints-subsec}
\indent

Throughout our analysis, we applied the most updated limits 
on the supersymmetric parameters and bounds on superpartner
masses coming from searches at LEP1,
as well as the more recent run with $\sqrt{s} = 130.3$
and $136.3$~GeV (collectively denoted `LEP130-136') where
integrated luminosities of about $2.8$ and $2.3$~pb$^{-1}$
were accumulated~\cite{LEP130}.  We also show the combined 
effect of the LEP limits and kinematical constraints on the 
selectron and light neutralino masses 
in the selectron interpretation of the $\eegg + \Et$ event,
and the derived ranges of $\mu$, $M_1$ and $M_2$ values. 
The somewhat conservative LEP1 bounds we imposed 
are~\cite{L3,Feng}: 
\begin{eqnarray}
\BR_{\rm invisible}(Z \ra {\rm SUSY}) &<& 2.3 \times 10^{-3} \nonumber \\
\Delta \Gamma_{\rm tot} (Z \ra {\rm SUSY}) &<& 23 \;\> {\rm MeV}  \\
\BR(Z \rightarrow \tilde{N}_1\tilde{N}_2) &<& 1.2 \times 10^{-5} \nonumber \\
\BR(Z \rightarrow \tilde{N}_2\tilde{N}_2) &<& 3.5 \times 10^{-5} . \nonumber
\end{eqnarray}
The evaluation of the supersymmetric contribution 
to the invisible Z width included
not only the contribution from the direct LSP production
$Z\rightarrow\tilde{N}_1\tilde{N}_1$, but also the
contribution from other channels
$Z\rightarrow \tilde{N}_i (\rightarrow \tilde{N}_1 \nu\bar{\nu}) 
\tilde{N}_j (\rightarrow \tilde{N}_1 \nu \bar{\nu})$. 
These contributions were then subtracted when calculating the 
supersymmetric contributions to the visible Z width. 

The constraints we applied at LEP130-136 are
\begin{eqnarray}
\sigma(e^+e^- \ra \; \mbox{visible SUSY}) &<& 1.8 \;\> {\rm pb} \;\;
    {\rm for} \;\> \sqrt{s} = 130.3 \;\> {\rm GeV} \nonumber \\
\sigma(e^+e^- \ra \; \mbox{visible SUSY}) &<& 2.2 \;\> {\rm pb} \;\; 
    {\rm for} \;\> \sqrt{s} = 136.3 \;\> {\rm GeV} 
\end{eqnarray}
corresponding to the 5 visible event level (before detector cuts) 
for each of the two runs~\cite{LEP130}.
A few remarks on the calculation of 
the expected total visible supersymmetric cross section are
in order.  First, we considered only the contribution 
from chargino/neutralino production, since charged sleptons 
relevant to the $\eegg + \Et$ event need to be heavier than 
$75$~GeV just to satisfy the kinematics 
(see Table~\ref{kinematics-cuts-table}).  We require squarks 
to be heavier than can be produced at LEP, except possibly
a light stop whose production cross section is always 
too small to see any events at LEP130-136 with the data
sample collected.  The total visible supersymmetric cross section
obviously does not include processes like 
$e^+e^- \ra \NI\NI$, and $e^+e^- \ra \Ni\Nj$ when both
$\tilde{N}_{i,j} \rightarrow \tilde{N}_1 \nu \bar{\nu}$.
This was achieved by doing a complete calculation of the 
branching ratios for chargino/neutralino decays for every
model.  To ensure the visibility of the signal, we also
required large enough phase space in the decay
of the produced $\Ni$, $\Ci$, which in
practice implied the mass 
difference $m_{\CI, \NII} - m_{\NI} > 10$~GeV, 
in accord with~\cite{LEP130}. 

The following observations are useful to understand 
in some detail how the LEP constraints affect our analysis 
in a general low energy supersymmetric framework 
(without assuming any relation between $M_1$ and $M_2$).  
Combining the bounds arising from 
neutralino searches at LEP with the need for
a next-to-lightest neutralino $m_{\NII} > 30$~GeV from
the $\eegg + \Et$ event kinematics (see Sec.~\ref{kinematics-sec}),
one finds the ``light Higgsino-gaugino window'' 
with $M_1$, $M_2$, $|\mu| \ll M_Z$ and
$\tbeta \approx 1$~\cite{Feng} is excluded.  
This also implies $|\mu| \gsim 33$~GeV,
at least for small $\tbeta$.
Further, given the light Higgsino-gaugino window is
excluded for our purposes, only $\mu < 0$ survives LEP 
constraints such that a large radiative neutralino 
branching ratio is present~\cite{AmbrosMele2}, 
thus we are left with $\mu < -33$~GeV\@.
For $\tbeta \gsim 1.3$ either 
the LEP chargino mass bound or the direct search for 
neutralinos begin to exclude regions with small negative
$\mu$, irrespective of $M_1$ and $M_2$ values.
Given a value of $\mu$, one can find rough regions
in the \mbox{$M_1$--$M_2$} plane that are allowed by LEP constraints,
generally independent of $\tbeta$.  In our framework, the constraints 
we listed above exclude $M_1 \lsim 30$~GeV
and, for instance, when $\mu = -45$~GeV then 
$M_1 \lsim 55$~GeV is not allowed if $M_2 \lsim 20$~GeV\@.
The region in $M_1$--$M_2$
space excluded by LEP is indicated in Fig.~\ref{radiative-M1-M2-fig}
for $\mu = -45$ GeV, etc.  Notice that since
the $\eegg + \Et$ event requires a suitable 
slepton decay, then the neutralino
contents in Eq.~(\ref{neutralino-content-eq}) can exclude
a comparable region (see Sec.~\ref{slepton-decay-subsection}, 
and in particular Fig.~\ref{radiative-M1-M2-fig}).
In contrast, the requirement $(m_{\NII}-m_{\NI}) > 21$~GeV 
of Observation 2 in Sec.~\ref{kinematics-sec}
combined with the LEP constraints
effectively sets a minimum suitable value of $M_1$ around 
$52$~GeV for any values of the other parameters. 
Only weaker bounds on $M_2$ can be identified in 
a similar way. 

In addition to the constraints from chargino and neutralino 
production, we also imposed 
\begin{equation}  
m_h  \; > \;  \left\{ \begin{array}{l}
                      44 \\
                      58.4 \> \sin^2(\beta-\alpha)
                      \end{array} \right. \quad {\rm GeV} ,
\end{equation}
on our models from LEP constraints.  Since the inputs to our 
model building to calculate the Higgs sector include $m_A$ and 
$\tbeta$, the above mass bounds impose a constraint on $m_A$ and 
higher order corrections from the stop sector.  This will be 
important for the discussion about models with a light stop in 
Sec.~\ref{light-stop-sec}.  
Small $\tbeta$ also suffers from possible non-perturbativity 
constraints, that have been discussed recently in e.g. 
Ref.~\cite{Feng} for the light Higgsino-gaugino 
window that requires small $\tbeta$.  However, the constraint 
is relatively weak ($\tbeta \gsim 1.2$), since as we shall 
see the allowed region of $\tbeta$ extends up to 
$\tbeta \sim 2.0 \ra 2.8$.

\section{Numerical results -- selectron interpretation}
\label{numerical-results-section}
\indent

To ensure a large branching ratio for the decay $\NII \ra \NI\ph$, 
pure photino and Higgsino
lightest neutralinos are sufficient, but not necessary conditions.  The
extent of the allowable impurity determines the character
of the models, but that is by no means the only degree
of freedom.  As we have seen, the branching ratios of the sleptons 
are also determined by the gaugino-Higgsino content of the neutralinos 
and charginos.  Further, the allowed sets of masses must satisfy
the $\eegg + \Et$ event kinematics, and proper experimental constraints 
are not trivial mass exclusions, etc.  What we present here are complete
low energy models constructed using the framework built up
in Sec.~\ref{model-building-section} using a randomized parameter 
selection scheme~\cite{oldwork}, and imposing 
all of the above constraints.

\subsection{Preliminaries}
\indent

Interpreting one event as a cross section is a tenuous procedure,
although some general methodology can be applied.  First, 
we establish a minimum threshold in the Tevatron selectron 
cross section times branching ratio into two electrons and 
two photons, 
\begin{equation}
\sigma \times \BR^2 \equiv \sigma( p\overline{p} \ra {\tilde e}^+{\tilde e}^- )
    \times \left[ \BR( {\tilde e} \ra \NII e) \BR( \NII \ra \NI\ph ) \right]^2 
    > {\cal A},
\end{equation}
where ${\cal A} \equiv (\sigma \times \BR^2)|_{\rm min}$ is the 
minimum threshold value.
Since the choice of the threshold ${\cal A}$ is somewhat arbitrary, we 
show the effect of increasing the threshold from $5$ to $7.5$
to $10$~fb to give at least some indication as to how sensitive
the constraints are to the value.  Imposing
${\cal A} = 20$~fb excludes all of our models, 
so there is a non-trivial importance of the precise numerical 
value of the threshold for phenomenology.

The quantity $\sigma \times \BR^2$ used in the 
general analysis does not include detector cuts, 
but we have simulated particular models to get indicative 
efficiencies (see Sec.~\ref{tevatron-subsec}).
For a detection efficiency
of $0.2$, the lowest threshold cut ${\cal A} = 5$~fb corresponds 
to assuming a cut on the effective $\eegg$ rate of 
$s = \sigma \times \BR^2 \times {\rm EFF} = 1$~fb or
$1/10$ of an event.
Given an expected number of events $s$, the probability of 
observing exactly $n$ events is from Poisson statistics
\begin{equation}
P = \frac{e^{-s} s^n}{n!} .
\end{equation}
For $s = 0.1$ corresponding to $1$~fb 
cross section at the Tevatron, one still has a $9\%$ chance of seeing 
exactly one event. 

The results are presented assuming a branching ratio into
only one family, although it is straightforward to compute
the total two lepton plus two photon rate including smuon and/or 
stau production.  The effect is of course to increase our
calculated rate by a factor of $2$ or $3$.  (Our results
remain unchanged if the threshold ${\cal A}$ is increased
by the same factor.)  Note that including more
than one family is of course crucially dependent
on the assumption of slepton mass degeneracy.

In the selectron interpretation there is no a priori 
requirement of having $\eL$ or $\eR$ production.  
We consider three cases:  A selectron interpretation from
$\eL$ production, where the kinematics of the $\eegg + \Et$ event
must be satisfied for $m_{\eL}$, but must {\em not}\/ be 
satisfied for $m_{\eR}$.  In this way, $\eR\eR$ production 
can still give an $\eegg$ signal but the kinematics are 
not consistent with the $\eegg + \Et$ event; 
hence only the rate from $\eL\eL$
production ought to be considered.
Second, the opposite scenario
with $\eR$ production where the kinematics must be
satisfied for $m_{\eR}$ but not for $m_{\eL}$.
Finally, we consider a set of models with the
simultaneous $\eL\eL$ and $\eR\eR$ production 
(denoted `$\eL + \eR$ models'), where
the kinematics are satisfied for {\em both}\/ 
$m_{\eL}$ and $m_{\eR}$.  The threshold ${\cal A}$ 
is applied as follows,
\begin{eqnarray}
\sigma_L \times \BR_L^2 \> &>& \> {\cal A} \;\; 
    {\rm for} \;\; \eL \;\> {\rm models} \nonumber \\
\sigma_R \times \BR_R^2 \> &>& \> {\cal A} \;\;
    {\rm for} \;\; \eR \;\> {\rm models} \\
\sigma_L \times \BR_L^2 + \sigma_R \times \BR_R^2 \> &>& \> {\cal A} \;\; 
    {\rm for} \;\; \eL + \eR \;\> {\rm models} , \nonumber
\end{eqnarray}
where $\sigma_{L,R} \equiv \sigma( p\overline{p} \ra {\tilde e}^+_{L,R}
{\tilde e}^-_{L,R})$ and $\BR_{L,R} \equiv
\BR( {\tilde e}_{L,R} \ra \NII e) \BR( \NII \ra \NI\ph )$.
The case of $\eL + \eR$ models assumes that the
contributions to the $\eegg$ cross section from $\eL$ 
and $\eR$ production can be summed,
hence the requirement that the kinematics of the event 
is satisfied for both contributions.
Further, for $\eL + \eR$ models we enforce $\sigma_{L,R} > 1$~fb
to avoid the difficulty of one of
$\sigma_{L,R} \times \BR_{L,R}^2$ being arbitrarily
close, but below the threshold ${\cal A}$ while the 
other contribution can be very small.  In such a case
the model could still pass the cut on
the sum $\sigma_L \times \BR_L^2 + \sigma_R \times 
\BR_R^2 > {\cal A}$, but would be on the borderline 
of classification as either an $\eL$, $\eR$, or $\eL + \eR$ model.  
We will show that this loose requirement on the 
cross section does not affect our results.
Finally, note that since $\BR(\NII \ra \NI \ph)$ depends in
general on both selectron masses $m_{\eL}$ and $m_{\eR}$, then
$\eL$, $\eR$ and $\eL + \eR$ models can each be considered 
a distinct class of models. 

We impose no restriction on the squared branching ratio
$\BR^2$ (unlike Ref.~\cite{PRL}), 
nor any restriction on associated phenomenology.
In practice, the cut on $\sigma \times \BR^2$ does
provide an effective lower limit on the branching ratio 
based on the largest allowed cross section $\sigma$, 
obtained from the smallest selectron mass allowed
from $\eegg + \Et$ event kinematics.
This avoids generating a disproportionate 
number of non-$\eegg$ events from
$\eL\eL$ production in $\eL$ models, and $\eR\eR$
production in $\eR$ models.  However, we do not
constrain possible non-standard visible phenomenology
from the other selectron.
The absence of knowledge of both the experimental
data and the efficiency of detection of such phenomenology 
prevents explicitly restricting our models in this regard.  
As an example, slepton mass degeneracy implies the rate
for two smuons or staus plus two photons is at the
same rate as selectrons.   But, without a fully analyzed,
statistically significant sample of two lepton plus two 
photon events, one cannot use the lack of reported events to 
exclude such a scenario.

\subsection{Model building results}
\label{model-building-results-subsec}
\indent

In Table~\ref{parameter-ranges-table}, we present the
parameters that enter our analysis common to all selectron 
interpretations, and the relevant ranges.
For the $\eL$ and $\eR$ interpretations, the allowed
range of $m_{\tilde e}$ is determined by the
lower bound from kinematics $m_{\tilde e} \gsim 100$~GeV
using Observation 4 in Sec.~\ref{kinematics-sec} 
(indeed $|\mu| \sim m_{\NI} \gsim 33$~GeV, from 
Sec.~\ref{LEP-constraints-subsec}).
The upper bound is obtained from the minimum threshold
in the cross section times branching ratio 
${\cal A}$.
For ${\cal A} = 5$, $7.5$, $10$ fb, the upper bound on 
the slepton mass is $m_{\eL} < 137$, $125$, $118$ GeV, 
and $m_{\eR} < 115$, $105$, $97$ GeV, in the $\eL$ and $\eR$
interpretations.  Notice that $\eR$ models always fail
the highest threshold, since the cross section never
exceeds $10$ fb in the allowed mass range.
The mass of the other slepton that is not the 
source of the $\eegg + \Et$ event 
(hence $\eegg + \Et$ event kinematics do not apply)
is allowed to take on a much 
wider mass range $60$--$500$ GeV\@.
For the $\eL + \eR$ interpretation,
both sleptons still must be greater than $100$ GeV by 
$\eegg + \Et$ event kinematics, 
but the upper limits are somewhat relaxed since each
individual rate $\sigma_L \times \BR_L^2$ or $\sigma_R \times \BR_R^2$ 
need not be larger than the threshold; only the sum must satisfy
the $\sigma \times \BR^2$ constraint.

\begin{table}
\renewcommand{\baselinestretch}{1.2}\small\normalsize
\begin{center}
\begin{tabular}{ccc} \hline\hline
Parameter & Range \\ \hline
$M_1$, $M_2$, $\mu$, $\tbeta$ & 
    \mbox{randomized throughout allowed range}                      \\
$m_{\tilde q} = m_{{\tilde t}_2}$ &  $250$, $500$, $1000$ GeV       \\
$m_{\tI}$                &  $> 150$ GeV, $< m_{\tilde q}$  \\
$\theta_{\tilde t}$               &  [$-\pi$,$\pi$]                 \\
$m_A$                             &  $50$, $100$, $200$, $400$ GeV  \\ 
  \hline\hline
\end{tabular}
\end{center}
\caption{Parameter ranges common to all selectron interpretations
with a heavier stop.  Models with a light stop are discussed in 
Sec.~\ref{light-stop-sec}.
}
\label{parameter-ranges-table}
\end{table}

\begin{figure}
\centering
\epsfxsize=4in
\hspace*{0in}
\epsffile{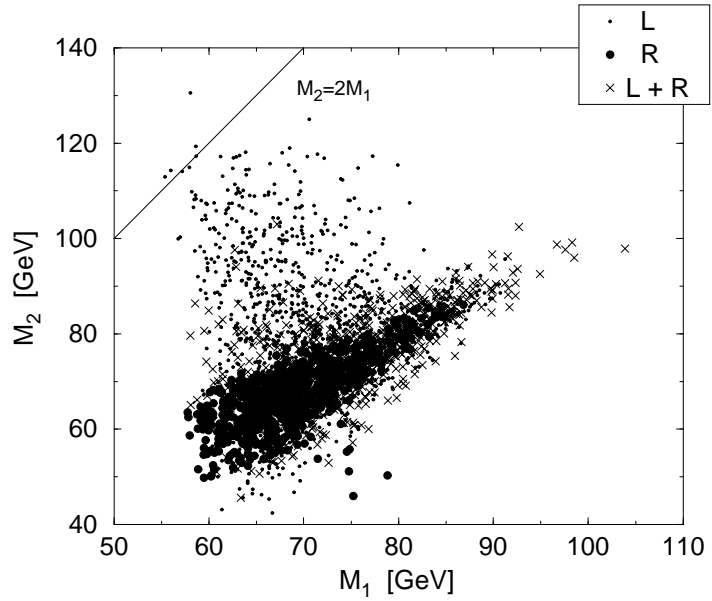}
\caption{The models satisfying the $\eegg + \Et$ event kinematics 
and the minimum threshold cut ${\cal A} = 5$~fb 
are shown in the $M_1$--$M_2$ plane.  In this figure,
$\eL$ (L), $\eR$ (R) and $\eL + \eR$ (L+R) have been separated
to show the varying restrictions on either type of model.}
\label{M1-M2-LRB-fig}
\end{figure}

\begin{figure}
\centering
\epsfxsize=4in
\hspace*{0in}
\epsffile{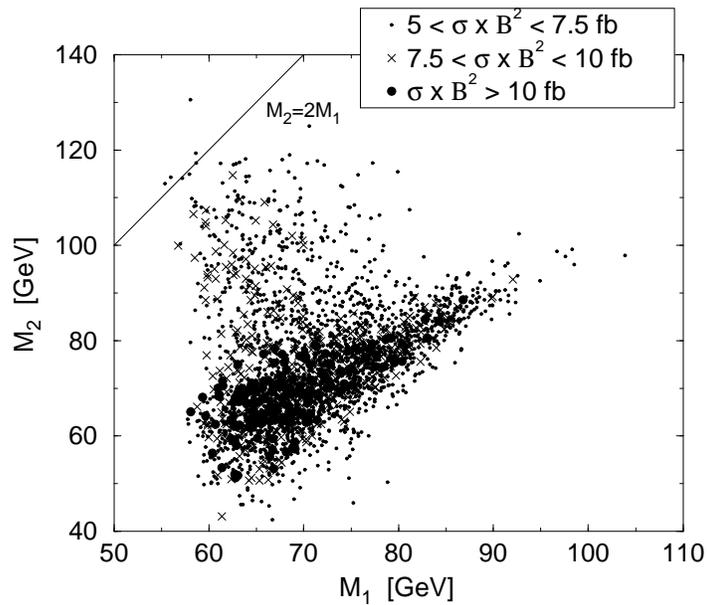}
\caption{As in Fig.~\protect\ref{M1-M2-LRB-fig}, except the models are
distinguished by the cut on 
${\cal A} = 5$, $7.5$, $10$~fb.}
\label{M1-M2-sigBR-fig}
\end{figure}

\begin{figure}
\centering
\epsfxsize=4in
\hspace*{0in}
\epsffile{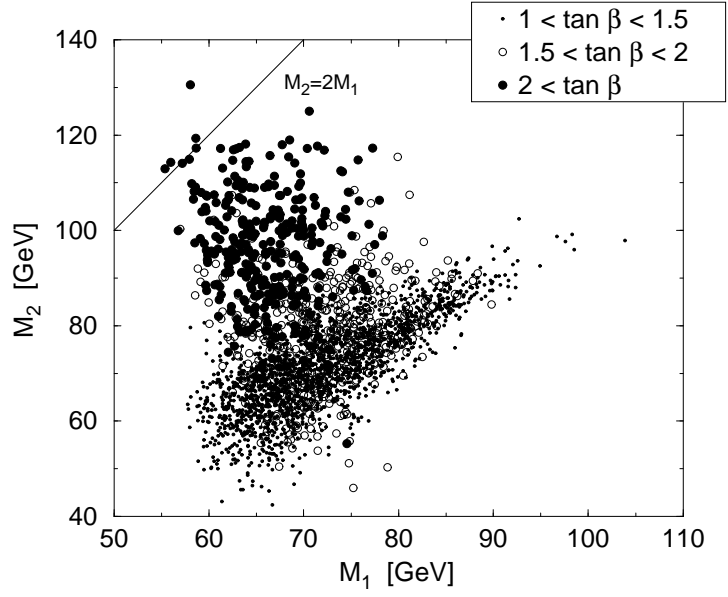}
\caption{As in Fig.~\protect\ref{M1-M2-LRB-fig} except that the models
are distinguished by the value of $\tbeta$.}
\label{M1-M2-tbeta-fig}
\end{figure}

We have explicitly constructed roughly 2500 models in total, with somewhat
more $\eL$ models than $\eR$ or $\eL + \eR$.  The results
are shown in a series of scatter plots and bar graphs
that are intended to give the general character of the models.
Figures~\ref{M1-M2-LRB-fig},~\ref{M1-M2-sigBR-fig},~\ref{M1-M2-tbeta-fig}
show the distribution of all the allowed models
in the $M_1$--$M_2$ plane, with groupings of models split up into
three plots.  All of the models pass the $\eegg + \Et$ event 
kinematic cuts 
for one or both sleptons (defined by the model type), and all
models pass the minimum threshold cut ${\cal A} = 5$ fb.
In Fig.~\ref{M1-M2-LRB-fig}, the models are
grouped by the type $\eL$, $\eR$, or $\eL + \eR$ according to
which slepton(s) passed the $\eegg + \Et$ event kinematic cuts.  
In Fig.~\ref{M1-M2-sigBR-fig}, the models are grouped by
the rate, $5 < \sigma \times \BR^2 < 7.5$, 
$7.5 < \sigma \times \BR^2 < 10$, 
and $\sigma \times \BR^2 > 10$ fb.  In Fig.~\ref{M1-M2-tbeta-fig}
the models are grouped by $\tbeta$ into the (arbitrary)
ranges $1 < \tbeta < 1.5$, 
$1.5 < \tbeta < 2$, and $\tbeta > 2$.
There are perhaps four regions with distinct character, 
and we will discuss each of them in the following.

Region 1 defined by roughly $0.8 \lsim M_2/M_1 \lsim 1.2$
represents the anticipated $M_1 \sim M_2$ region.
All three types of models $\eL$, $\eR$
and $\eL + \eR$ fall into this range, with $\eR$ models 
almost contained within the $M_2/M_1$ limits.
This is the region where the dynamical enhancement of the radiative
neutralino branching ratio is present, with the limiting case 
$(M_1 - M_2) \ra 0$, $\tbeta \ra 1$ giving the largest value.
Hence, the highest $\sigma \times \BR^2$ can be found in this 
region, but the rate need not be high since the slepton cross 
section can be low independent of the branching ratio.  
For example, $\eR$ models always have $\sigma_R \times \BR_R^2 
\lsim 8.2$ fb with $\BR_R^2 \lsim 98\%$, whereas $\eL$ models 
have $\sigma_L \times \BR_L^2 \lsim 16.2$ fb with
$\BR_L^2 \lsim 88\%$.  Since the decay $\eL \ra \CI \nu_e$
is always present, the maximum branching ratio $\BR_L^2$ 
is always less than the maximum for $\BR_R^2$.

\begin{figure}
\centering
\epsfxsize=4in
\hspace*{0in}
\epsffile{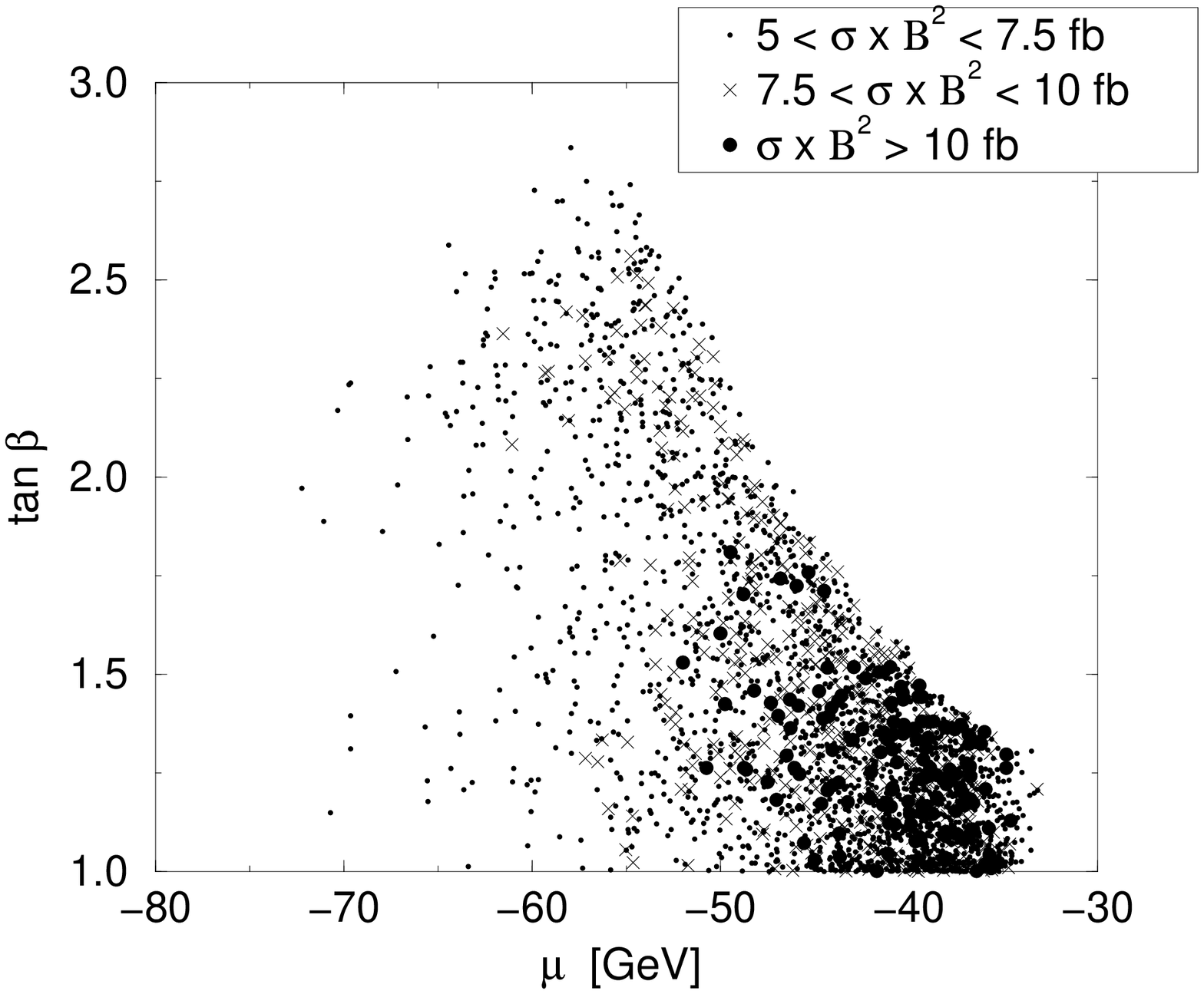}
\caption{All of the models are shown in the $\mu$--$\tbeta$ 
plane, distinguished by the cut on
${\cal A} = 
5$, $7.5$, $10$~fb.}
\label{mu-tbeta-fig}
\end{figure}

Region 2 defined by roughly $M_2/M_1 \gsim 1.2$ is populated
with mostly $\eL$ models, extending barely up to
the $M_2$ = $2 M_1$ line near $M_1 \sim 60$ GeV\@.
The reason for the much larger range in $M_2$ values
for $\eL$ models is a direct consequence of the higher
cross section $\sigma_L \sim 2.2 \sigma_R$ for a given slepton mass.
With a higher cross section, the total squared branching ratio 
can be lower, which translates into looser restrictions on the radiative
neutralino branching ratio.  For $\eL$ and $\eR$ models, 
the minimum $\BR^2$ is 25\% and 56\%,
which corresponds to a minimum $\BR( \NII \ra \NI\ph )$ of
50\% and 75\% respectively (when $\BR( {\tilde e} \ra \NII e ) = 100\%$).
Fig.~\ref{radiative-M1-M2-fig} 
already showed (for a specific set of $\mu$, $\tbeta$, 
$m_{\tilde q}$, $m_A$ values) that a looser restriction 
on the radiative branching ratio admits a larger region in 
the $M_1$--$M_2$ plane.  The models observed 
with $M_2/M_1 \gsim 1.2$ lie in just this extended
region which benefit from the kinematical 
mechanism (in addition to the dynamical mechanism) for the 
radiative neutralino decay enhancement.  This can be deduced
by examining the slepton masses for the $\eL$ models in
this region, where one finds $m_{\eR} \gg m_{\eL}$
by a factor of $2$ or more.  This is necessary
to obtain a large radiative neutralino decay, since
the branching ratio for 3-body decays $\NII \ra \NI l^+l^-$ through 
sleptons cannot be reduced to zero when the kinematical mechanism
for a large $\BR(\NII \ra \NI \ph)$ operates~\cite{AmbrosMele2}.  
In addition, the squark
masses must also be heavy to prevent the analogous
3-body decays mediated by squarks, although
the choice of $m_{\tilde q} \ge 250$~GeV in our
models is sufficient.  Finally, the existence
of only $\eL$ models in this region is due to
the fact that kinematical enhancement of the
radiative neutralino decay cannot be maximized
simultaneously with the $m_{\NII} - m_{\NI} > 21$ GeV,
and so $\BR^2$ cannot be very large.  Thus, one
needs a large cross section to supplement a lower
$\BR^2$, which can only be achieved with $\eL$ models. 

The character of the `extended' $\eL$ models in Region 2
is more clearly visible in Fig.~\ref{M1-M2-tbeta-fig}, 
where all of the models have been plotted in the $M_1$--$M_2$ 
plane distinguished only by the $\tbeta$ value.  The
models with $M_2/M_1 \gsim 1.2$ always have 
$1.5 \lsim \tbeta \lsim 2.8$, where the upper limit 
in $\tbeta$ (and $M_1$, $M_2$) is established by the 
smallest allowed radiative neutralino branching ratio.
Indeed, the kinematical mechanism that contributes to the enhanced 
radiative neutralino decay in this region does not 
necessarily require
$\tbeta \simeq 1$~\cite{AmbrosMele2}.
In Fig.~\ref{M1-M2-sigBR-fig} it is clear that increasing
the threshold ${\cal A}$ to $7.5$, $10$~pb restricts
$M_2/M_1 \lsim 1.9$, $1.2$, and so the existence
of models with $M_2 = 2 M_1$ is sensitive to
the choice of the minimum threshold.  Further,
while $M_2 = 2 M_1$ seemingly admits gaugino mass
unification, we noted above that for the extended $\eL$ models
$m_{\eR} \gg m_{\eL}$.  Hence, scalar
mass unification probably cannot be achieved, 
at least in the slepton sector, and a completely
unified scenario seems not to be compatible with
the $\eegg + \Et$ event.

In Region 3 loosely defined as $M_2/M_1 \lsim 0.8$, $\eR$ models 
appear near $M_1 \sim 75$ GeV and $M_2 \sim 50$ GeV\@.  These
models have $\sigma_R \times \BR_R^2 \sim 5.5$ fb
and $\tbeta \sim 2$.  This is the only region where 
the usual mass hierarchy $|\mu| < M_2$ can be slightly violated. 
On closer inspection one finds
the chargino mass is about $\sim 68$ GeV\@.  We found 
no $\eL$ models in this region, due to the light chargino
that induces a large branching ratio for $\eL \ra \C \nu_e$ 
over $\eL \ra \NII e$.  Also, the width for the 
3-body decay $\NII \ra \NI e^+ e^-$ turns out to be
considerably enhanced when the $\eL$ is light.
Hence the radiative neutralino decay is strongly
suppressed in such a case, and thus $\eL$ models
cannot be constructed in Region 3.  As $\tbeta$ 
is increased, the chargino mass becomes smaller and thus
is excluded by LEP130-136 constraints.  Lowering
$\tbeta$ decreases the radiative neutralino
branching ratio, and so is excluded by the $\sigma \times
\BR^2$ cut.  This localized region is basically due to a hybrid
of the dynamical and kinematical enhancement of the radiative
neutralino decay.  One can use an argument analogous to that
used for Region 2, to observe that $m_{\eL} \gg m_{\eR}$
in all of the models.  The absence of light $\eL$ is 
a consequence of the kinematical mechanism at least partly
at work.  Thus, these models sit at the edge of exclusion,
between a multitude of constraints.

Finally, the voids with no models found for 
$M_1 \gsim 85$ GeV with $M_2/M_1 \lsim 0.8$ or $M_2/M_1 \gsim 1.2$ 
are excluded by a low radiative neutralino branching ratio.
This behavior can be discerned from Fig.~\ref{radiative-M1-M2-fig},
but of course the numerical result here encompasses a full range of
$\mu$ and $\tbeta$ values.

Naively one might think that $\eL + \eR$ models can
always be constructed from $\eL$ or $\eR$ models, by
simply shifting the other slepton mass such that
$m_{\eL} \approx m_{\eR}$.  This construction always
satisfies the $\eegg + \Et$ event kinematics, which
are of course invariant under $L \leftrightarrow R$.
Indeed, such a construction can work in the region with 
a dominant dynamical enhancement of the radiative neutralino decay.
However, the construction need {\em not} work in 
the region where a kinematic enhancement of
the radiative neutralino decay occurs, such as
in Region 2 populated by $\eL$ models.  As discussed above,
$m_{\eR} \gg m_{\eL}$ in
this region which prevented 3-body decays 
$\NII \ra \NI l^+ l^-$ mediated by $\lR$ 
to overwhelm the radiative decay $\NII \ra \NI \ph$.

In general, $\eL + \eR$ models tend to be constrained 
similar to $\eR$ models, but looser bounds on $M_2/M_1$ are present 
and larger $M_1$ values accessible.  
The region with $\eL + \eR$ models that is devoid 
of $\eL$ or $\eR$ models, defined as Region 4, has
the properties that the $\sigma \times \BR^2 < 7.5$~fb
and $\tbeta \lsim 1.5$, while simultaneously 
$\sigma_L \times \BR_L^2 < 5$~fb and $\sigma_R \times \BR_R^2 < 5$~fb.
For $M_1 (\approx M_2) \gsim 90$~GeV, 
larger chargino and neutralino masses are allowed
than in either $\eL$ or $\eR$ models.
In particular, $m_{\NII}$ is near the upper bound from 
$\eegg + \Et$ event kinematics, so presumably 
values of $M_1$ higher than obtained 
in $\eL + \eR$ models are not accessible.
As for the size of the $\eegg$ rate, 
the maximum (summed) $\sigma \times \BR^2 \lsim 19$ fb, 
so it would appear one does not gain more than a factor
of about $1.2$ over the maximum $\eegg$ rate for $\eL$ models alone.  
Further, since $\eL + \eR$ models enlarge the allowed region
of parameter space by reducing the minimum
$\sigma_{L,R} \times \BR_{L,R}^2$, 
one can use the results as an indication of the region 
resulting from relaxing the ${\cal A} = 5$~fb cut 
in $\eL$ or $\eR$ models separately.  
It is clear that $\eL + \eR$ models have a distinct
character separate from $\eL$ or $\eR$ models.

In Fig.~\ref{mu-tbeta-fig} we show the models in the 
$\mu$--$\tbeta$ plane to completely specify the parameters.
Three features are worthy of explanation:  First, the upper and lower
limits on $|\mu|$ are approximately the upper and lower limits on
$m_{\NI}$, since $\NI \approx \Hb$.  From Observation 1
in Sec.~\ref{kinematics-sec}, we know the upper limit on
$m_{\NI}$ is $50$, $74$ GeV for $\eR$ and $\eL$ models, 
and this can be translated into rough upper limits on $|\mu|$.
The lower limit on $m_{\NI} \sim |\mu| \gsim 33$ GeV, 
and the region devoid of models in the upper right-hand
corner (larger $\tbeta$, smaller $|\mu|$),
come from a confluence of LEP1, LEP130-136 and 
$\eegg + \Et$ constraints as explained in
in Section~\ref{LEP-constraints-subsec}.  For example,
the LEP constraints on chargino and neutralino 
production forbid models with $|\mu| < 40 \> (50)$~GeV
for $\tbeta > 1.5 \> (2)$, once very small $|\mu|$ are
excluded by $\eegg + \Et$ event kinematics.

\begin{figure}
\centering
\epsfxsize=4in
\hspace*{0in}
\epsffile{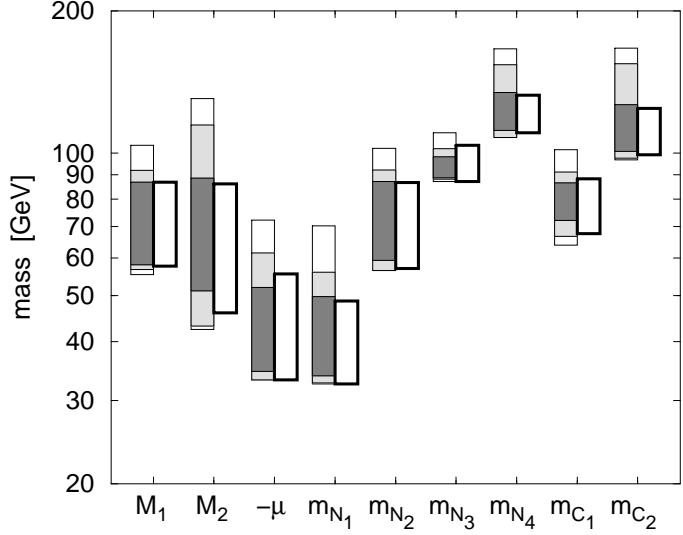}
\caption{The allowed mass spectrum is shown for all models 
(shaded bands on the left) and for $\protect\eR$ models only 
(thick solid outline on the right).  
The increasingly darker shades in 
the left-hand column correspond to increasing stricter
cuts on ${\cal A} = 5$, $7.5$, $10$~fb.
As for $\tbeta$, the allowed range in all models
is $1.0 < \tbeta < (2.8, \> 2.6, \> 1.8)$ for 
${\cal A} = 5$, $7.5$, $10$~fb respectively.
The allowed range of $\tbeta$ in $\eR$ models only
is $1.0 < \tbeta < 2.0$.
}
\label{mass-spectrum-fig}
\end{figure}

\begin{figure}
\centering
\epsfxsize=4in
\hspace*{0in}
\epsffile{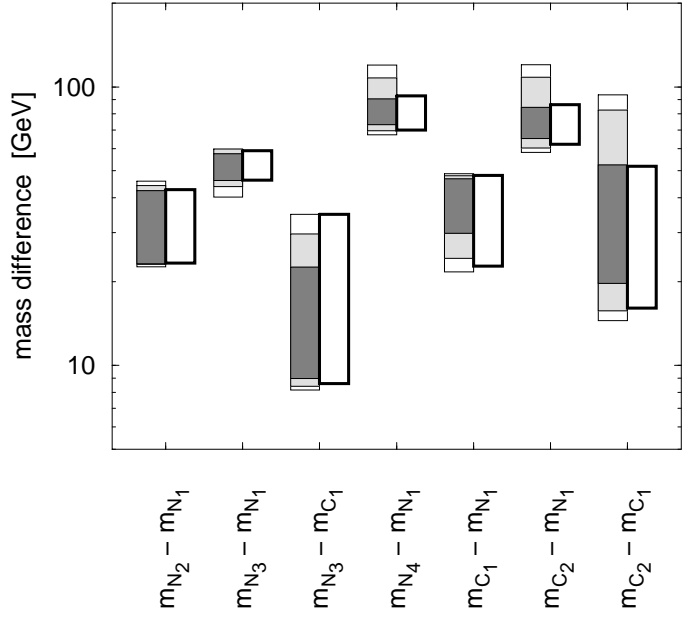}
\caption{As in Fig.~\ref{mass-spectrum-fig}, except that
mass differences between certain charginos and neutralinos 
are shown.}
\label{mass-difference-fig}
\end{figure}

The final allowed ranges of $M_1$, $M_2$, $\mu$ and the ranges of 
masses $m_{\NI}$, $m_{\NII}$, 
$m_{\NIII}$, $m_{\NIIII}$, $m_{\CI}$, $m_{\CII}$
derived from them
are presented in Fig.~\ref{mass-spectrum-fig}. 
The effect of imposing a stricter cut ${\cal A} = 5$, $7.5$,
$10$~fb is shown, in addition to the ranges for $\eR$ models only.
The latter is to give an idea of the stronger constraints
that exist when a specific origin of the $\eegg + \Et$
event is assumed.
Correlations between a selection of chargino/neutralino masses can
be discerned from Fig.~\ref{mass-difference-fig}.
Sleptons can also have correlations with chargino/neutralino
masses, which are relevant for the branching ratios.  
We present these mass ranges in Table~\ref{slepton-correlations-table}.
For example, notice that the mass of the slepton satisfying 
the $\eegg + \Et$ event kinematics always 
obeys $m_{\tilde e} > m_{\CI}$.

\begin{table}
\renewcommand{\baselinestretch}{1.2}\small\normalsize
\begin{center}
\begin{tabular}{ccrcr} \hline\hline
Model Type  &  Mass difference          &  
   \multicolumn{3}{c}{Range (in GeV)} \\ \hline
$\eL$       &  $m_{\eL}  - m_{\NI}$     &  $64$  & $\ra$ & $87$ \\
            &  $m_{\eL}  - m_{\NII}$    &  $23$  & $\ra$ & $63$ \\
            &  $m_{\eL}  - m_{\NIII}$   &   $7$  & $\ra$ & $35$ \\
            &  $m_{\eL}  - m_{\NIIII}$  &  $-50$ & $\ra$ & $6$  \\
            &  $m_{\eL}  - m_{\CI}$     &  $18$  & $\ra$ & $61$ \\
            &  $m_{\eL}  - m_{\CII}$    &  $-51$ & $\ra$ & $14$ \\
            &  $m_{\eL}  - m_{\veL}$    &   $0$  & $\ra$ & $26$ \\
            &  $m_{\veL} - m_{\NI}$     &  $39$  & $\ra$ & $79$ \\
            &  $m_{\veL} - m_{\NII}$    &   $9$  & $\ra$ & $55$ \\
            &  $m_{\veL} - m_{\NIII}$   &  $-17$ & $\ra$ & $27$ \\
            &  $m_{\veL} - m_{\NIIII}$  &  $-71$ & $\ra$ & $1$  \\
            &  $m_{\veL} - m_{\CI}$     &  $14$  & $\ra$ & $43$ \\
            &  $m_{\veL} - m_{\CII}$    &  $-71$ & $\ra$ & $11$ \\ \hline
$\eR$       &  $m_{\eR}  - m_{\NI}$     &  $64$  & $\ra$ & $77$ \\
            &  $m_{\eR}  - m_{\NII}$    &  $23$  & $\ra$ & $53$ \\
            &  $m_{\eR}  - m_{\NIII}$   &   $6$  & $\ra$ & $25$ \\
            &  $m_{\eR}  - m_{\NIIII}$  &  $-27$ & $\ra$ & $-2$ \\
            &  $m_{\eR}  - m_{\CI}$     &  $18$  & $\ra$ & $44$ \\
            &  $m_{\eR}  - m_{\CII}$    &  $-21$ & $\ra$ & $8$  \\ 
  \hline\hline
\end{tabular}
\end{center}
\caption{Ranges of selected mass differences between $\eL$, $\veL$, and $\eR$
and chargino/neutralinos in $\eL$ and $\eR$ models.}
\label{slepton-correlations-table}
\end{table}

Squarks do not play a large role in our analysis,
since they are assumed to be heavier than 
charginos and neutralinos.  However, two effects
for a given value of the squark mass persist:  
First, in 3-body decays
of neutralinos, the $t$-channel exchange of squarks
can lower the branching ratio of $\NII \ra \NI \ph$, 
hence the rate $\sigma \times \BR^2$.  Second,
the stops enter in the loops of the one-loop 
radiative neutralino decay width (since the Yukawa 
coupling of ${\tilde H}_b$ to ${\tilde t}$ is significant), 
and also tend to slightly decrease the radiative neutralino
branching ratio for lighter $m_{{\tilde t}_{1,2}}$~\cite{AmbrosMele2}.
With $m_{\tilde q} = m_{\tII} = 250$~GeV, we 
found no $\eR$ models satisfying the ${\cal A} = 5$~fb
cut, and $\eL$ or $\eL + \eR$ models always 
have $\sigma \times \BR^2 \lsim 8$~fb.

The effect of different neutral CP-odd Higgs masses $m_A$
is primarily confined to the neutralino branching ratios,
although $H^\pm$ does enter the one-loop 
radiative neutralino decay width.  We find that
varying $m_A$ from $50$ to $400$~GeV does not significantly change
the size of the radiative neutralino 
branching ratio, hence the $\sigma \times \BR^2$ for
the $\eegg + \Et$ event.

\subsection{Neutralino composition and branching ratios}
\label{neut-BR-subsec}
\indent

\begin{figure}
\centering
\epsfxsize=4in
\hspace*{0in}
\epsffile{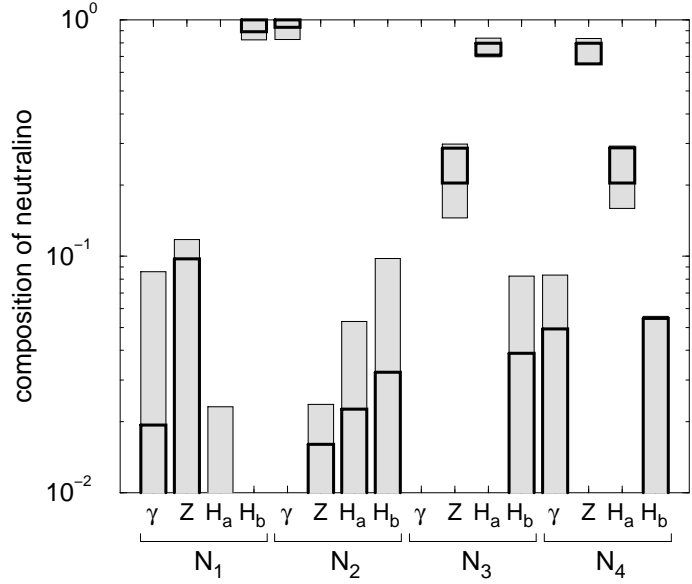}
\caption{The allowed range of all four neutralinos' composition 
$\langle \N | {\tilde \chi} \rangle^2$ in terms of the
interaction eigenstates ${\tilde \chi} = \phino, \; \Zino, \; \Ha, \; \Hb$ 
is shown for all of the models.  The thick solid
outline corresponds to $\protect\eR$ models only.
Bars that touch the x-axis correspond to a neutralino component that
can be lower than $10^{-2}$; the absence of a bar for the $\phino$ 
component of $\NIII$ implies $\langle \NIII | \phino \rangle^2 < 10^{-2}$
for all models.}
\label{neut-composition-fig}
\end{figure}

In Fig.~\ref{neut-composition-fig} we show the (maximum) allowed 
range of the neutralino composition 
$\langle \Ni | {\tilde \chi} \rangle^2$
of all of the models in the 
${\tilde \chi} = \phino, \; \Zino, 
\; \Ha, \; \Hb$
basis.  For a given threshold in $\sigma \times \BR^2$ 
applied to all models, 
the minimum radiative neutralino branching ratio is 
always larger for $\eR$ than for $\eL$ models.  
A larger minimum radiative neutralino branching ratio
implies the constraints on the neutralino composition
must be similarly stronger, hence the differing notation 
for all models and $\eR$ models in the plot.  We make 
three observations:  First, we find that 
\begin{equation}
\NI \simeq \Hb, \;\; \NII \simeq \phino,
\end{equation}
so the lightest two neutralinos are composed of exactly the content
expected from Eq.~(\ref{neutralino-content-eq}).  
To a lesser extent,
\begin{equation}
\NIII \sim \Ha, \;\; \NIIII \sim \Zino,
\end{equation}
the heavier two neutralinos turn out have a fairly 
specific composition as well.  This will be relevant to the 
branching ratios and cross sections for associated phenomenology.  
Second, $\NI$ tends to have a much 
larger $\Zino$ component than $\NII$.
Third, the required purity of the lightest neutralinos 
in $\eR$ models is significant compared with $\eL$ models, 
and this is perhaps most easily observed by looking 
at for example the photino content of $\NI$ and $\NII$
in Fig.~\ref{neut-composition-fig}.

In the following
discussion of the branching ratios (and the discussion 
in subsequent sections), we discuss only
the distinctions between $\eL$ and $\eR$ models, since 
the branching ratios in $\eL + \eR$ models are 
a relatively simple extension of $\eL$ and $\eR$ separately.
The range of branching ratios of $\NII$ are shown in 
Table~\ref{BR_N2-table}.  
In the pure state limit $\NII = \phino$, only the
radiative channel is open for $\NII$.  However, the impurity of $\NII$
(see Fig.~\ref{neut-composition-fig}) causes other modes
to have non-negligible branching fractions ($\NII$ is
somewhat of a special case since the radiative decay branching
ratio is required to be large).  The possible
decays for $\NII$ in our models are:  $\NI \ph$, $\NI$``$Z$'', 
$\CI$``$W$'', $\vL \nu$, $\lL l$, $\lR l$.  
We use ``$Z$'' and ``$W$'' to mean the 3-body decay
mediated by an on- or off-shell $Z$ and $W$, 
plus off-shell sleptons and squarks.
The rate for the final states 
``$Z$''$ \ra l^+l^-, \nu\overline{\nu}, q\overline{q}$ and 
``$W$''$ \ra l \nu, q\overline{q}'$
are determined roughly by the corresponding 
SM gauge boson branching ratios.
The only significant deviation from the SM gauge
boson branching fractions is modes that involve 
sleptons, since the $\eegg + \Et$ event requires at
least one slepton is light.  The presence of some modes
depends on the particular class of models; for example 
in $\eL$ models,
the mode $\NII \ra \lR l$ is open if $m_{\lR} < m_{\NII}$. 
This never happens in $\eR$ models since $\eR \ra \NII e$ is
required to obtain the $\eegg + \Et$ event!  
The 2-body mode $\NII \ra \vL \nu$ is open
if $m_{\vL} < m_{\NII}$, which happens in $\eR$ models 
and could potentially happen in
$\eL$ models.  However, for $\eL$ models 
one never finds decays $\NII \ra \vL \nu$
because the mass splitting between $\vL$ and $\lL$ 
is never more than about $25$~GeV 
(see Table~\ref{slepton-correlations-table}).  Since there always
must be a large mass difference between $m_{\eL}$ and
$m_{\NII}$ from $\eegg + \Et$ event kinematics, 
then the 2-body mode into a sneutrino is
always closed.

The $\NIII$ and $\NIIII$ branching ratio pattern is
progressively more complicated than for the lighter neutralino
due to possible 2-body decays into sleptons and Higgs bosons.  
For $\NIII$, there are several distinct classes of final states:
${\tilde N}_{1,2}$``$Z$'', $\CI$``$W$'', $\NI h (A)$,
$\lL l$, $\lR l$, $\vL \nu$; all other possible channels 
are strongly suppressed or forbidden.  For example, for 
the heavier chargino one has
$m_{\CII} > m_{\NIII}$ in all of our models, hence
$\NIII$ decay into $\CII$ is forbidden.

The upper limits on the mass differences 
$m_{\NIII} - m_{\NI} < 60$~GeV and $m_{\NIII} - m_{\CI} < 35$~GeV
in our models are crucial to determining the allowed decays of $\NIII$.  
In particular, the decay $\NIII \ra \NI h$ or $\NIII \ra \NI A$
will only occur when $m_h$ or $m_A < 60$~GeV, with constraints 
from LEP that exclude $m_h < 44$~GeV and the coupling 
$\sin^2(\beta-\alpha) \lsim \frac{m_h}{60}$~GeV\@.
The restriction on the mass of $A$ from LEP that excludes 
$m_A < 22$~GeV is considerably weaker than the one on $m_h$, 
and so decays $\NIII \ra \NI A$
are always possible with an appropriate choice of $m_A$
(provided this does not imply an excluded $m_h$ value). 
The situation is actually considerably more subtle.
We find decays into $\NI$``$Z$'' are not suppressed even if decays
into the light Higgs $h$ are open, with a maximum 
$\BR(\NIII \ra \NI h) \sim 35\%$ while $\NIII \ra \NI A$ decay
is closed.  However, with low $\tbeta$ the mass splitting 
between $m_A$ and $m_h$ tends to be small for $m_A \sim 50$ GeV, 
and because of the couplings, decays into $A$ typically dominate 
over $h$ if kinematically accessible.  In $\eL$ ($\eR$) models, the 
decay $\NIII \ra \lL l$ ($\NIII \ra \lR l$) is always kinematically
forbidden.  Thus, it is only when the other slepton 
($\lR$ in $\eL$ models, $\lL$ in $\eR$ models) has a
mass $m_{\tilde l} < m_{\NIII}$ that 2-body decays 
into sleptons can dominate.  When kinematically accessible,
the branching ratio for the 2-body decay $\NIII \ra \vL \nu$ 
can be $\sim 100\%$, and is always larger than decays 
into $\lL l$ by a factor of at least $10$.
This is due to the larger $\Zino$ impurity in $\NIII$, i.e.\
$\langle \NIII | \Zino \rangle^2 \gg \langle \NIII | \phino \rangle^2$,
and Eq.~(\ref{sum-rule-eq}) requiring $m_{\vL} < m_{\lL}$.
The 3-body decays into the lightest chargino $\NIII \ra \CI$``$W$'' 
depend on the chargino mixings, but are always smaller than the 
3-body decays $\NIII \ra \NI$``$Z$'' mainly due to phase space.  The
presence of decays into $\CI$``$W$'' can suppress the
branching ratio for decays into $\NI$``$Z$'' by at most a factor 
of $2$, but even then the branching ratios for $\NIII$ 
are still larger into $\NI$``$Z$''.  Also, $\NIII$ decays
into $\NII$ are strongly suppressed, because of the
particular neutralino composition in our models.

The branching ratios of $\NIIII$ are quite intricate, however
a few features can be discerned.  The main possible decays include: 
${\tilde N}_{1,2,3}$``$Z$'', ${\tilde C}_{1,2}$``$W$'', 
$\NI h (A)$, and possibly open 2-body modes $\lL l$, $\lR l$,
$\vL \nu$.  Since the mass difference
$m_{\NIIII} - m_{\NI} \gsim 67$~GeV and can be as large as
$100$ ($120$)~GeV in $\eR$ ($\eL$) models, then the decay
$\NIIII \ra \NI h$ is a prominent possibility if 
kinematically allowed.  We find that even if $\NIIII \ra \NI A$ is also 
open, it is always suppressed to of order $\sim 5\%$ compared with
a much larger $\NI h$ mode.  This is because the $\NIIII$ composition
is roughly inverted with respect to the $\NIII$ one, which
feeds into the $\NIIII$ couplings to the Higgs sector.  
In $\eR$ models, the 2-body slepton 
decay $\NIIII \ra \lR l$ is always open, and can be $\sim 100\%$.
In $\eL$ models, the decay $\NIIII \ra \vL \nu$ is typically open, 
but sometimes can be kinematically inaccessible.  Note that if both
$\NIIII \ra \vL \nu$ and $\NIIII \ra \lL l$ are accessible, 
then $\NIIII \ra \vL \nu$ always overwhelms $\NIIII \ra \lL l$ 
by at least a factor of $5$ due to the 
large $\Zino$ component of $\NIIII$ 
(see Fig.~\ref{neut-composition-fig}) and phase space.  
Similarly, if
none of the 2-body modes are open, then the neutralino
composition of $\NIIII$ implies $\NI$``$Z$'' 
dominates over all other 3-body decays.

\begin{table}
\renewcommand{\baselinestretch}{1.2}\small\normalsize
\begin{center}
\begin{tabular}{ccrclrcl} \hline\hline
Final State   &  Kinematic Condition  &  
    \multicolumn{3}{c}{Range in $\eL$ models} & 
    \multicolumn{3}{c}{Range in $\eR$ models}  \\ \hline
$\NI$``$Z$''  &  -- 
    & \phantom{00000} & $\ra$ & $47$  & \phantom{00000} & $\ra$ & $26$  \\
$\NI \ph$     &  -- 
    & $53$ & $\ra$ & $100$ & $74$ & $\ra$ & $100$   \\
$\CI$``$W$''  &  $m_{\NII} > m_{\CI}$
    &      & $\ra$ & $10$  & & $\ra$ & $2$     \\
$\lL \overline{l} + \overline{\lL} l$       &  $m_{\NII} > m_{\lL}$
    &      &  --   &       & &  --   &         \\
$\lR \overline{l} + \overline{\lR} l$       &  $m_{\NII} > m_{\lR}$
    &      & $\ra$ & $3$   & &  --   &         \\
$\vL \overline{\nu} + \overline{\vL} \nu$     &  $m_{\NII} > m_{\vL}$
    &      &  --   &       & & $\ra$ & $8$     \\ \hline\hline
\end{tabular}
\end{center}
\caption{Ranges of selected $\NII$ branching ratios (in \%) in our
models.  The notation `$\ra X$' denotes a range from less than
1\% up to $X\%$.  The kinematic condition must be satisfied 
for the mode to be open; no kinematic condition implies
the mode always open.  Note that the 2-body decays into
sleptons sums over all three families, because of the
assumption of slepton mass degeneracy.
The final state into $\NI e^+ e^-$
can be enhanced over that expected from $\NI$``$Z$'' because of light 
slepton exchange.
}
\label{BR_N2-table}
\end{table}

\begin{table}
\renewcommand{\baselinestretch}{1.2}\small\normalsize
\begin{center}
\begin{tabular}{ccrclrcl} \hline\hline
Final State   &  Kinematic Condition  &  
    \multicolumn{3}{c}{Range in $\eL$ models} & 
    \multicolumn{3}{c}{Range in $\eR$ models}  \\ \hline
$\NI$``$Z$''  &  -- 
    & \phantom{000} & $\ra$ & $99$  & \phantom{000} & $\ra$ & $99$        \\
$\NI h$       &  $m_{\NIII} - m_{\NI} > m_h$
    & & $\ra$ & $29$  & & $\ra$ & $31$        \\
$\NI A$       &  $m_{\NIII} - m_{\NI} > m_A$
    & & $\ra$ & $66$  & & $\ra$ & $71$        \\
$\CI$``$W$''  &  -- 
    & & $\ra$ & $34$  & & $\ra$ & $29$        \\
$\NII$``$Z$'' &  -- 
    & & $\ra$ & $1.5$ & & $\ra$ & $1.5$       \\
$\lL \overline{l} + \overline{\lL} l$       &  $m_{\NIII} > m_{\lR}$
    & &  --   &       & & $\ra$ & $22$        \\
$\lR \overline{l} + \overline{\lR} l$       &  $m_{\NIII} > m_{\lR}$
    & & $\ra$ & $99$  & &  --   &             \\
$\vL \overline{\nu} + \overline{\vL} \nu$       &  $m_{\NIII} > m_{\vL}$
    & & $\ra$ & $99$  & & $\ra$ & $99$        \\ \hline\hline
\end{tabular}
\end{center}
\caption{Ranges of selected $\NIII$ branching ratios (in \%), as 
in Table~\ref{BR_N2-table}.
}
\label{BR_N3-table}
\end{table}

\subsection{Chargino composition and branching ratios}
\label{chargino-BR-subsec}
\indent

\begin{figure}
\centering
\epsfxsize=4in
\hspace*{0in}
\epsffile{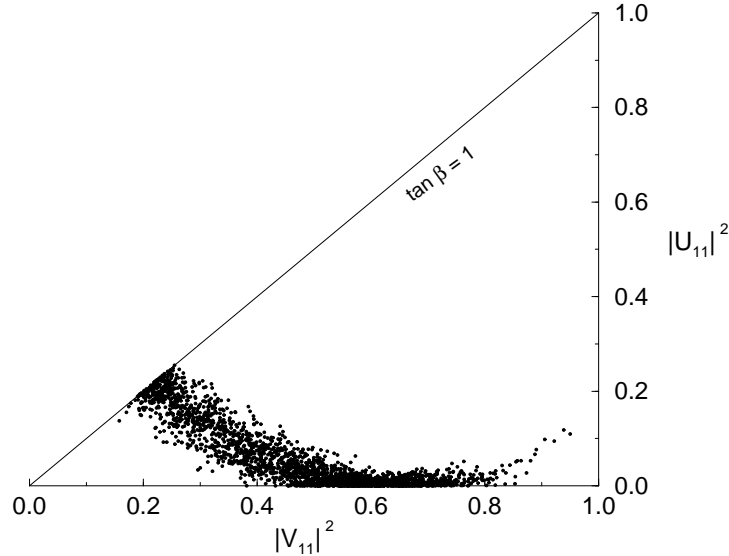}
\caption{Scatter plot of the chargino mixing matrix 
elements $U_{11}$, $V_{11}$ for all models.  The narrow band of
points indicates the presence of strong constraints
in our models from the $\eegg + \Et$ event.}
\label{char-composition-fig}
\end{figure}

The chargino composition is determined by the mixing matrices
$U$ and $V$, as defined in Sec.~\ref{susy-parameters-subsection}.
$U$ and $V$ (real and orthogonal in our conventions) can be expressed 
in terms of two independent rotation 
angles $\phi_{\mp}$ (see e.g.\ Ref.~\cite{BBO2}), however
the Dirac nature of the chargino spinors does not allow
an intuitive identification of their Wino and Higgsino
components.  Nevertheless, in Fig.~\ref{char-composition-fig} we present
the elements $|V_{11}|^2 = \cos^2 \phi_{+}$ vs. 
$|U_{11}|^2 = \cos^2 \phi_{-}$ to give
a sense of the constraints that the $\eegg + \Et$ event
imposes on the chargino composition.  In the limit
$\tbeta \ra 1$, the chargino mass matrix is symmetric which
implies $\phi_{-} = \phi_{+}$, and so $U = V$.
In Fig.~\ref{char-composition-fig}, this is the diagonal 
line where $|U_{11}|^2 = |V_{11}|^2$, and note that 
along this line our models lie in the region 
$0.15 \lsim |U_{11}|^2 \lsim 0.25$, due to the 
mass hierarchy $M_2 > |\mu|$.  Here, one can
identify $\CI$ as mostly a charged Higgsino.
For larger $\tbeta$ values, $|V_{11}|^2$ tends to
increase, while $|U_{11}|^2 \lsim 0.25$ throughout
our models.

The branching ratios of $\CI$ are displayed in Table~\ref{BR_C1-table},
that assumes $m_{\tI}$ is heavier than both charginos as in
the discussion below.
There are only a few possible channels:  ${\tilde N}_{1,2}$``$W$'',
$\lL \nu$, $\vL l$.  Further, the 3-body decays into $\NII$``$W$'' 
are always $\lsim 5\%$ due to the photino nature of $\NII$,
the Higgsino nature of $\CI$ and phase space.  Thus, the 3-body 
decays into $\NI$``$W$''
are the typical decay pattern.  In $\eL$ models 
$\lL$, $\vL$ are always heavier than $\CI$,
thus it is only in $\eR$ models that 2-body channels 
into $\lL \nu$ and $\vL l$ can possibly be open.
When both are allowed, these 2-body decays can sum to
a branching ratio of $\sim 100\%$ 
(when summed over three families).

The branching ratios of $\CII$ are displayed in Table~\ref{BR_C2-table}.
The possible decays include:  ${\tilde N}_{1,2,3}$``$W$'',
$\CI f\overline{f}$ ($f = l$, $\nu$, $q$), 
$\lL \nu$, $\vL l$, and the Higgs channels
${\tilde N}_{1,2} H^\pm$, $\CI h (A)$.  When only 3-body decays are open,
$\NI$``$W$'' dominates over all other decays.  However, 
$\NII l \nu_l$ is roughly $1$--$5\%$, and can be larger
than the decays into $\NII q\overline{q}'$ due to the possible 
enhancement from light slepton exchange in the 3-body decay.
The 2-body decay $\CII \ra \vL l$ summed over three families 
can have a branching ratio up to $95\%$, when it is the only slepton 
mode open (the remainder is distributed to the 3-body decays as above).
When both $\CII \ra \vL l$ and $\CII \ra \lL \nu$ are simultaneously 
open, the sum can be nearly $100\%$.
Finally, the 2-body decay into $\NI H^\pm$ is also 
possible when $m_{H^\pm} \lsim 90 \> (120)$~GeV, for $\eR$ ($\eL$) 
models, which requires $m_A \lsim 50 \> (100)$~GeV\@.  In addition,
decays into neutral Higgs bosons are possible when
$m_{\CII} - m_{\CI} < m_{h,A}$.

\begin{table}
\renewcommand{\baselinestretch}{1.2}\small\normalsize
\begin{center}
\begin{tabular}{ccrclrcl} \hline\hline
Final State   &  Kinematic Condition  &  
    \multicolumn{3}{c}{Range in $\eL$ models} & 
    \multicolumn{3}{c}{Range in $\eR$ models} \\ \hline
$\NI$``$W$''  &  --
    & $95$ & $\ra$ & $100$ & \phantom{000} & $\ra$ & $100$     \\
$\NII$``$W$'' &  $m_{\CI} > m_{\NII}$
    & \phantom{00000} & $\ra$ & $5$   & & $\ra$ & $5$       \\
$\vL l$       &  $m_{\CI} > m_{\vL}$
    &      &  --   &       & & $\ra$ & $100$     \\
$\lL \nu$     &  $m_{\CI} > m_{\lL}$
    &      &  --   &       & & $\ra$ & $50$      \\ \hline\hline
\end{tabular}
\end{center}
\caption{Ranges of selected $\CI$ branching ratios (in \%), as 
in Table~\ref{BR_N2-table}.  $m_{\tI} > m_{\CI}$ is assumed here.
}
\label{BR_C1-table}
\end{table}

\begin{table}
\renewcommand{\baselinestretch}{1.2}\small\normalsize
\begin{center}
\begin{tabular}{ccrclrcl} \hline\hline
Final State   &  Kinematic Condition  &  
    \multicolumn{3}{c}{Range in $\eL$ models} & 
    \multicolumn{3}{c}{Range in $\eR$ models} \\ \hline
$\NI$``$W$''  &  -- 
    & \phantom{000} & $\ra$ & $92$   & \phantom{000} & $\ra$ & $100$   \\
$\NII$``$W$'' &  -- 
    & & $\ra$ & $23$   & & $\ra$ & $17$       \\
$\NIII$``$W$'' &  --
    & & $\ra$ & $0.7$  & & $\ra$ & $0.3$      \\
$\CI f\overline{f}$  &  -- 
    & & $\ra$ & $4$    & & $\ra$ & $1$        \\
$\vL l$       &  $m_{\CII} > m_{\vL}$
    & & $\ra$ & $95$   & & $\ra$ & $69$       \\
$\lL \nu$     &  $m_{\CII} > m_{\lL}$
    & & $\ra$ & $52$   & & $\ra$ & $59$       \\ \hline\hline
\end{tabular}
\end{center}
\caption{Ranges of selected $\CII$ branching ratios (in \%)
assuming $m_A \protect\gsim 100$~GeV, as in Table~\ref{BR_N2-table}.
$m_{\tI} > m_{\CII}$ is assumed here.}
\label{BR_C2-table}
\end{table}

\subsection{Sneutrino branching ratios}
\label{sneutrino-BR-subsec}
\indent

In the selectron interpretation, 
sneutrinos do not directly enter the branching ratios relevant 
for the $\eegg + \Et$ event, however the mass of the 
sneutrino $\veL$ is necessarily smaller than $m_{\eL}$ 
due to the sum rule in Eq.~(\ref{sum-rule-eq}), and so
the sneutrino is certainly relevant in $\eL$ and $\eL + \eR$ 
models.
In Ref.~\cite{PRL} it was shown that the cross section for
sneutrino production $p\overline{p} \ra \veL\veL$ is comparable 
to $\eL\eL$ production, and $\eL\veL$ production is larger 
by a factor of 2--3 for a fixed value of $m_{\eL}$.  
Thus, the viability of the $\eegg + \Et$ event as $\eL$ production 
(and the ability to distinguish $\eL$ from $\eR$) depends 
in part on the phenomenology associated with sneutrinos.

The dominant branching fraction of sneutrinos depends 
on the size of the $\Zino$ component of the neutralinos 
and the gaugino mixings of the chargino, in addition to
the mass hierarchy.  There are
$4 \times 3$ kinematic possibilities, where $m_{\vL}$
is lighter or heavier than $m_{{\tilde N}_{2,3,4}}$ and
$m_{{\tilde C}_{1,2}}$.  In the limit of pure neutralino
states $\NI = \Hb$ and $\NII = \phino$, the sneutrino
has no coupling to the lightest two neutralinos 
since it does not couple to either pure state.  Thus, 
in the case where $m_{\vL} < m_{{\tilde C}_{1,2}}$, 
the dominant decay of $\vL$ will be to the kinematically
accessible neutralino with the largest $\Zino$ component.
The relative branching fraction into $\NI$ or $\NII$
is therefore determined by the size of their $\Zino$ component
{\em impurity}\/.  The branching ratios are shown 
in Table~\ref{BR_sv-table}.

For $\eL$ models, $m_{\vL} > m_{\CI}$, so that
decays into the lightest chargino are always possible.
The branching ratio for $\vL \ra \CI l$ is always larger 
than $53\%$, while the branching ratio for the $\vL \ra \CII l$ 
channel (if open) can reach $26\%$.  The next largest
channel is $\vL \ra \NI \nu$, with a branching ratio
up to $36\%$.  The decay
$\vL \ra \NII \nu$ is always open, but with 
a branching ratio below $6\%$
due to the small $\Zino$ component in $\NII$.

For $\eR$ models $m_{\vL}$ is unconstrained, so the
decay $\vL \ra \NI \nu$ is the only mode that is always open.
If decays into $\NII$ are also allowed, then the
dominant decay of $\vL$ can be into either 
$\NI \nu$ or $\NII \nu$.
In special cases, we found it is possible for the $\Zino$ 
impurity to be larger in $\NII$ than $\NI$,
thus the dominant decay could be $\vL \ra \NII \nu$.
This is possible when $m_{\NII} < m_{\vL} < m_{\CI}$, i.e.\
decays into charginos must be kinematically forbidden
(an impossible scenario in $\eL$ models).  When
a channel into a chargino is sufficiently open, it dominates 
over decays into the lightest two neutralinos
by a factor of more than $10$.  However, if the sneutrino
is heavy $m_{\vL} > m_{{\tilde N}_{3,4}}$, decays into
the heavier neutralinos can be moderately large 
(branching ratio $10$--$30\%$), with decays 
$\vL \ra \NIIII \nu$ dominating over $\vL \ra \NIII \nu$ 
due to the larger $\Zino$ component in $\NIIII$.

\subsection{Selectron branching ratios}
\label{selectron-BRs-subsec}
\indent

We have already discussed $\eL$ branching ratios
for $\eL$ models, and $\eR$ branching ratios for 
$\eR$ models in Sec.~\ref{slepton-decay-subsection},
since they are a fundamental part of the model building.
The other slepton ($\eR$ in $\eL$ models, and
$\eL$ in $\eR$ models), will have branching ratios similar
to $\eL$ (in $\eL$ models), or $\eR$ (in $\eR$ models) 
if its mass is roughly included in the $\eegg + \Et$ allowed
range.  In general, $\eR$, $\eL$ will decay into the kinematically
allowed final states with neutralinos, with the largest branching
ratio for the channel $\NII e$, if open.  $\eL$ can also decay into 
${\tilde C}_{1,2} \nu_e$ if open, with a maximum branching ratio of 
$27\%$ and $59\%$ respectively.

\begin{table}
\renewcommand{\baselinestretch}{1.2}\small\normalsize
\begin{center}
\begin{tabular}{ccrclrcl} \hline\hline
Final State   &  Kinematic Condition  &  
    \multicolumn{3}{c}{Range in $\eL$ models} & 
    \multicolumn{3}{c}{Range in $\eR$ models} \\ \hline
$\NI \nu_e$  &  -- 
    & \phantom{00000} & $\ra$ & $36$    & \phantom{000} & $\ra$ & $100$  \\
$\NII \nu_e$ &  $m_{\veL} > m_{\NII}$
    &      & $\ra$ & $5.5$   & & $\ra$ & $97$           \\
$\NIII \nu_e$ &  $m_{\veL} > m_{\NIII}$
    &      & $\ra$ & $29$    & & $\ra$ & $22$           \\
$\CI e$       &  $m_{\veL} > m_{\CI}$
    & $53$ & $\ra$ & $94$ & & $\ra$ & $100$          \\
$\CII e$       &  $m_{\veL} > m_{\CII}$
    &      & $\ra$ & $26$    & & $\ra$ & $48$           \\ \hline\hline
\end{tabular}
\end{center}
\caption{Ranges of selected $\veL$ branching ratios (in \%), as 
in Table~\ref{BR_N2-table}.
}
\label{BR_sv-table}
\end{table}

\subsection{Predictions for LEP}
\label{LEP-subsec}
\indent

The imminent upgrade of LEP to $\sqrt{s} = 161$~GeV (LEP161) 
and the forthcoming upgrade to $\sqrt{s} = 190$~GeV (LEP190) 
provide a potential testing ground for the models constructed.
With expected integrated luminosities of $25$~pb$^{-1}$
and $500$~pb$^{-1}$ (per detector), the one event level 
is at $40$~fb and $2$~fb for 
$\sqrt{s} = 161$, $190$~GeV respectively.
The first priority is to identify which processes have
non-negligible production cross sections, then determine
the possible signatures that depend on the branching ratios
of the produced sparticles.
It is important to emphasize that the following
predictions assume the minimum cut ${\cal A} = 5$~fb
is placed on the $\sigma \times \BR^2$ for the $\eegg + \Et$
event.  For instance, in some cases we are able
to predict a non-negligible minimum number of events with a 
particular signature must be produced, although we do not necessarily
give detector efficiencies.
In principle, if one could demonstrate that failure 
to detect such events implies they do not occur at all, 
then only two possibilities remain:  (1)  
A supersymmetric explanation of the $\eegg + \Et$ event
in our framework must rely on an upward fluctuation from 
$\sigma \times \BR^2$ even lower than $5$~fb, 
or (2) a supersymmetric explanation 
in our framework is not possible.

Based on Observation~4 in Sec.~\ref{kinematics-sec},
selectron production is always kinematically forbidden
at LEP161 and LEP190 for the selectron that satisfies
the kinematics.  The other slepton ($\eR$ in $\eL$ models, 
or $\eL$ in $\eR$ models) can potentially be kinematically 
accessible at LEP161 or LEP190 by simply requiring 
its mass be less than the threshold.  This is obviously
not a requirement (nor a constraint) of the selectron 
interpretation of the $\eegg + \Et$ event, and so we ignore 
selectron production at LEP.  However, in $\eL$ models it
was shown in Eq.~(\ref{sum-rule-eq}) that $m_{\vL}$
must be less than $m_{\eL}$, and so sneutrino
production could be a visible signal at LEP190
(since $m_{\veL} > 81$~GeV for all $\eL$ models),
as will be discussed below.

\subsubsection{LEP161}
\label{LEP161-subsec}
\indent

\begin{figure}
\centering
\epsfxsize=4in
\hspace*{0in}
\epsffile{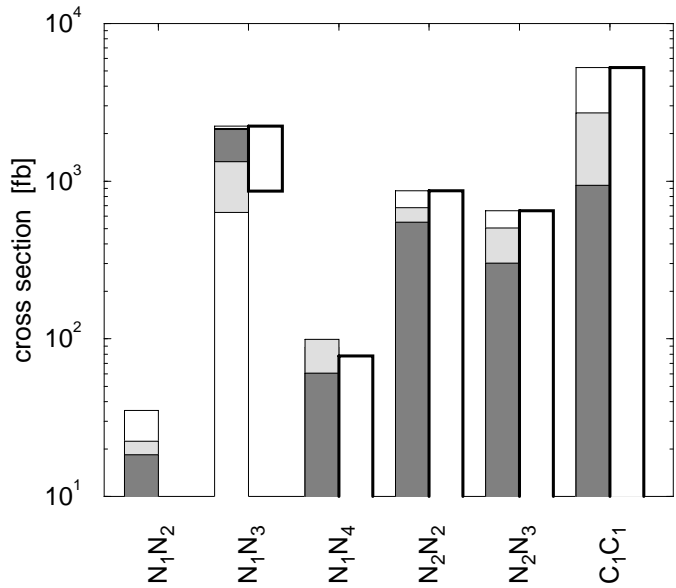}
\caption{The range of the non-negligible cross sections at
LEP161, for all models (shaded bar on left) and only $\eR$ models 
(thick solid line on right).  Each bar represents a particular 
production cross section, where the maximum and minimum height 
of the bar (or thick solid line) is the maximum and minimum 
cross section respectively.  The shading on the left bars 
indicates the range of cross section for all models passing
the cut ${\cal A} = 5$, $7.5$, $10$ fb.
Bars that touch the x-axis correspond to cross sections 
that can be smaller than $1$~fb.}
\label{sigma_LEP161-fig}
\end{figure}

In Fig.~\ref{sigma_LEP161-fig} we present all of the 
chargino/neutralino production processes that have cross
sections above about $10$~fb.
The cross sections were computed 
with initial state radiation effects included.
In $\eL$ or $\eL + \eR$ models, none
of the processes need to have large cross sections, 
although if it were possible to establish 
an upper bound on $\sigma(\NI\NIII) 
\lsim 600$~fb, then an upper bound on $\sigma \times \BR^2$ for the
$\eegg + \Et$ event can be established at $7.5$~fb, and 
in $\eR$ models $\sigma \times \BR^2 < 5$~fb
(i.e.\ all of our $\eR$ models would be excluded).
Given the cut ${\cal A} = 5$~fb, then in $\eR$ models one
expects a minimum of $22$ $\NI\NIII$ pairs to be produced, 
but no other process (nor any processes in $\eL$ or
$\eL + \eR$ models) can have non-negligible minimum rates 
at LEP161.  There are only four processes that could have 
large rates, which have the following maximum
\begin{equation}
\begin{array}{rll}
e^+e^- \ra & \NI\NIII   & (55, \; 56, \; 49)  \\
           & \NII\NII   & (19, \; 22, \; 12) \\
           & \NII\NIII  & (11, \; 16, \; 7)  \\
           & \CIp\CIm & (48, \; 132, \; 42) \;\; 
           {\rm pairs \; \> produced} \\
\end{array}
\end{equation}
for ($\eL$, $\eR$, $\eL + \eR$) models.  Notice that the maximum
pair production rates are always largest for $\eR$ models, then
$\eL$ models, then $\eL + \eR$ models.  The rate for 
$\NI\NIII$ is roughly the same in all models since 
the cross section is dominated by $Z$ exchange.  For the 
other processes, differing interference contributions between 
the $Z$ exchange and light slepton exchange cause the differences 
in the production cross sections (see Table~\ref{nn_cc_nc-table}).  
In addition, stop pairs could be produced at 
LEP161 (see Table~\ref{LEP-stops-table}).

The character of the signal from $\NI\NIII$ production is 
completely dependent on the
decay of $\NIII$ which was described in Sec.~\ref{neut-BR-subsec}
(see also Table~\ref{BR_N3-table}).
The dominant decay possibilities are $\NIII \ra \NI$``$Z$'', 
$\NIII \ra \NI A (h)$ (if $m_{A} (m_{h}) < 60$~GeV), $\NIII \ra \lR l$ in 
$\eL$ models (if $m_{\lR} < m_{\NIII}$), and 
$\NIII \ra \vL \nu$ in $\eR$ models (if $m_{\vL} < m_{\NIII}$).
The general signature is therefore ``$Z$''$+ \E$.  Extra
$b\overline{b} + \E$ occurs if the mass difference
$m_{\NIII} - m_{\NI}$ is larger than $m_h$ or $m_A$.
Some other signatures are possible in special cases:  
In $\eL$ models
one could have excess $l^+l^- + \E$ (if $m_{\lR} < m_{\NII}$), 
or $\ph l^+l^- + \E$ (if $m_{\lR} < m_{\NIII}$). 
In $\eR$ models the decay $\NIII \ra \vL \nu$ becomes the 
dominant decay if the sneutrino (and necessarily $\lL$) 
are light.  Thus the dominant signature could be invisible, 
or $\ph + \E$, or $\l^+l^- + \E$, if the mass hierarchy is 
$m_{\vL} < (m_{\NII}, \> m_{\CI})$, or $m_{\NII} < m_{\vL} < m_{\CI}$,
or $m_{\C} < m_{\vL}$ respectively.  However, in these cases
the cross section for $\lL\vL$ at the Tevatron would be 
quite large (see below). 

The dominant signal of $\NII\NII$ production is
$\ph\ph + \E$ in all models.  Note that the process
$\sigma(\NII\NII)$ is always accompanied by
$\sigma(\NII\NIII)$ at a comparable rate 
(when kinematically allowed), 
which has the same signatures 
as $\NI\NIII$ production (as above) plus one photon.

$\CIp\CIm$ production can be present with
a large rate, the decay signature of $\CI$ being the usual
``$W$''$+ \E$ in all models (see Table~\ref{BR_C1-table}).  
The exception is if $m_{\vL}$ (and possibly $m_{\lL}$) is
lighter than $m_{\CI}$, which can happen only
in $\eR$ models.  In this special $\eR$ model scenario, 
if $m_{\vL} < m_{\CI}$, then the decay signature is 
likely invisible.  However, 
if the decay $\vL \ra \NII \nu$ is large, then
the signature is $\ph\ph + \E$.  If 
$m_{\lL} < m_{\CI}$, then additional possible
signatures are $l^+l^- + \E$ (if $m_{\lL} < m_{\NII}$), 
or $l^+l^-\ph\ph + \E$ (if $m_{\lL} > m_{\NII}$).
Notice that the latter could be an additional
source of $\eegg$ events 
(see Appendix~\ref{char-sec}).
These remarks assume the stop is heavier than $\CI$.  

As an aside, we find that a maximum of (14, 13, 12) 
$\NI\NI$ pairs can be produced, which can be observed
as a $\ph + \E$ signal once visible initial state 
radiation is attached.
Although the SM background is severe,
there are other possibly important contributions 
from e.g.\ $\ph\NI\NIII (\ra \nu\overline{\nu}\NI)$.

\begin{figure}
\centering
\epsfxsize=4in
\hspace*{0in}
\epsffile{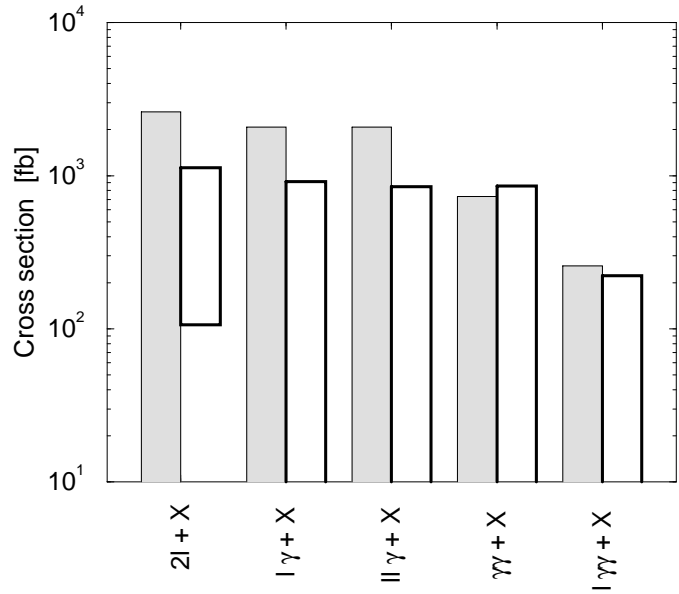}
\caption{Range of inclusive cross sections for selected signatures
at LEP161 without detection efficiencies; all signatures 
necessarily have missing energy
in addition to that above.  The shaded bar on the left
corresponds to $\eL$ models and the thick solid outline
on the right corresponds to $\eR$ models.
Here, $X =$ leptons, photons, jets, or nothing, and $l = e$, $\mu$, 
or $\tau$ summed over both charges and all three families.
See the text for details.}
\label{signature-LEP161-fig}
\end{figure}

In Fig.~\ref{signature-LEP161-fig} we present the
ranges of the inclusive production of particular signals
at LEP161 for $\eL$ and $\eR$ models.  
These signatures were generated by 
searching all possible decay paths.  No efficiencies resulting 
from detector geometry or lepton/photon energy cuts are
included.  If the signals are the result of decays with 
moderate mass splittings, then presumably some of the events
could be detected after applying reasonable cuts.
A lepton $l$ can be either $e$, $\mu$ or $\tau$, 
with either charge $\pm 1$.  In particular, when referring 
to a ``$2l$'' signal, we sum over all family and charge
possibilities (including, e.g., like-sign dileptons).
$X$ can be any combination of leptons, photons, jets, or nothing.
In addition, all the signals implicitly include missing 
energy in their signature.
We only include chargino/neutralino production processes
in the inclusive sum, since $\eL$ in $\eL$ models and 
$\eR$ in $\eR$ models is too heavy to be produced.
If the other slepton ($\eR$ in $\eL$ models, $\eL$ in $\eR$ 
models) is light, then the maximum cross section
for particular signatures can be higher.

Jet production is also an important signal.
If $\NI\NIII$ production is kinematically allowed and if
only 3-body decays of $\NIII$ occur, 
then the rate into the $jj + X + \E$ signal is between
roughly $400$--$1800$ fb for both $\eL$ and $\eR$ models.  
If chargino production is open then the rate can be larger.  
But, if 2-body decays into sleptons are open for $\NIII$, 
then the rate can be near zero.

Notice that in $\eR$ models only the $2 l + X (+ \E)$ must 
be produced, the rate being between $\sim 2$ to $20$
events in $25$~pb$^{-1}$ of integrated luminosity.
The reason that the $2 l$ rate always has a non-negligible minimum 
is due to a combination of effects:  $\eR$ models have
a minimum $\sigma( \NI\NIII ) \gsim 850$~fb, and
decays of $\NIII \ra \NI l^+ l^-$ are always non-zero,
even if 2-body decays operate.  If only 3-body 
decays occur, then $\NIII \ra \NI$``$Z$''$(\ra l^+l^-)$
occur, with a rate of nearly $10\%$ (summed over families).  
Alternatively,
if $\NIII \ra \vL \nu$ is open, then $\vL \ra l \C (\ra l \nu)$
is the decay pattern.  If $m_{\vL} < m_{\C}$, then
it turns out that $m_{\lL} < m_{\NIII}$, and so
decays $\NIII \ra l \lL (\ra l {\tilde N}_{1,2})$ 
are non-zero, giving an appreciable $2 l$ signal.
All of the other inclusive signals could have rates
smaller than the one event level.  If one of these
signatures were found (and deduced to be above background),
then looking in the other channels might serve to confirm
the signal.

\begin{figure}
\centering
\epsfxsize=4in
\hspace*{0in}
\epsffile{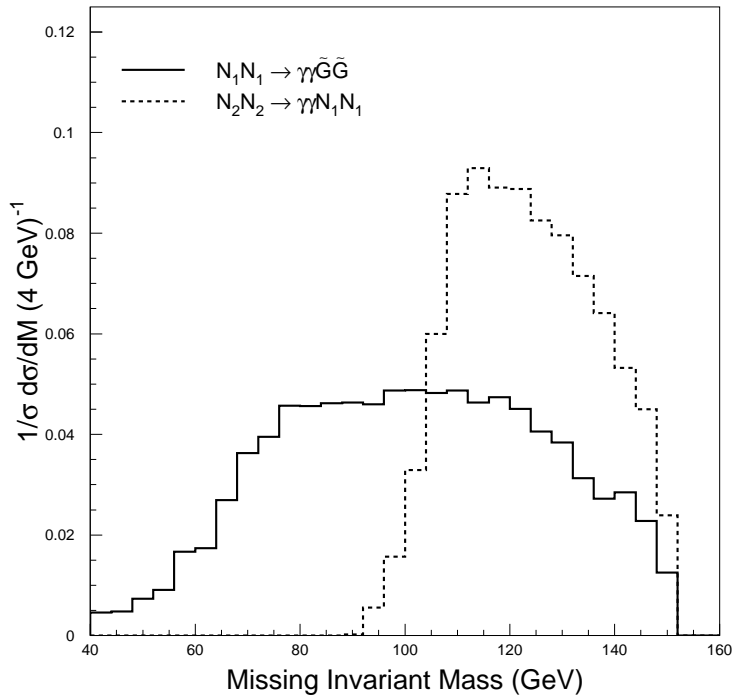}
\caption{Comparison of the missing invariant mass distribution in 
the $\ph\ph + \E$ signal at LEP161 from two different selectron
interpretation models, (a) a sample $\NI=$LSP model 
with $m_{{\tilde N}_{1,2}} = 37, 65$~GeV (dashed line), and
(b) a model with a gravitino LSP and $m_{\NI} = 65$~GeV (solid line).}
\label{gravitino-compare-fig}
\end{figure}

One promising signal is $\ph\ph + \E$ without any other event
activity, which primarily originates from $\NII\NII$ production in 
the selectron interpretation.  (This is part of the inclusive
signal $\ph\ph + X + \E$ described above.)  In a scenario with
a gravitino LSP, we found that the standard model background 
for $\ph\ph + \E$ is distinguishable from the gravitino signal
$e^+ e^- \ra \NI \NI \ra \ph\ph \G\G$ 
using the missing invariant mass distribution~\cite{Grav}.  
Here, we point out that a selectron interpretation
with a neutralino LSP can be distinguished from one with
a gravitino LSP using the missing invariant mass distribution,
assuming the SM background is small (see Ref.~\cite{Grav} for 
a discussion of the background).
In Fig.~\ref{gravitino-compare-fig} we show the missing
invariant mass distribution 
$M^2_{\rm inv} = (p_{e^+} + p_{e^-} - p_{\ph_1} - p_{\ph_2})^2$
at LEP161 for two different models:
(a) The $\eL$ sample model in Appendix~\ref{sample-models-sec} 
with $m_{{\tilde N}_{1,2}} = 37, 65$~GeV, and
as usual $\NI \simeq {\tilde H}_b$, $\NII \simeq \phino$.
(b) A model with a (very light) gravitino LSP with 
$m_{\NI} = 65$~GeV, and $\NI \simeq \phino$.  The difference in
the missing invariant mass distribution illustrates how the scenarios 
might be distinguished using the $\ph\ph + \E$ signal.  
It should be noted the general character
of the missing invariant mass distribution for the gravitino LSP model
in Fig.~\ref{gravitino-compare-fig} is not particularly sensitive 
to $m_{\NI}$, but simply that $m_{\G}$ is very small compared to
the neutralino or selectron masses.

\subsubsection{LEP190}
\label{LEP190-subsec}
\indent

\begin{figure}
\centering
\epsfxsize=4in
\hspace*{0in}
\epsffile{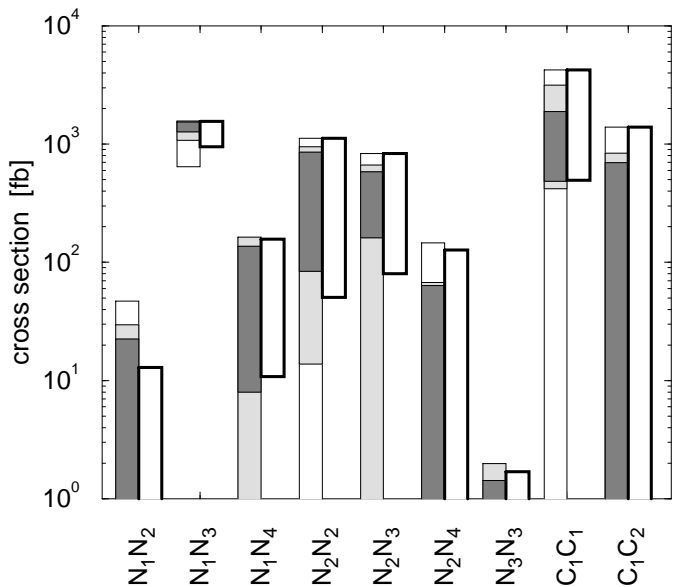}
\caption{As in Fig.~\ref{sigma_LEP161-fig}, for LEP190.}
\label{sigma_LEP190-fig}
\end{figure}

In Fig.~\ref{sigma_LEP190-fig}, we present all the 
chargino/neutralino production processes with cross sections 
possibly larger than about $1$~fb for LEP with $\sqrt{s} = 190$~GeV\@.  
As above, the cross sections were computed with initial 
state radiation effects included.  Now, $\NI\NIII$ production 
must be large in all models, and many other processes can 
easily give large rates.  The processes with large
rates include all of the ones at LEP161, and also $\NI\NII$, 
$\NI\NIIII$, $\NII\NIIII$, $\CIpm\CIImp$.  The maximum
rates are as follows:
\begin{equation}
\begin{array}{rll}
e^+e^- \ra & \NI\NII      & (20, \; 6, \; 24)        \\
           & \NI\NIII     & (785, \; 780, \; 780 )   \\
           & \NI\NIIII    & (82, \; 79, \; 78)       \\
           & \NII\NII     & (505, \; 560, \; 346)    \\
           & \NII\NIII    & (335, \; 416, \; 230)    \\
           & \NII\NIIII   & (73, \; 64, \; 34)       \\
           & \CIp\CIm   & (965, \; 2120, \; 1195)  \\
           & \CIpm\CIImp  & (409, \; 695, \; 350) 
                            \; \; {\rm pairs \; \> produced}, \\
\end{array}
\end{equation}
for ($\eL$, $\eR$, $\eL + \eR$) models.  For 
$\NI\NIII$ production, the minimum number of pairs produced
is (400, 475, 320) for ($\eL$, $\eR$, $\eL + \eR$) models
given the minimum threshold ${\cal A} = 5$~fb.
For $\eR$ models only, a minimum of $5$ $\NI\NIIII$ pairs, 
$25$ $\NII\NII$ pairs, $40$ $\NII\NIII$ pairs, and $250$ $\CIp\CIm$
pairs must be produced given the minimum threshold
${\cal A} = 5$~fb.  As for $\NI\NI$ pair production we found
a maximum of (177, 164, 152) pairs can be produced.

The detection signatures for the chargino/neutralino pairs
common to LEP161 are the same as above.  Here we discuss
the processes that are different.  First, the 
process $\NI\NII$ gives a $\ph + \E$ signature.  The
signatures for $\NI\NIIII$ and $\NII\NIIII$ are entirely dependent
on the $\NIIII$ branching ratio; $\NIIII$ can decay in
a variety of ways outlined in Sec.~\ref{neut-BR-subsec}.
Perhaps the most striking signature is when $\NIIII \ra \NI h (A)$,
giving a $b\overline{b} + \E$ signature for $\NI\NIIII$ 
production and $b\overline{b}\ph + \E$ signature for $\NII\NIIII$
production.
The signature of the process $\CIpm\CIImp$ also depends
crucially on the branching ratio of $\CII$, but
one lepton with perhaps one photon plus missing energy
is typical (assuming the stop is heavier than $\CI$).
Thus, a reasonable expectation for $\CIpm\CIImp$
production is $l^+l^- (+ \ph) + \E$.  It is also possible
that only 3-body decays of $\CII$ are open, in which case
no photon would appear in the final state.
The final states from $\CII$ decay are summarized
in Table~\ref{BR_C2-table}.

In addition, sneutrino pair production (if open) is another
process that is relevant for $\eL$ models. 
To have $\veL\veL$ production kinematically 
accessible with $m_{\veL} < 95$~GeV, then the sum rule in
Eq.~(\ref{sum-rule-eq}) implies $\tbeta \gsim 1.2$ 
is required for $m_{\eL} > 100$~GeV (as needed 
by the kinematics of the $\eegg + \Et$ event),
and for $m_{\eL} = 107 \, (118)$~GeV, then $\tbeta > 1.5 \, (2.8)$.
Hence sneutrino production in $\eL$ models 
never occurs at LEP190 if $m_{\eL} > 118$~GeV\@.
The signature of $\veL\veL$ production depends
on the sneutrino branching ratio, but it was
already established in Sec.~\ref{sneutrino-BR-subsec} that
$\veL \ra e \CI (\ra \NI$``$W$''$)$ is the
dominant decay pattern.  Thus the signature
is $ee$``$W$''``$W$''$ + \E$, which is indeed
quite prominent.

\begin{figure}
\centering
\epsfxsize=4in
\hspace*{0in}
\epsffile{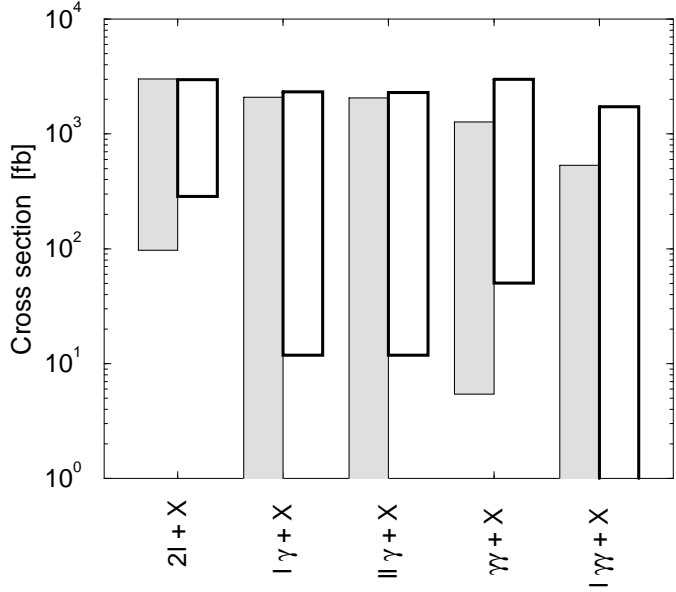}
\caption{Range of inclusive cross sections for selected signatures
without detection efficiencies, as in 
Fig.~\ref{signature-LEP161-fig}, but for LEP190.}
\label{signature-LEP190-fig}
\end{figure}

In Fig.~\ref{signature-LEP190-fig} we present the
ranges of the inclusive production of particular signals
at LEP190 for $\eL$ and $\eR$ models.  As in 
Fig.~\ref{signature-LEP161-fig}, no detection efficiencies 
are included.  Notice that
while only $\NI\NIII$ production had a non-negligible minimum rate 
(see Fig.~\ref{sigma_LEP190-fig}), both the signals
$2l + X + \E$ and $\ph\ph + X + \E$ (which rarely
comes from $\NI\NIII$ production) are always larger 
than one event.  Further, inclusive production 
of $l\ph + X + \E$ and $ll\ph + X + \E$ are
always larger than the $10$ event level for
$\eR$ models only.  
All of the other inclusive signals could have rates
smaller than the one event level.  As in LEP161, if one of these
signatures were found (and deduced to be above background),
then looking in the other channels might serve to confirm
the signal.  Another important search strategy would be 
inclusive signatures with jets ($+$ photon(s)) that can have 
significantly larger rates than the lepton(s) ($+$ photon(s)) 
signatures.

\subsection{Predictions for Tevatron}
\label{tevatron-subsec}
\indent

The assumption underlying the selectron interpretation 
is that the Tevatron has already observed a candidate 
selectron pair production event.  
Because many more states of the underlying supersymmetric 
model are accessible at a hadron collider, we here focus on the 
associated signals that should be observed in the present 
data set (100 pb$^{-1}$ per detector) or in the next scheduled 
upgrade (1--2 fb$^{-1}$ per detector).  As in Sec.~\ref{LEP-subsec}, 
we identify the processes that have non-negligible production
cross sections, then determine the possible signatures
that depend on the branching ratios.  Again, 
it is important to emphasize that the following
predictions assume the minimum cut ${\cal A} = 5$~fb
is placed on the $\sigma \times \BR^2$ for the $\eegg + \Et$
event.

\begin{figure}
\centerline{
\epsfxsize=4in
\epsffile{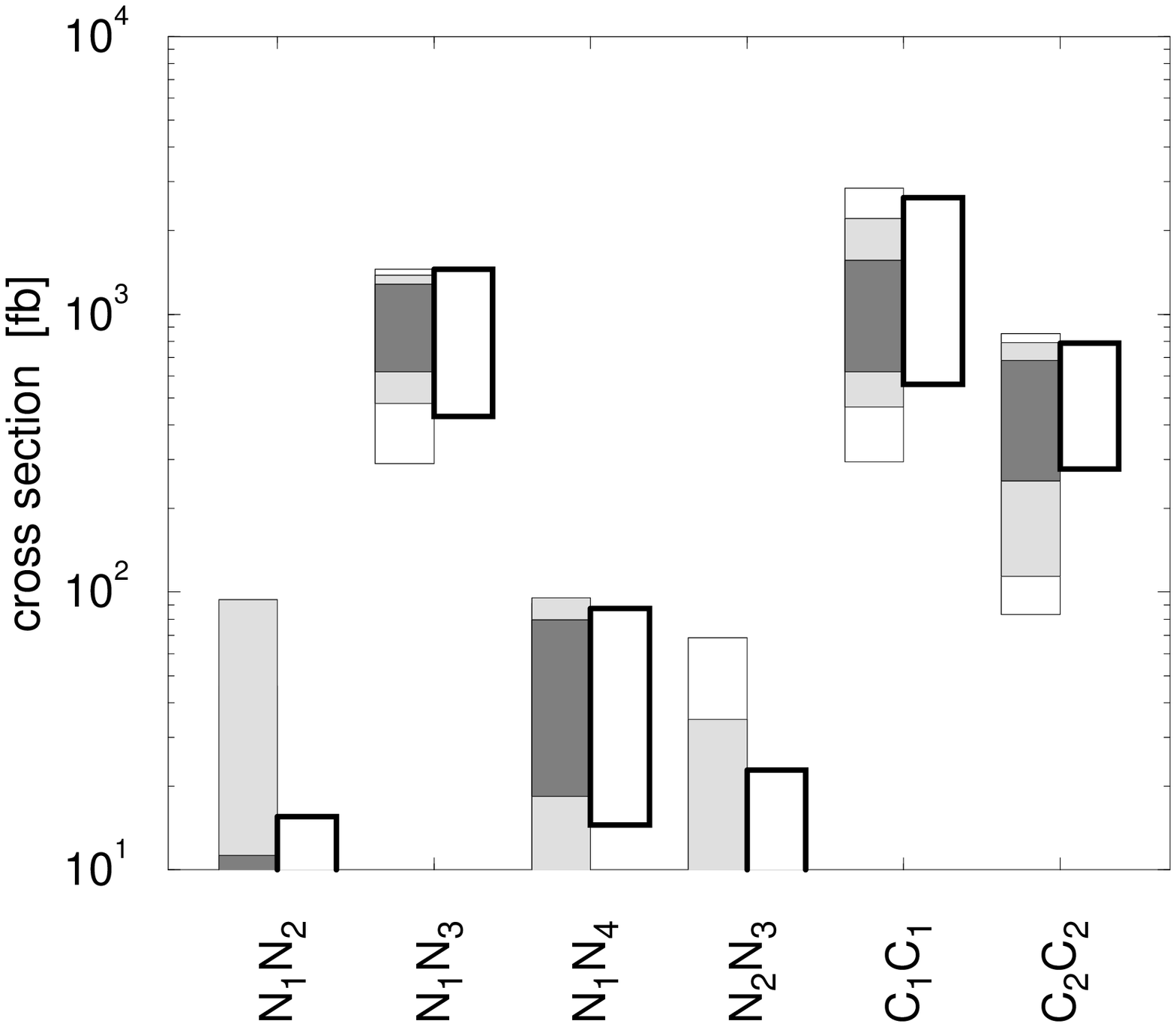}}
\centerline{
\epsfxsize=4in
\epsffile{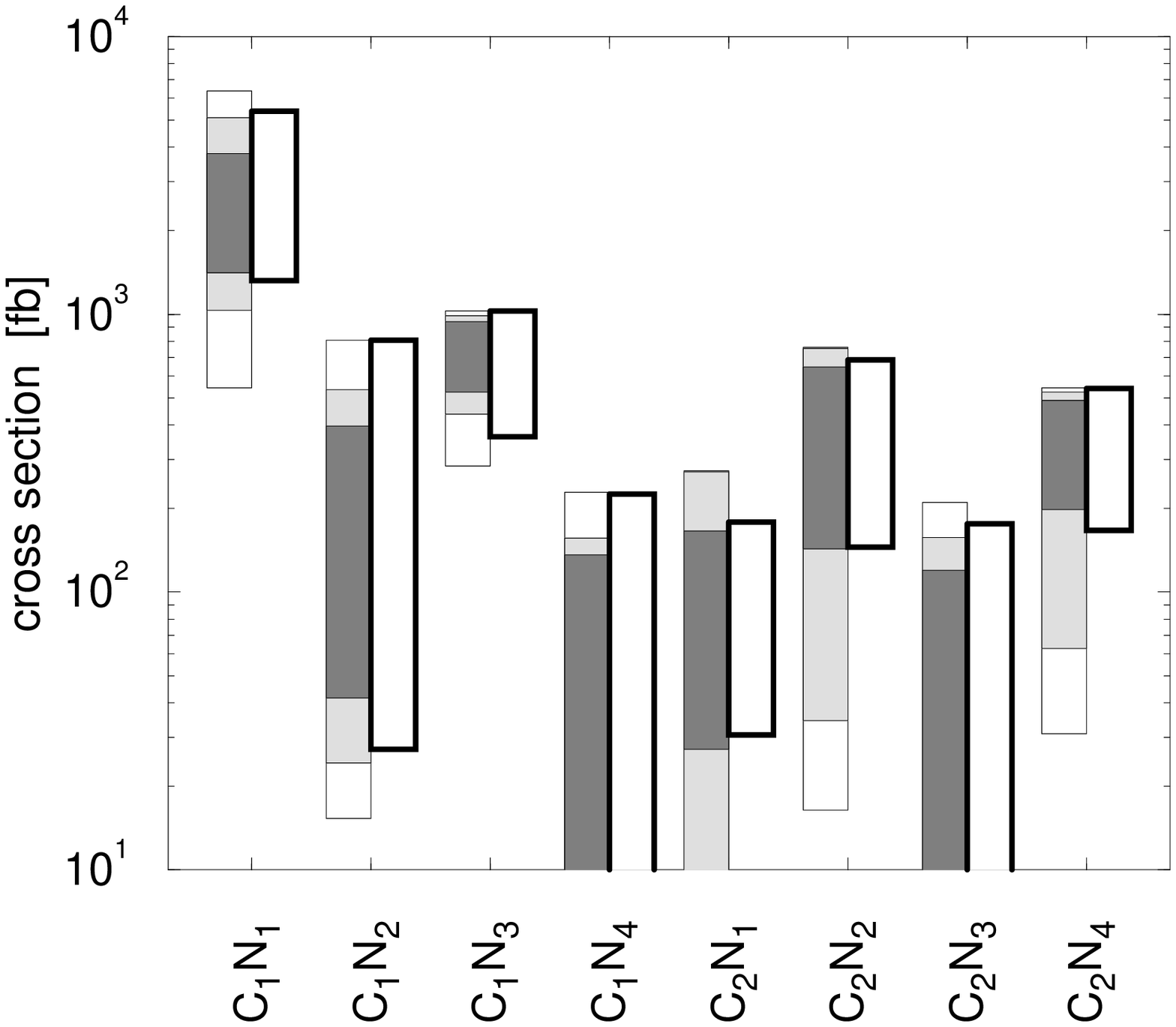}}
\caption{As in Fig.~\ref{sigma_LEP161-fig}, for Tevatron 
$\protect\sqrt{s} = 1.8$~TeV with all $\Ci\Cjmp$, $\Ni\Nj$, $\Ci\Nj$ 
processes shown that can have cross sections larger than about $50$~fb.}
\label{sigma_tevatron-fig}
\end{figure}

In Fig.~\ref{sigma_tevatron-fig} we present all of the 
chargino/neutralino production processes that can have cross
sections above about $50$~fb.  We use leading order CTEQ3L~\cite{CTEQ3}
structure functions evaluated at $Q^2 = \hat{s}$.
At the Tevatron the
cross sections do have a contribution from $t$-channel squark 
exchange (see Table~\ref{nn_cc_nc-table}), but the dependence
on the squark mass is usually weak for the squark masses
in our models.  
If only $\eR$ is light, then these 
are the only necessary associated processes to the $\eegg$ 
event.  If $\eL$ is light, however, then
there must also be $\eL\veL$ and $\veL\veL$ production.
In Ref.~\cite{PRL} we found $\sigma(\veL\veL) \sim \sigma(\eL\eL)$
and $\sigma(\eL\veL) \sim (2 \ra 3) \sigma(\eL\eL)$ for 
the same $m_{\eL}$, i.e.\ the cross sections are typically tens of fb.
It is also possible that both $\eL$ and $\eR$ can be light;
in particular, the other slepton 
($\eR$ in $\eL$ models, $\eL$ in $\eR$ models) can be lighter
than the one giving the $\eegg + \Et$ event, which can 
dramatically affect the signatures.
The pair production processes that have the largest
cross sections and also have a non-negligible minimum cross section 
are given in Table~\ref{pairs-produced-tevatron-table},
where the full range from the minimum to the maximum number of 
pairs produced for an integrated luminosity of $100$~pb$^{-1}$ 
are shown.

\begin{table}
\renewcommand{\baselinestretch}{1.2}\small\normalsize
\begin{center}
\begin{tabular}{crclrclrcl} \hline\hline
Process &  \multicolumn{3}{c}{Range in $\eL$ models} & 
    \multicolumn{3}{c}{Range in $\eR$ models} &
    \multicolumn{3}{c}{Range in $\eL + \eR$ models}  \\ \hline
$\NI\NIII$     & \phantom{000} $31$ & $\ra$ & $129$ 
               & \phantom{000} $43$ & $\ra$ & $145$  
               & \phantom{0000} $29$ & $\ra$ & $128$   \\
$\CIp\CIm$   &  $40$ & $\ra$ & $285$ 
               &  $56$ & $\ra$ & $264$  
               &  $29$ & $\ra$ & $258$   \\
$\CIIp\CIIm$ &   $8$ & $\ra$ &  $85$  
               &  $28$ & $\ra$ &  $79$   
               &  $15$ & $\ra$ &  $77$   \\
$\CIpm\NI$     &  $75$ & $\ra$ & $638$ 
               & $132$ & $\ra$ & $540$  
               &  $54$ & $\ra$ & $552$   \\
$\CIpm\NII$    &   $2$ & $\ra$ &  $75$  
               &   $3$ & $\ra$ &  $80$   
               &   $1$ & $\ra$ &  $75$   \\
$\CIpm\NIII$   &  $32$ & $\ra$ &  $98$  
               &  $36$ & $\ra$ & $103$  
               &  $28$ & $\ra$ &  $96$   \\
$\CIIpm\NII$   &   $2$ & $\ra$ &  $76$  
               &  $15$ & $\ra$ &  $69$   
               &   $5$ & $\ra$ &  $74$   \\
$\CIIpm\NIIII$ &   $3$ & $\ra$ &  $51$  
               &  $17$ & $\ra$ &  $54$   
               &   $8$ & $\ra$ &  $55$   \\ \hline\hline
\end{tabular}
\end{center}
\caption{The range of the number of chargino/neutralino pairs 
produced at the Tevatron assuming an integrated luminosity 
of $100$~pb$^{-1}$.  The processes displayed here include
those that have both a large production rate {\em and}\/ 
a non-negligible minimum production rate.}
\label{pairs-produced-tevatron-table}
\end{table}

The signatures for $\NI\NIII$ and $\CIp\CIm$ are the same
as for LEP (described in Sec.~\ref{LEP161-subsec}) and
the decays of $\CII$ were also discussed in Sec.~\ref{LEP190-subsec}. 
For completeness we list the possible signatures of
all of these processes here:
$\NI\NIII$ will mainly give ``$Z$''$+\Et$, or $b\overline{b} + \Et$
if $m_A < 60$~GeV\@.  If 2-body slepton decays are allowed,
then in $\eL$ models one can have $l^+l^- + \Et$,
or in $\eR$ models one of invisible,
$\ph + \Et$, or $l^+l^- + \Et$.  $\CIp\CIm$ production
gives typically $l^+l^- + \Et$, or if 2-body decays
into $\lL$, $\vL$ occur (in $\eR$ models only), then depending
on the mass hierarchy one can have $\ph\ph + \Et$, or $l^+l^- + \Et$,
or $l^+l^-\ph\ph + \Et$.  $\CIIp\CIIm$ production
gives similar signatures as $\CIp\CIm$ production, 
given $m_{\CI} \ra m_{\CII}$ and allowing for the possibility
of 2-body decays in the context of both $\eL$ and $\eR$ models
as above.

The processes $\Ci\Nj$ are unique to the Tevatron, 
with $\CIpm\NI$, $\CIpm\NII$, $\CIpm\NIII$, $\CIIpm\NII$
and $\CIIpm\NIIII$ giving the largest rates.  As described
above, the chargino typically gives $jj + \Et$ and $l^\pm + \Et$, 
although possible 2-body decays into sleptons can
give $\ph + \Et$, or $l^\pm + \Et$, or $l^\pm \ph + \Et$.
Thus the signature of $\CIpm\NI$ production is one
of the above signatures for a single chargino.  The
signatures of $\CIpm\NII$ and $\CIIpm\NII$ are as above
plus one photon.  Finally, the decays $\CIpm\NIII$ and
$\CIIpm\NIIII$ are one of the above signatures coupled
with $\NIII$ or $\NIIII$ decay.  Here again we can utilize
Secs.~\ref{neut-BR-subsec},~\ref{LEP-subsec}, to obtain
the possible decay signatures.  For $\NIII$, the decay
signature is ``$Z$''$+ \Et$, $b\overline{b} + \Et$ (if
$m_A < 60$~GeV), and if 2-body decays to sleptons
are open then for $\eL$ models the signature could be 
$l^+l^- + \Et$ or $\ph l^+l^- + \Et$, while for $\eR$ models
the signature could be invisible,
or $\ph + \Et$, or $\l^+l^- + \Et$.  Thus, if only 
3-body decays were open for charginos and neutralinos
the signature of $\CIpm\NIII$ and $\CIIpm\NIIII$ would 
be ``$W$''``$Z$''$+ \Et$, which gives the well-studied 
trilepton signal~\cite{FermilabPaper}.  If 2-body 
decays of the charginos or heavier neutralinos
are present, then one or more photons could be present
in the final state, with possibly fewer leptons.

\begin{figure}
\centering
\epsfxsize=4in
\hspace*{0in}
\epsffile{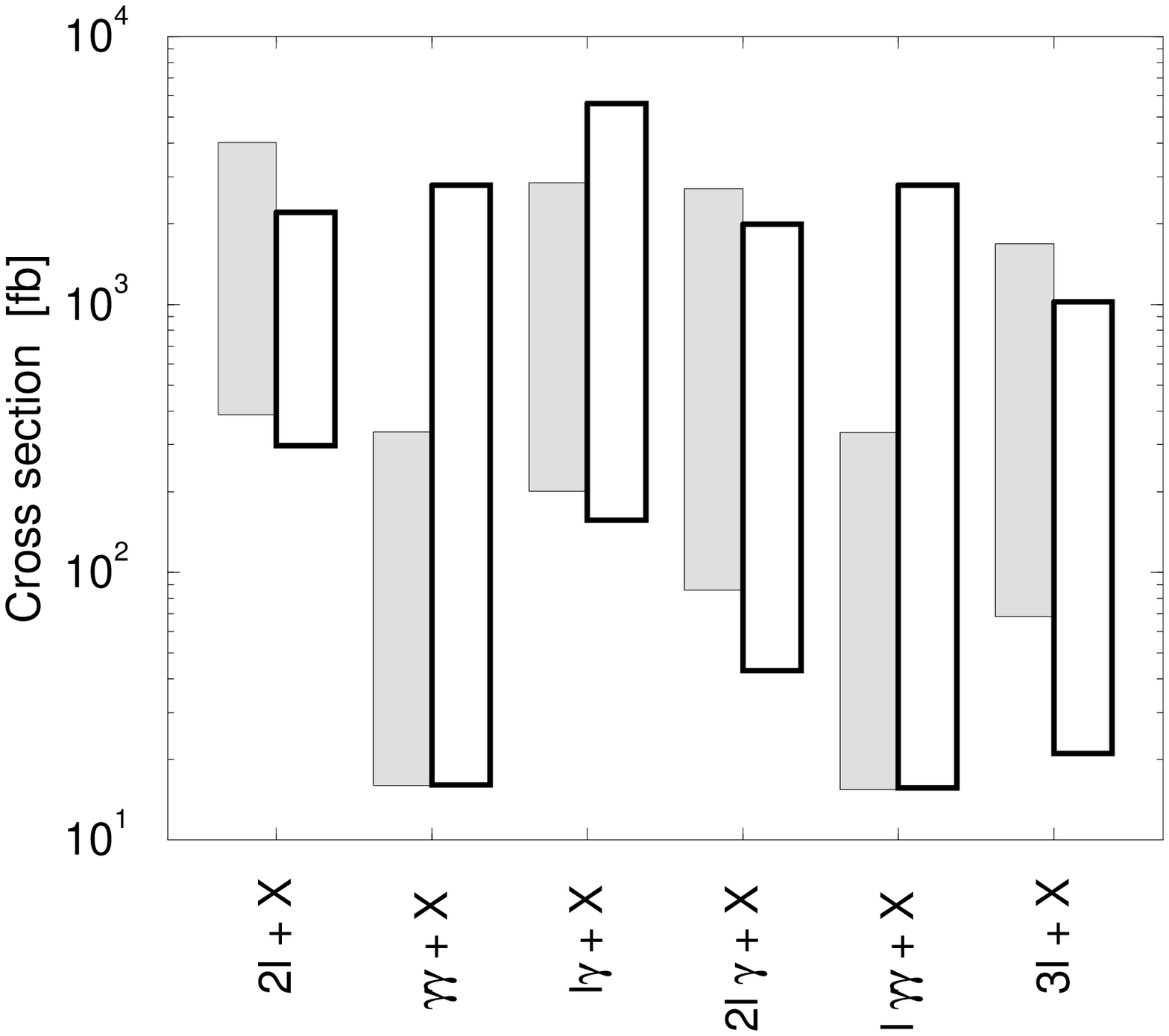}
\caption{Range of inclusive cross sections for selected signatures
without detection efficiencies, as in 
Fig.~\ref{signature-LEP161-fig}, but for Tevatron.}
\label{signature-tevatron-fig}
\end{figure}

In Fig.~\ref{signature-tevatron-fig} we present the cross section
for many promising signatures at the Tevatron.  
As in Figs.~\ref{signature-LEP161-fig} and \ref{signature-LEP190-fig},
no detection efficiencies have been included.  
We include chargino/neutralino processes in the sum, 
as well as $\eR\eR$ production in $\eR$ models, and
$\eL\eL$, $\veL\veL$, $\eL\veL$ production in $\eL$ models.
We see that all six inclusive signatures involving leptons
or photons are expected to have minimum rates of roughly
$2$ to $30$ events, regardless of the type of model
($\eL$ or $\eR$).  The $\ph\ph + X$, $\l\ph + X$
and $l\ph\ph + X$ signatures can be much larger in $\eR$
models, but this only happens in the particular kinematic
scenario with $m_{\NII} < m_{\lL}, m_{\vL} < m_{\CI} (< m_{\eR})$. 
In this case, charginos always decay through the 2-body
channels $\C \ra \lL \nu$ and $\C \ra \vL l$, with
$\lL, \vL \ra \NII (\ra \NI \ph)$.  Thus, processes with
intrinsically large cross sections such as $\CI\NI$ production 
can lead to a large $l\ph + \Et$ signal, and similarly for
other processes involving charginos.  

The $l\ph + \Et$ (and $jj\ph + \Et$) signals are 
important~\cite{KaneMrennaGluino} and can 
arise from: $\CI\NII$ and $\CII\NII$ production in models 
with $\C \ra l\nu \NI$; 
$\C_i \NI$ production in models with 
$\C_i \ra \nu \lL (\ra l \NII (\ra \NI \ph))$ or 
$\C_i \ra l \vL (\ra \nu \NII (\ra \NI \ph))$;
and $\lL \vL$ production with $\lL \ra l \NII (\ra \NI \ph)$.
The chargino decays assume $m_{\tI} > m_{\C}$.  For just
${\tilde C}_i \NII$ production there are
roughly $10$--$130$ pairs produced in the present CDF and D0 
samples (each) with the probable signatures $\ph +$``$W$''$ + \E$ 
(before cuts); ``$W$'' decays to $jj$ or $l^\pm \nu$ as usual.  
For ``$W$''$\ra jj$, these events have no parton-level SM 
background.

Many of these signatures should be detectable, since the 
mass differences between superpartners is often constrained 
to be small but non-zero, as in Fig.~\ref{mass-difference-fig}. 
For example, in decays such as $\NIII \ra \NI$``$Z$''
and ${\tilde C}_{1,2} \ra \NI$``$W$'', the invariant
mass of the virtual ``$Z$'' or ``$W$'' can be large. 
In particular, the invariant mass of the ``$Z$'' from
$\NIII$ decay is between $0$ to $40$--$60$~GeV, thus an excess
in pairs of leptons (or jets) that reconstruct to
an invariant mass $m_{l^+l^-} \lsim 60$~GeV accompanied
by a large missing energy is a distinctive
signature of $\NI\NIII$ production in our models.

In addition to classifying the most promising signatures,
we have also performed a number of event level simulations for
a limited subset of our $\eL$ and $\eR$ models, with the
other slepton heavy.  The purpose is to get a feeling
for the efficiency of detecting multi--lepton and/or 
photon signatures.  First, we address
the issue of efficiencies for the $\eegg + \Et$
event, since this is important for interpreting the
threshold ${\cal A}$ in the $\eegg$ rate.
An efficiency represents the probability that a certain 
class of events passes a particular set of cuts defined 
before the data is analyzed.  We chose a set of cuts such that:
(1) the event would be triggered on and analyzed, and
(2) the event would not suffer from obvious detector backgrounds
like jets faking leptons or photons.  To show the dependence
of our efficiencies on the particular set of cuts, we choose
a loose set with $|\eta^{e}| < 2$, $|\eta^\ph| < 1$, 
$(p_T^{(e,\ph)}, \Et) > E_T^{\rm min} = 10$~GeV 
and a tight set identical to the loose set except 
$E_T^{\rm min} = 20$~GeV.  The efficiencies we
found range from $0.02$--$0.23$
for the loose cuts, and from $0.01$--$0.12$
for the tight cuts, but efficiencies outside these ranges
(from models not covered in the subset) are possible.  
If $E_T^{\rm min}$ is 
increased to $25$~GeV, the mean efficiency is $0.04$.
The loose cuts are sufficient for CDF to have triggered on
the $\eegg$ event.
 
We have also studied $l\ph$, $ll$, and $ll\ph$ signatures 
using a similar set of cuts ($\Et > 20$~GeV and 
$p_T^{(l,\ph)} > E_T^{\rm min}$), where for the purposes
of detection $l$ is summed over $e$ and $\mu$ only.
Typically, when $E_T^{\rm min} = 10$~GeV
one expects between $1$--$5$ ($2$--$12$) $l\ph + \Et$ events 
in 100 pb$^{-1}$ for $\eR$ models ($\eL$ models) from 
chargino/neutralino production alone.  
An additional $1$--$2$ events are expected from $\eL\veL$ 
production in $\eL$ models.  This result is essentially 
unchanged for the simulation subset of models 
if $E_T^{\rm min} = 20$~GeV\@.  This is expected at least
for the photons since the kinematics enforce hard photons
in the final state from slepton decay.
The SM background from $W\ph$ production yields
$105$ and $37$ events for each set of cuts respectively.  
With tighter cuts, it is possible to achieve a signal to 
background ratio near one for some models.
The expected $ll$ signature, resulting mainly from 
${\tilde C}_i {\tilde C}_j$ production, is between $0$--$6$ 
events for $E_T^{\rm min} = 10$~GeV, and $0$--$2$ 
for $E_T^{\rm min} = 20$~GeV\@.  Similarly, the expected 
$ll\ph$ signature is between $0$--$5$ events 
for $E_T^{\rm min} = 10$~GeV, and $0$--$2$ events
for $E_T^{\rm min} = 20$~GeV\@.  Other signatures, such 
as $\ph\ph$, $l\ph\ph$
and $3l$, produce at most $1$ or $2$ events for 
$E_T^{\rm min} = 10$ or $20$~GeV\@.  Therefore, it would appear that
the $l\ph$ channel is the most promising for confirming the
supersymmetric interpretation of the $\eegg$ event
(assuming $m_{\tI} > m_{\C}$), though other signals 
with limited backgrounds are clearly possible.

\subsection{Alternative interpretation}
\indent

Throughout this section we have described the constraints
and predictions in the selectron interpretation.  However,
in Sec.~\ref{kinematics-sec} we described an alternative 
interpretation involving chargino production that could
explain the $\eegg + \Et$ event.  Those readers interested in 
the model building associated with the chargino interpretation 
are referred to Appendix~\ref{char-sec}, which provides many 
details and an example model.

\section{Comments on models with a light stop}
\label{light-stop-sec}
\indent

We have seen that the effect of requiring 
a large $\sigma \times \BR^2$ for the $\eegg + \Et$ event 
is to strongly constrain the chargino, neutralino and 
slepton sections.  Up to now, we have assumed the squarks 
are sufficiently heavy so as not to directly interfere 
with the necessary decay chain.  However, it is possible
that a light stop $\tI$ can exist simultaneously
with the needed hierarchy in the other sectors.  
In particular, neutralino decays $\N \ra \tI t$
are absent in our models (with the cut ${\cal A} = 5$~fb), 
since all neutralinos are lighter than the top quark.
Therefore, the decay chain in the selectron interpretation
need not be disrupted, if the radiative neutralino decay
can be large with a light stop.  

The chargino interpretation described in Appendix~\ref{char-sec}
is a different matter, since charginos would always 
decay to the light stop $\C \ra \tI b$
if kinematically accessible.
This is true regardless of the 
mixing angle $\theta_{\tilde t}$ that determines the 
${\tilde W}^\pm$--${\tilde t}$ coupling, since the Yukawa 
coupling ${\tilde H}^\pm$--${\tilde t}$ is large.
Thus, it would seem that a chargino interpretation of the 
$\eegg + \Et$ event from $p\overline{p} \ra \Ci\Cjmp$ is not 
possible unless $m_{\tI} > m_{\C}$.  This is
basically the scenario described in Appendix~\ref{char-sec}.

To construct models with a large $\eegg + \Et$ event rate and a 
light stop, one must consider the effects of a small $m_{\tI}$ 
on the radiative neutralino decay width and on the mass hierarchy.
As we have remarked in Sec.~\ref{model-building-results-subsec}, 
the dynamical mechanism for a large radiative neutralino branching 
ratio appears not to be strongly dependent on 
$m_{\tI}$~\cite{AmbrosMele2}.  For instance, 
models can be constructed with $m_{\tI} = 50$~GeV, 
$m_{{\tilde t}_2} \gsim 250$~GeV, and a large radiative 
neutralino branching ratio arising from the dynamical mechanism.
However, some suppression to the radiative neutralino
branching ratio from light stops is present,
so the $\eegg + \Et$ rate is maximized in the limit
of all squark masses large.  For example, the largest
$\eegg + \Et$ rate in $\eL$ and $\eR$ models 
with a light stop is $13.8$~fb and $6.1$~fb 
respectively.
Since $m_{\NI} < m_{\tI}$ must be obeyed so that
$\NI=$LSP, the upper limit on
$m_{\NI}$ can be more restrictive than found above 
if $m_{\tI} \lsim 74 \; (50)$~GeV
in $\eL$~($\eR$) models by Observation~1 in Sec.~\ref{kinematics-sec}.
This induces a rough upper limit on $|\mu|$, which
also has implications for the chargino masses.

There is an additional degree of freedom in the value of
$\theta_{\tilde t}$, which determines the SU(2) couplings
of ${\tilde t}_{1,2}$ with the gaugino components of 
charginos and neutralinos.  Maintaining a hierarchy
between $m_{\tI} \ll m_{{\tilde t}_2} \> (\simeq m_{\tilde q})$
would seem difficult without giving a large 
$\delta \rho$~\cite{DreesDeltarho}, but this can be avoided 
if $\tI \simeq \tR$ (or $\theta_{\tilde t} \simeq \pi/2$ 
by our definition).  However, requiring $m_{\tI} \simeq 50$~GeV,
$\theta_{\tilde t} \simeq \pi/2$ implies $m_{{\tilde t}_2}$
must be large (perhaps of order $1$~TeV or more) if the light Higgs 
$h$ is to have a mass that 
is not excluded by LEP.
In general this implies that $m_h$ will 
lie within the region accessible to LEP, though further 
analysis is needed to be precise; $\sin^2(\beta - \alpha)$ 
can be below one, and $m_h$ can be near its present lower limit
from LEP1.  Note also $\theta_{\tilde t}$ slightly
affects the radiative neutralino decay~\cite{AmbrosMele2}.

Another constraint on models with a light stop 
comes about if $m_{\tI} + m_b < m_t$.  Then 
top quarks must decay into stops with a
branching fraction of about $1/2$ if $m_{\tI} \sim 50$~GeV\@.
It was observed in
Ref.~\cite{KaneMrennaGluino} that a branching ratio of
$t \ra \tI b$ of $50\%$ is not excluded by 
Tevatron data, if gluinos and squarks
with masses of roughly ${\cal O}(250)$~GeV\@ exist, giving
additional top production to supplement the SM
contribution while half the top quarks decay into
the lightest stop.  For our purposes we note
that if the masses of non-stop squarks are 
greater than roughly $250$~GeV, then they are not
crucial in maintaining a large radiative neutralino
branching fraction.

The simultaneous existence of a light right stop,
a heavy left stop, other squarks (except $b_L$)
and the gluino with masses ${\cal O}(250)$~GeV, 
and a large $\eegg + \Et$ rate is therefore 
an interesting possibility.  We explicitly constructed
nearly 200 models, mostly of the $\eL$ class
due to their larger cross section.  We did not
find significant differences in the models' 
distribution in $M_1$--$M_2$ plane, nor 
in the $\mu$--$\tbeta$ plane.
However, regions in these planes that were
populated by heavy stop models with $\sigma \times \BR^2$ near
the ${\cal A} = 5$~fb cut are no longer are allowed.
For instance, no light stop models approached the 
gaugino mass unification ($M_2 = 2 M_1$) line,
and $|\mu|$ was restricted to be less than $62$~GeV\@.
Hence, there are a number of phenomenological consequences 
of assuming a light stop ($m_{\tI} = 50$~GeV).
First, as noted above, the branching ratio 
of $\Ci \ra \tI b$ is virtually $100\%$
(when kinematically accessible), 
followed by the one loop decay $\tI \ra c \NI$
if $m_{\tI} < m_{\C}$.  Thus all the 
signatures as noted in Secs.~\ref{LEP-subsec}
and \ref{tevatron-subsec} arising from charginos
become $bc + \Et$.  For example, while the
dilepton signal from $\NI\NIII$ is unchanged,
the dilepton signal from $\Ci\Cjmp$ becomes
$b\overline{b}c\overline{c} + \Et$.  Also,
the restriction $m_{\NI} < m_{\tI} (= 50 \; {\rm GeV})$
results in somewhat tighter restrictions on the
upper bounds of the other chargino and neutralino masses.
In particular, $m_{\CI} \lsim 90$~GeV and $m_{\NIII} \lsim 100$~GeV 
in our light stop models, and the
sum $(m_{\NI} + m_{\NIII}) \lsim 150$~GeV\@. 
One consequence is that $\NI\NIII$ production
is always kinematically allowed at LEP161, with
a cross section in the range $1.1 < \sigma( \NI\NIII ) < 2.1$~pb.

Stop production at LEP161 may be directly visible
with the expected integrated luminosity 
if $2 m_{\tI}$ is below threshold~\cite{stop-production}.  
In Table~\ref{LEP-stops-table} we present the
cross section for stop production at LEP161 and LEP190
for a selection of light stop masses.  At LEP161 
one would expect roughly $20$~($5$) stop pairs produced
per detector, for $m_{\tI} = 50 \; (70)$~GeV\@.  
At LEP190 one would expect roughly $380$~($95$) 
stop pairs produced per detector, for 
$m_{\tI} = 50 \; (80)$~GeV\@.  All of the
cross sections were calculated with approximate 
final state QCD corrections
and QED initial state radiation effects included, 
and assuming $\tI = \tR$.  Also, $\tI\tI^*$ bound state
effects can be important close to the threshold.

\begin{table}
\renewcommand{\baselinestretch}{1.2}\small\normalsize
\begin{center}
\begin{tabular}{ccc} \hline\hline
$m_{\tI = \tR}$ &  \multicolumn{2}{c}{Cross section (in pb)} \\ \hline
    (GeV)       &  LEP161  &  LEP190   \\ \hline
    $50$        &  $0.85$  &  $0.76$   \\
    $60$        &  $0.50$  &  $0.56$   \\
    $70$        &  $0.20$  &  $0.37$   \\
    $80$        &    --    &  $0.19$   \\ \hline\hline
\end{tabular}
\end{center}
\caption{
Cross sections for light stop $\tI \> (= \tR)$ 
production at LEP161 and LEP190 with approximate final state 
QCD corrections and QED initial state radiation effects included.
Close to the threshold the cross section values may receive
large corrections due to $\tI\tI^*$ bound state effects.
}
\label{LEP-stops-table}
\end{table}

It has been noted~\cite{KaneMrennaGluino} that when there is a light 
stop (so that $\Ci \ra \tI b$ and $t \ra \tI \Ni$), there is a 
large set of events predicted at the Tevatron by supersymmetry 
that has no parton-level SM background.  Even after all branching
ratios and detection efficiencies are included, tens of events 
remain in the present $100$~pb$^{-1}$ at Tevatron.  These events 
arise from three sources, (1) $\Ci (\ra \tI b) \NII$, see 
Table~\ref{pairs-produced-tevatron-table}; 
(ii) $t (\ra W b) \overline{t} (\ra \tI \NII)$; 
(iii) ${\tilde q}( \ra q \NII) {\tilde q} 
(\ra q {\tilde g} (\ra t(\ra W b) \tI))$.  
In all cases, $\NII \ra \NI\ph$, $\tI \ra c \NI$, and typically 
$W \ra jj$.  
After branching ratios and cuts there should be approximately 
$35$--$100$ events with the signature $b\ph + \E +$~jets.  
`Jets' means $1$--$5$ parton level jets, including $1$--$2$ 
charm jets (an average of $1.5$/event).  
This prediction could lead to a sample that allowed 
a robust (rather than one event level) detection of 
superpartners in the present CDF and D0 data.  
When $W \ra l \nu$ for these events, additional good signatures 
arise and one expects an excess of ``$W$''$bc$ events that would 
appear in the top sample, and $l^\pm\ph + \E +$~jets 
events.

The simultaneous existence of a light stop and 
a light chargino (as necessarily arises in $\eegg + \Et$ models)
can give rise to a shift in $R_b$~\cite{Rb}.
We have analyzed models with $m_{\tI} = 50$~GeV,
$\tI = \tR$ and find that the maximum shift in
$R_b$ is $\delta R_b^{\rm max} \lsim 0.003$ 
from chargino-stop loops only.  Charged Higgs-top 
loops can also be significant, with a shift 
$\delta R_b \lsim -0.0005$ depending on $m_A$.
In all cases $\tbeta$ must be near $1$ for a maximal
shift in $R_b$.  For example, $\tbeta = 1.1$, $1.5$, $2.0$
can all give a large $\eegg + \Et$ rate, while the 
shift in $R_b$ is at best $0.0028$, $0.0021$, $0.0018$
for chargino-stop loops only.  Further, $R_b$ is sensitive
more to the parameter $\tbeta$ than $m_{\CI}$, as is
clear since the chargino mass is inversely related
to $\tbeta$; in the above three cases 
$m_{\CI}$ is roughly $83$, $80$, $70$~GeV\@.
We note that these calculations have been done assuming
$m_{\tI} = 50$~GeV, $\theta_{\tilde t} = \pi/2$, which is
nearly optimal since the maximum shift 
in $R_b$ decreases as either the stop mass
is increased, or $\theta_{\tilde t}$ is taken 
far from $\pi/2$.

As has been emphasized, getting a significant shift in $R_b$ 
requires a chargino that has a large Higgsino component, 
and the related result that $\mu$ is small and negative.  
It is interesting that the value of $\mu$ and the chargino 
properties coming from the analysis of the $\eegg + \Et$ event 
have the properties needed to give such an effect.  Finally, 
we note that a shift in $R_b$ necessarily implies a shift in $\alpha_s$ 
extracted from the LEP Z lineshape, through the relation 
$\delta \alpha_s(M_Z) \sim -4 \delta R_b$~\cite{alphas}.  
This limits the maximum shift in $R_b$ to about $0.0025$, 
consistent with the above numbers and giving $R_b \lsim 0.2182$.  
It is worth emphasizing that a significant shift in $R_b$ (and $\alpha_s$) 
is only possible simultaneously with a supersymmetric 
interpretation of the $\eegg + \Et$ event if $\NI$ is 
the LSP~\cite{Grav}.

\section{Concluding remarks}
\label{conclusions-sec}
\indent

We have seen that supersymmetry with $\NI=$LSP is a viable 
explanation of the CDF $\eegg + \Et$ event.  The primary constraints 
are the kinematics of the $\eegg + \Et$ event, the radiative 
neutralino branching ratio $\NII \ra \NI \ph$, the selectron decay 
${\tilde e} \ra e \NII$ and LEP1--LEP130 data.  Given a minimum 
threshold on the cross section times branching ratio of 
$p\overline{p} \ra {\tilde e}^+ {\tilde e}^- \ra e^+ e^- \NII \NII 
\ra e^+ e^- \ph\ph \NI \NI$ at the Tevatron, a selectron 
interpretation requires $M_1$, $M_2$, $\mu$, $\tbeta$, $m_{\tilde e}$ 
in tight ranges (see Table~\ref{slepton-correlations-table} and 
Figs.~\ref{mass-spectrum-fig},~\ref{mass-difference-fig}).  The corresponding 
chargino and neutralino masses and the cross sections at LEP and 
Tevatron are similarity constrained.  This is the origin of the 
predictions made for both LEP and Tevatron based solely on
the $\eegg + \Et$ event, where many signals can be large, 
and some must be produced.  
These signals are deduced from the cross sections and branching 
ratios without efficiencies, although in many cases the mass differences 
between sparticles cannot be arbitrarily small, and so presumably 
the signals are detectable.  For example, $\NI\NIII$ production 
must occur at LEP190 with the mass difference $40 < m_{\NIII} - 
m_{\NI} < 60$~GeV in all models, which implies a pair of leptons 
or jets from the decay $\NIII \ra \NI f\overline{f}$ would have 
an invariant mass up to roughly $60$~GeV\@.  The inclusive signals that 
must be produced at LEP190 with an integrated luminosity of 
$500$~pb$^{-1}$ are 
$\sigma( 2l + X + \E ) \gsim 50$ events, and
$\sigma( \ph\ph + X + \E ) \gsim 3$ events.
At the Tevatron, the inclusive signals that should have been produced 
(with an integrated luminosity of $100$~pb$^{-1}$) are
$\sigma( 2l + X + \Et) \gsim 30$ events, 
$\sigma( \ph\ph + X + \Et ) \gsim 2$ events,
$\sigma( l\ph + X + \Et ) \gsim 15$ events,
$\sigma( 2l\ph + X + \Et ) \gsim 4$ events,
$\sigma( l\ph\ph + X + \Et ) \gsim 2$ events, and
$\sigma( 3l + X + \Et ) \gsim 2$ events.
All of these signals assume $X=$ anything (leptons, photons, jets), 
and are valid for $\eL$ or $\eR$ models.  For $\eR$ models only,
the inclusive signal that must be produced at
LEP161 with an integrated luminosity of $25$~pb$^{-1}$ is
$\sigma( 2l + X + \E ) \gsim 2$ events.  Also for $\eR$ models
only, the inclusive signals that must be produced at LEP190 
(in addition to the ones above) are
$\sigma( l\ph + X + \E ) \gsim 5$ events, and
$\sigma( 2l\ph + X + \E ) \gsim 5$ events.
We have examined many inclusive signals with leptons and photons,
but of course inclusive signals with jets ($+$ photons) are also 
important and in some cases can be larger.  

The selectron interpretation can be made with the selectron 
${\tilde e}$ being either ${\eL}$, ${\eR}$, or a 
sum over ${\eL}$ and ${\eR}$ contributions.  The difference
between $\eL$ and $\eR$ is in the SU(2)$_L$ couplings
of $\eL$, causing for example the cross section at the Tevatron 
$\sigma( p\overline{p} \ra \eL\eL) \approx 2.2 
\sigma( p\overline{p} \ra \eR\eR)$ (in the mass range 
of interest), and the presence of $\eL$ couplings to charginos.
Thus one way to distinguish $\eL$ (and $\eL + \eR$) models 
from $\eR$ models is with the associated charged current channel 
$p\overline{p} \ra \eL\veL$ that gives at least $l\ph + \Et$, 
with possibly more leptons or photons depending on the decay of $\veL$.
Studies of such signals are relevant for $l=e$, $\mu$, $\tau$.  
A further source of $l^\pm \ph + X + \Et$ events comes 
from ${\tilde C}_{1,2} (\ra \NI l\nu) \NII (\ra \NI \ph)$, 
as well as $\Ci\Cjmp$ with $\C \ra \eL \nu_e$ or $\C \ra \veL e$ 
if $m_{\eL}$ or $m_{\veL}$ is lighter than the chargino.
Thus if no excess of associated events can be attributed to
the absence of $\eL\veL$ production, then it becomes 
less likely that the original selectron was $\eL$, though it 
cannot be definitive until a clean result is published.  
In addition, particular signals must be produced at LEP161/190
for $\eR$ models that are not necessarily present for
$\eL$ models, and thus if these associated events were not
found, then it becomes less likely that the original selectron
was $\eR$, with the same caveat as above.
At LEP it is necessary to study the relative rates of 
different channels to distinguish $\eL$ from $\eR$, 
unless selectron pair production is actually observed there.
In fact, if $\eL$ or $\eR$ production is observed (and the
LSP can be established to be $\NI$), then we immediately
know which charged slepton is {\em not}\/ responsible for 
the $\eegg + \Et$ event, since as we have shown in this
paper the slepton giving the $\eegg + \Et$ is kinematically
forbidden at LEP161 and LEP190.  Thus there is no unique
signal to discriminate $\eL$ from $\eR$ (from $\eL + \eR$)
models; only through the pattern of multiple signals 
can the nature of the selectron be determined.

We have also seen that a chargino interpretation of the
$\eegg + \Et$ event is a distinct possibility.
In either the selectron or chargino interpretation 
we expect at least the constraints
from radiative neutralino decay to hold, and light sleptons
are probably also a shared requirement for either interpretation
(see Appendix~\ref{char-sec}).  One way to
eventually distinguish the selectron interpretation from the 
chargino interpretation is to compare the rates of $\eegg$,
$\mu\mu\ph\ph$ and $e\mu\ph\ph$.  Assuming a mass degeneracy
among the sleptons of different families, the selectron
interpretation predicts roughly an equal number of 
$\eegg$ and $\mu\mu\ph\ph$ events, with a significantly
depleted $e\mu\ph\ph$ signal originating only from 
${\tilde \tau}^+ {\tilde \tau}^-$ production followed
by $\tau^+ \tau^- \ra e^\pm \mu^\mp + X$. 
Alternatively, in the chargino interpretation one would expect 
roughly double the number of $e\mu\ph\ph$ events as compared
with either $ee\ph\ph$ or $\mu\mu\ph\ph$ events.
Thus comparing the $e\mu\ph\ph$ rate with either $\eegg$ 
or $\mu\mu\ph\ph$ would provide a useful means to
discriminate between the two interpretations.
Notice also that events of the type $l^+l'^-\ph\ph + \Et$ 
can be produced only from $\Ci\Cj$ and 
${\tilde \tau}^+{\tilde \tau}^-$ production.

It is important to remark that the $\eegg + \Et$ event phenomenology 
could be connected with other phenomena.  If the LSP$=\NI$ is stable,
then it could provide a cosmologically significant relic
density even if it is mostly a Higgsino~\cite{HiggsinoLSP} 
(as required by the $\eegg + \Et$ event).  
For a given value of $\Omega h^2$ the mass of $\NI$ is 
correlated with $\tbeta$, and so gives
a subset of the models constructed here.  
The predictions for associated phenomenology
are tighter; and generally the signals can be larger.
Also, we have described in detail the effect of
assuming a light stop in addition to the $\eegg + \Et$ event,
in particular its connection to $R_b$~\cite{Rb} 
(and $\alpha_s$~\cite{alphas}).  A light stop has
many other consequences~\cite{KaneMrennaGluino}, that
we will not go into detail about here.

However, it is perhaps useful to remark on how can we learn 
if there is a light stop.  The easiest way would 
be to observe it at LEP.  The cross section ranges from about 
$0.2$--$0.8$~pb over the range $50 < m_{\tI} < 70 \; (80)$~GeV 
of most interest for LEP161 (LEP190).  LEP190 with tens of pb$^{-1}$ 
will be definitive.  For such light stops and even for somewhat 
heavier ones up to 
$m_{t} - m_{\NI}$ ($\lsim 100$~GeV in models considered here) 
searches in $t (\ra \tI \NI) \overline{t} (\ra W b)$ 
or stop pair production can be definitive.  Indirect evidence 
for a light stop before there is definitive collider data could 
come from a convincing $R_b$ excess, from slepton pair production 
at the Tevatron without associated leptons, and photons from 
chargino channels because $\Ci \ra \tI b$, from anomalous 
behavior of top properties~\cite{KaneMrennaGluino}, and from the 
$\ph + b +$~jets events~\cite{KaneMrennaGluino} commented on in
Sec.~\ref{light-stop-sec}.  Note that $\tI$ 
could be near $\NI$ in mass, and therefore give very soft
fermions plus large missing energy.

We have stated that certain signals must be produced at LEP
and Tevatron, and some signals might be produced if kinematically
accessible.  For example, at LEP161 three neutralino and one 
chargino pair cross sections are large enough to give a signal 
if about $25$~pb$^{-1}$ is collected.  The signatures 
are described in Sec.~\ref{LEP161-subsec} and can sometimes be 
somewhat unusual.  At LEP190 many more processes can be open,
which can all give signals with possibly unusual signatures
(see Sec.~\ref{LEP190-subsec}).
It is important to emphasize that the predictions assume 
the minimum cut ${\cal A} = 5$~fb is placed on the 
$\sigma \times \BR^2$ for the $\eegg + \Et$ event.
In principle, if one could demonstrate that failure 
to detect the signals implies they do not occur at all, 
then only two possibilities remain:  (1)  
A supersymmetric explanation of the $\eegg + \Et$ event
in our framework must rely on an upward fluctuation from 
$\sigma \times \BR^2$ even lower than $5$~fb, 
or (2) a supersymmetric explanation 
in our framework is not possible.
We note that even if the cut ${\cal A} = 5$~fb needs 
to be relaxed, there are still constraints from requiring 
a moderate branching ratio for $\NII \ra \NI\ph$
as demonstrated in Fig.~\ref{radiative-M1-M2-fig}.

The $\eegg + \Et$ event has given us a profound example
of how low energy supersymmetry could be discovered with 
one event.  It is not obvious that such an event {\em could}\/
be explained by supersymmetry, and we emphasize here the
predictability of the theory once such an explanation is adopted.
In particular, we have shown that assuming the $\eegg + \Et$ 
event {\em is}\/ due to supersymmetry with a $\NI=$LSP imposes 
strong constraints on the supersymmetric parameters, and predicts
much associated phenomenology.  Confirmation at LEP or Tevatron 
from the myriad of associated signals described in this paper 
is necessary to be definitive.  It is remarkable how much can
be learned from the Tevatron data, if the signal is confirmed.

\section*{Acknowledgments}

This work was supported in part by the U.S. Department of Energy.
S.~A. is supported mainly by an INFN postdoctoral fellowship, Italy.

\begin{appendix}
\refstepcounter{section}

\section*{Appendix~\thesection:~~Models in the chargino interpretation}
\label{char-sec}
\indent

The chargino interpretation purports to explain the 
$\eegg + \Et$ event through chargino production and
decay, a priori sharing only the requirement 
of radiative neutralino decay with the selectron
interpretation.  The possible sources of $\eegg$ in the 
chargino interpretation are from $p\overline{p} \ra \Ci\Cjmp$
with ${\tilde C}_{i,j} \ra \NII e \nu$, followed by $\NII \ra \NI \ph$.
The decay ${\tilde C}_{i,j} \ra \NII e \nu$ can proceed through 
either on-shell or off-shell $W$, $\eL$ and $\veL$.  
However, the 2-body decay $\CI \ra W \NII$ is not
possible, since $m_{\CI} - m_{\NII} \lsim 25$~GeV
when the radiative neutralino branching ratio 
$\BR( \NII \ra \NI\ph )$ is required to be large.

\subsection{Chargino production and 3-body decays}
\indent

If $\eL$, $\veL$ are heavy, the branching ratio for
the decay ${\tilde C}_i \ra \NII e \nu_e$ is 
dominated by $W$-exchange, with a branching ratio the same as that
for the SM decay $W \ra e\nu_e$ equal to $11\%$.
Chargino production with heavy sleptons therefore implies
for every $l^+l^-\ph\ph$ event, roughly $20$ other events 
with jet activity (possibly accompanied by one charged 
lepton) or two different charged leptons (plus two photons).  
In addition, the channel ${\tilde C}_i \ra \NI e \nu$ is always 
open and it is generally favored by phase space, in particular 
in the case $i=1$, since the mass difference 
$m_{\CI} - m_{\NII}$ is never large.  Further, it seems
difficult to find a region of the parameter space allowed by LEP data, 
consistent with a large neutralino radiative decay branching ratio 
and the general kinematical $\eegg$ requirements, 
where the non-radiative channels into $\NI$ are dynamically 
suppressed.  This holds for both on- and off-shell $W$-exchange,
and as a result the branching ratio for the decay  
${\tilde C}_i \ra \NII e \nu$ hardly exceeds $6\%$ for $i=1$ and 
is even lower for $i=2$.  
Hence, to get $> 5$~fb $\eegg$ signal from $\Ci\Cjmp$ 
production and decay ${\tilde C}_{i,j} \ra W^{(*)} 
(\ra e \nu) \NII (\ra \NI\ph)$ 
one needs a cross section at least roughly $1.5$~pb
(even assuming $\BR( \NII \ra \NI \ph ) = 100\%$), 
since $\BR(\Ci\Cj \ra \NII\NII e^+e^- \nu_e \overline{\nu}_e)$
is well below 1\%.
This does not seem to be possible with an individual
chargino pair production process, given all the other $\eegg$
constraints.  However, a small but non-zero signal
can always arise from this source in models which are compatible 
with the selectron interpretation.  We have 
found models with up to $\sim 1$~fb $\eegg$ total 
signal from the sum of $\Ci\Cjmp$ production 
and 3-body $\Ci$ decay in our selectron interpretation models.
(These contributions were not included in the selectron 
interpretation.)

\subsection{Chargino production and 2-body decays}
\indent

We consider in the following chargino production followed
by 2-body decays into sleptons, which allows an enhancement
of the total possible branching ratio into the $\eegg$ final state.
The regions are somewhat different in the chargino interpretation 
with $\C \ra \tilde{l} l$ than 
in the selectron interpretation; in particular we found the
constraint $m_{\NII} - m_{\NI} \gsim 20$~GeV is no longer
required.  (We have checked that a neutralino mass difference 
of order $10$~GeV can be sufficient in the chargino interpretation.) 
This may in principle allow the kinematical mechanism for the enhancement 
of the radiative neutralino decay branching ratio to
operate simultaneously with the dynamical mechanism to obtain a 
large $\eegg$ rate.  In Sec.~\ref{model-building-results-subsec} 
we already encountered particular models in the selectron 
interpretation where the kinematical mechanism plays an 
important role, and this may be true for the chargino
interpretation to an even greater extent. 
However, a small mass difference $m_{\NII} - m_{\NI}$ 
seems only to be allowed when $m_{\NII}$ is small, so that 
it can presumably receive a large boost after the $\eL$ or $\veL$ decay 
and generate a hard photon.  The only way to construct a 
model with two very light neutralinos and a heavier chargino
is to enter the ``light gaugino-Higgsino window'' 
(see Sec.~\ref{model-building-results-subsec}), but even there 
it seems difficult to build a model which falls in the region
suggested by the $\eegg + \Et$ event kinematics with the constraints 
from the branching ratios.  Also, with a neutralino mass difference 
of order 10 GeV or more the radiative neutralino branching 
ratio never approaches $100\%$ from only the kinematical 
enhancement~\cite{AmbrosMele2}.  Hence, as in the selectron
interpretation it would appear that the dynamical mechanism
for a large radiative neutralino decay is required.
This, along with the following argument for the need 
of a mostly gaugino $\NII$, explains why it seems possible 
to build models with large $\eegg$ rates in the chargino interpretation 
only in regions of the gaugino-Higgsino parameter space similar
to that in the selectron interpretation.

The maximum
$\BR[\Ci \ra l^{\prime} \tilde{l}^{(*)} (\ra l \NII)]$ 
for ${\tilde l} = \veL$ is $1/3$, and for ${\tilde l} = \eL$ 
is $1/6$ due to the slepton mass degeneracy assumption 
and $m_{\veL} < m_{\eL}$ (assuming the decay into sneutrinos
is not strongly suppressed).  Also, the slepton decay 
channels with $\NI$ in the final state are always open and 
enhanced by phase space.  Thus to maximize the branching
ratio into $\NII$, one has to minimize the $\NII$ Higgsino
components (which do not couple with sleptons) and maximize 
the Higgsino component of $\NI$.  In this way, the branching ratio
for $\eL \ra \NII e$ is enhanced, analogous to the selectron
interpretation.
Typically, the branching ratio for the combined
decay $\Ci\Cjmp \ra \eegg$ though 2-body decays into sleptons
can reach at best $\sim 4\%$, 
assuming $\BR( \NII \ra \NI\ph ) = 100\%$.  In the $\CII$ case, 
a further source of suppression can come from the channel
$\veL \ra \CI e$ (if open), that always dominates over
$\veL \ra \NII \nu_e$ or $\veL \ra \NI \nu_e$.  
A similar suppression in the $\CII$ case can also come 
from $\eL \ra \CI \nu_e$.  Thus, the actual 
$\eegg$ rate depends strongly also on the mass hierarchy 
between $m_{\veL}$, $m_{\eL}$ and $m_{\CI}$. 

Maximizing the Higgsino component of $\NI$ and minimizing that 
of $\NII$, leads us to the conclusion that $\NII$ 
is mostly photino and $\NI$ is mostly Higgsino, 
analogous to the selectron interpretation.
However, differences do exist between the chargino interpretation
and $\eL$ models in the selectron interpretation.
For example, one needs $100 \lsim m_{\eL} \lsim 137$~GeV 
in the selectron interpretation, while 
in the chargino interpretation one only needs at least one of
$\eL$, $\veL$ heavier than roughly 60 GeV but lighter than 
at least one of the charginos.  Of course, additional constraints on 
$m_{\eL}$, $m_{\veL}$ are present, due to the particularly  
complicated decay chain and the large radiative neutralino
branching ratio needed.  The right selectron enters
the 3-body decay $\NII \ra \NI e^+ e^-$, but if its mass is
moderately large then the decay cannot be enhanced.
Squark masses are relatively unconstrained, 
although lighter squark masses increase the $\CII\CII$ cross 
section, but decrease the radiative branching ratio.

The absence of $\eegg + \Et$ event kinematical solutions 
with chargino masses less than $95$~GeV implies that
to construct a chargino interpretation that at least
possibly satisfies the kinematics one should
conservatively choose to search only for models with
$m_{\C} > 95$~GeV\@.  Restricting to $M_2$, $\mu$ and 
$\tbeta$ values roughly in the allowed ranges singled 
out in the selectron interpretation, one finds
a rough upper limit of $400$, $50$ and $1200$~fb for the 
cross section of $\CI\CI$, $\CI\CII$ and $\CII\CII$ production 
respectively.  Given at best a useful branching ratio of about 
$5\%$, then a $\CI\CII$ interpretation (alone) can be excluded. 
For $\CI\CI$ and $\CII\CII$ production, the $\eegg$ signal
could be up to roughly $20$ and $60$~fb, therefore the lower
bound on the radiative 
neutralino decay branching ratio is $50\%$ and $30\%$ 
respectively, to pass ${\cal A} = 5$~fb cut used
in the selectron interpretation.  The $\CI\CI$ cross section 
drops rapidly as $m_{\CI}$ is increased, and it appears
not to give a sizeable $\eegg + \Et$ signal when 
$m_{\CI} \gsim 110$~GeV\@.  Alternatively, the $\CII\CII$ cross 
section can still be large, and give a sizeable $\eegg$
signal for $m_{\CII} \sim 150$~GeV 
(if $m_{\tilde q} \sim 250$~GeV).
In practice, this sets rough upper limits for $M_2$ and $|\mu|$ 
which determine the chargino masses. 
Further, our analysis of the $\eegg + \Et$ event 
kinematics in the chargino interpretation gives an 
indication that large mass differences ($\gsim 30$~GeV) between 
${\tilde C}_i$ and $\NII$ may be required to reconstruct the 
$\eegg + \Et$ event.  In the $i=1$ case this is very difficult, 
if not impossible, given all the other constraints.  
Thus, we conclude that sizeable $\eegg + \Et$ signals can
probably only be achieved from $\CIIp\CIIm$ production, 
with the decay chain $\CII \ra \nu_e \eL (\ra e \NII)$ or
$\CII \ra e \veL (\ra \nu_e \NII)$, followed by $\NII \ra \NI \ph$.
This appears to happen only in a region of the parameter space
similar to the selectron interpretation.

A few final remarks on model building are in order.
The sneutrino always plays a role when the mass hierarchy
$m_{\C} > m_{\eL} (> m_{\veL})$ exists, and as 
a consequence the $\eegg$ signal is depleted from $\C \ra \veL e$
since $\veL$ tends to have comparable 
branching ratio into $\NI$ and $\NII$.
Further, if $m_{\veL} < m_{\NII}$, then a 2-body decay opens 
for $\NII \ra \veL \nu_e$, which often suppresses the 
radiative decay branching ratio.  Also, a sneutrino mass larger 
than $m_{\CI}$ implies a possibly large branching ratio 
for $\veL \ra \CI e$.  To ensure sufficient phase space for 
the decay $\CII \ra \eL \nu_e$ and to have the masses fall
in regions where we found kinematical solutions, the mass difference
$m_{\CII} - m_{\eL} \gsim {\cal O}(10)$~GeV probably should be enforced.
The selectron also must be larger than $m_{\NII}$ by at 
least $\sim 20$~GeV for analogous reasons, but not larger 
than $m_{\CI}$ otherwise the branching ratio for $\eL$ 
will be dominated by $\eL \ra \CI \nu_e$.  It is clear
that maintaining such a mass hierarchy between $m_{\CII}$,
$m_{\eL}$, $m_{\veL}$, $m_{\NII}$, $m_{\CI}$, $m_{\NI}$
is considerably more difficult than in the selectron interpretation,
and to some extent a fine-tuning of the masses of the 
particles involved is always required.  Also, the relevant 
branching ratio is always small and never exceeds a few percent
while in the selectron interpretation it can in principle reach $100\%$.  
All of these facts seem to render a chargino interpretation
problematic (in stark contrast to a scenario with the 
gravitino as the LSP~\cite{Grav}).

\subsection{Chargino interpretation -- an example}
\indent

We searched our model samples compatible with a selectron
interpretation of the $\eegg + \Et$ event for cases where $\CII\CII$
production could yield an additional $\eegg$ signal.  
We found several tens of candidate models: some in the 
$\eR$ samples, and a few in the $\eL$ sample.  However,
the general kinematical requirements for a chargino interpretation
of the $\eegg + \Et$ event slightly favor the $\eL$ models,
which are located roughly in Region 2 
(according to the classification of 
Sec.~\ref{model-building-results-subsec}).  Such models 
could give rise to a $\eegg$ signal with the 
kinematical characteristics of the event, from
simultaneously $\eL$ and $\CII$ pair production,
although the $\CII$ signal is generally below $6$~fb.
We report one model as an example of the above:
$M_1 = 65$~GeV, $M_2 \simeq M_Z$, $\mu = -53$~GeV $\tbeta = 2$, 
$m_{\eL} = 110$~GeV, $m_{\eR} = 350$~GeV, $m_{\vL} = 90$~GeV, 
$m_{\tI} = 150$ GeV, $m_{\tII} \approx m_{\tilde{q}} = 250$ GeV\@.
The neutralino masses $m_{{\tilde N}_{1,2,3,4}} =  65, 70, 96, 137$~GeV,  
and the chargino masses $m_{{\tilde C}_{1,2}} = 72, 137$~GeV\@. 
The $\CII\CII$ production cross section at the Tevatron
is $380$~fb, while the $\eL\eL$ cross section is $13$~fb. 
The $\BR( \NII \ra \NI \ph ) = 81\%$, 
the $\BR( \CII \ra \veL e ) = 17\%$, the $\BR( \CII \ra \eL \nu_e) = 16\%$,
the $\BR( \eL \ra \NII e) \sim 100\%$, and $\BR( \veL \ra \CI e ) = 77\%$.
The $\eegg$ rate is roughly $6$~fb from only chargino production, 
and so is slightly above the 
${\cal A} = 5$~fb cut imposed in the selectron interpretation.
It is worthwhile to remark on how sensitive the $\eegg$ rate
is to a change in the masses.  For example, one can attempt 
to raise the $\eegg$ rate from chargino production by slightly 
reducing the $\eL$ mass in such a way to get a sneutrino lighter 
than the $\CI$, and gain the additional signal from $\CII$ decays 
into on-shell sneutrinos and sneutrino decays into $\NII$. 
This would require $m_{\eL} \lsim 96$~GeV\@, although the
modified model would appear to be farther from the region of 
masses satisfying the $\eegg + \Et$ event kinematics.
However, the radiative neutralino decay branching ratio drops 
quite sensitively when the already light slepton masses are further
reduced.  Thus, constructing models in the chargino interpretation
is somewhat difficult, and it is not obvious how one ought to perturb
around any given model to increase the $\eegg$ rate.
However, we did find some models with interesting characteristics,
as shown above.  A more in-depth analysis is necessary to determine
if the chargino interpretation is tenable, and if so the ranges
of the parameters needed.

\refstepcounter{section}

\section*{Appendix \thesection:~~Sample Models}
\label{sample-models-sec}
\indent

Here four sample models from the set used in the selectron 
interpretation are provided in Tables~\ref{sample-models12-table}
and \ref{sample-models34-table}.  Input parameters and calculated
masses are given, along with many branching ratios and 
cross sections.  Notice that the four models' input parameters 
are similar (except for the slepton and stop masses), but the cross 
sections for both the $\eegg + \Et$ event and associated phenomenology
are quite different.  

\begin{table}
\begin{small}
\begin{center}
\begin{tabular}{lcc} \hline\hline
Model parameters        &  $\eL$ model & $\eR$ model \\ \hline\hline
$M_1$, $M_2$            
    &  $64.7$ , $64.3$  &  $74.4$ , $77.6$ \\
$\mu$, $\tbeta$         
    &  $-37.0$ , $1.18$  &  $-38.3$ , $1.11$ \\
$m_A$, $m_{\tilde q} = m_{\tII}$   
    & $200$ , $500$ & $400$ , $500$ \\
$m_{\tI}$, $\theta_{\tilde t}$     
    & $204$ , $-0.342$  &  $487$ , $-0.123$ \\
$m_{\lL}, m_{\lR}, m_{\vL}$
    & $105$ , $272$ , $99.6$  &  $391$ , $104$ , $390$ \\ \hline
$m_{\CI}, m_{\CII}$
    & $79.6$ , $110$  &  $78.9$ , $119$ \\
$m_{\NI}, m_{\NII}, m_{\NIII}, m_{\NIIII}$  
    & $36.6$ , $64.6$ , $90.5$ , $118$  &  $38.2$ , $75.1$ , $88.5$ , $127$ \\
$\langle \NI | {\tilde H}_b \rangle^2, \langle \NII | \phino \rangle^2$ 
    & $0.997$ , $1.000$  &  $0.999$ , $0.999$ \\
$m_h, m_H, m_{H^\pm}, \alpha_h$
    & $70.2$ , $229$ , $216$ , $-0.825$ & $69.2$ , $415$ , $408$ , $-0.765$ \\
$\sigma \times \BR^2$
    &  $13.2$  &  $6.6$ \\
$\BR(\NII \ra \NI \ph)$  
    &  $0.98$  &  $0.94$ \\
$\BR(\NIII \ra l^+l^-), \BR(\NIII \ra \nu\overline{\nu}), \BR(\NIII \ra 
    q\overline{q})$ & $0.10$ , $0.22$ , $0.67$ & $0.10$ , $0.20$ , $0.69$ \\
$\BR(\NIIII \ra \vL \overline{\nu} + \overline{\vL} \nu),
    \BR(\NIIII \ra \lL \overline{l} + \overline{\lL} l)$
    & $0.83$ , $0.13$ & -- , -- \\
$\BR(\NIIII \ra \lR \overline{l} + \overline{\lR} l)$ 
    & -- & $0.80$ \\
$\BR(\CI \ra \NI l\nu), \BR(\CI \ra \NI q\overline{q}')$
    & $0.34$ , $0.66$ & $0.33$ , $0.67$ \\
$\BR(\CII \ra \vL l), \BR(\CII \ra \lL \nu)$ 
    & $0.66$ , $0.28$ & -- , -- \\
$\BR(\CII \ra \NI l \nu), \BR(\CII \ra \NI q\overline{q}')$ 
    & $0.02$ , $0.03$ & $0.33$ , $0.66$ \\
$\BR(\eL \ra \NII e), \BR(\eL \ra \C \nu_e)$
    & $0.91$ , $0.07$ & $0.30$ , $0.59$ \\
$\BR(\eR \ra \NII e), \BR(\eR \ra \NIIII e)$
    & $0.81$ , $0.14$  & $0.98$ , -- \\
$\BR(\veL \ra \NIII \nu_e), \BR(\veL \ra \C e)$
    & $0.08$ , $0.90$ & $0.10$ , $0.61$ \\ \hline
\multicolumn{3}{l}{\underline{LEP161 cross sections:}} \\ 
$\sigma( \NI\NIII ), \sigma( \CI\CI )$  
    & $2010$ , $405$ & $2130$ , $1320$ \\
$\sigma( \NII\NII ), \sigma( \NII\NIII )$
    & $191$ , $123$ &  $40$ , -- \\
inclusive $\sigma( 2l + X ), \sigma( \ph\ph + X)$
    & $276$ , $184$ &  $365$ , $36$ \\ \hline
\multicolumn{3}{l}{\underline{LEP190 cross sections:}} \\ 
$\sigma( \NI\NIII ), \sigma( \NI\NIIII )$
    & $1450$ , $89$ & $1530$ , $49$ \\
$\sigma( \NII\NII ), \sigma( \NII\NIII )$
    & $342$ , $243$ & $199$ , $164$ \\
$\sigma( \CI\CI ), \sigma( \CI\CII )$
    & $1080$ , $167$ & $2760$ , -- \\
inclusive $\sigma( 2l + X ), \sigma( \ph\ph + X)$
    & $473$ , $331$ & $529$ , $177$ \\
inclusive $\sigma( l\ph + X), \sigma( ll\ph + X)$
    & $115$ , $73$ & $60$ , $59$ \\ \hline
\multicolumn{3}{l}{\underline{Tevatron cross sections:}} \\ 
$\sigma( \eL\eL ), \sigma( \eR\eR )$   
    & $16.5$ , -- & -- , $7.9$ \\
$\sigma( \veL\veL ), \sigma( \eL\veL )$  
    & $18.5$ , $45.0$ & -- , -- \\
$\sigma( \NI\NIII ), \sigma( \CI\CI ), \sigma( \CII\CII )$  
    & $1180$ , $907$ , $552$ & $1270$ , $887$ , $415$ \\
$\sigma( \CI\NI ), \sigma( \CI\NII ), \sigma( \CI\NIII )$
    & $2690$ , $113$ , $840$ & $2710$ , $55$ , $915$ \\
$\sigma( \CII\NII ), \sigma( \CII\NIII ), \sigma( \CII\NIIII )$
    & $324$ , $28$ , $332$ & $190$ , $8.4$ , $241$ \\
inclusive $\sigma( 2l + X ), \sigma( \ph\ph + X )$ 
    & $1700$ , $174$ & $631$ , $24$ \\
inclusive $\sigma( l \ph + X ), \sigma( 2l \ph + X )$
    & $954$ , $714$ & $318$ , $237$ \\
inclusive $\sigma( l \ph\ph + X ), \sigma( 3l + X )$
    & $171$ , $892$ & $22$ , $101$ \\ \hline\hline
\end{tabular}
\end{center}
\end{small}
\caption{Two sample models in the selectron interpretation.  
All masses are in GeV, all cross sections are in fb.
Only the largest branching ratios and cross sections
are displayed.  $l$ is summed over $e$, $\mu$, and 
$\tau$ in the branching ratios and inclusive
cross sections (which have no detector efficiencies included).
In the branching ratios $\C$ refers to a sum over $\CI$ and $\CII$.
}
\label{sample-models12-table}
\end{table}

\begin{table}
\begin{small}
\begin{center}
\begin{tabular}{lcc} \hline\hline
Model parameters        &  $\eL + \eR$ model  &  
    $\eR$ model (with light $\tI$)  \\ \hline\hline
$M_1$, $M_2$            
    &  $70.2$ , $76.2$  &  $76.5$ , $77.0$ \\
$\mu$, $\tbeta$         
    &  $-48.8$ , $1.26$  &  $-38.9$ , $1.39$ \\
$m_A$, $m_{\tilde q} = m_{\tII}$   
    &  $200$ , $500$  &  $400$ , $2000$ \\
$m_{\tI}$, $\theta_{\tilde t}$     
    & $488$ , $0.263$  &  $50$ , $\pi/2$ \\
$m_{\lL}, m_{\lR}, m_{\vL}$
    & $119$ , $121$ , $113$  &  $439$ , $105$ , $437$ \\ \hline
$m_{\CI}, m_{\CII}$
    & $84.8$ , $118$  &  $75.2$ , $121$ \\
$m_{\NI}, m_{\NII}, m_{\NIII}, m_{\NIIII}$  
    & $47.8$ , $71.5$ , $96.8$ , $124$  &  $37.4$ , $76.6$ , $88.4$ , $128$ \\
$\langle \NI | {\tilde H}_b \rangle^2, \langle \NII | \phino \rangle^2$ 
    & $0.990$ , $0.998$  &  $0.988$ , $0.999$ \\
$m_h, m_H, m_{H^\pm}, \alpha_h$
    & $67.8$ , $227$ , $216$ , $-0.792$ & $59.1$ , $411$ , $408$ , $-0.651$ \\
$\sigma \times \BR^2$
    &  $10.2$  &  $5.1$ \\
$\BR(\NII \ra \NI \ph)$  
    &  $0.92$  &  $0.86$ \\
$\BR(\NIII \ra l^+l^-), \BR(\NIII \ra \nu\overline{\nu}), \BR(\NIII \ra 
    q\overline{q})$ & $0.10$ , $0.22$ , $0.67$ & $0.10$ , $0.20$ , $0.68$ \\
$\BR(\NIIII \ra \vL \overline{\nu} + \overline{\vL} \nu),
    \BR(\NIIII \ra \lL \overline{l} + \overline{\lL} l)$
    & $0.85$ , $0.05$ & -- , -- \\
$\BR(\NIIII \ra \lR \overline{l} + \overline{\lR} l)$ 
    & $0.01$ & $0.74$ \\
$\BR(\CI \ra \NI l\nu), \BR(\CI \ra \NI q\overline{q}'), \BR(\CI \ra \tI b)$
    & $0.34$ , $0.66$ , -- & $0.00$ , -- , $1.00$ \\
$\BR(\CII \ra \vL l), \BR(\CII \ra \tI b)$ 
    & $0.78$ , -- & -- , $0.98$ \\
$\BR(\CII \ra \NI l \nu), \BR(\CII \ra \NI q\overline{q}')$ 
    & $0.06$ , $0.11$ & $0.01$ , $0.01$ \\
$\BR(\eL \ra \NII e), \BR(\eL \ra \C \nu_e)$
    & $0.94$ , $0.03$ & $0.30$ , $0.59$ \\
$\BR(\eR \ra \NII e), \BR(\eR \ra \NIIII e)$
    & $0.97$ , --  & $0.96$ , -- \\
$\BR(\veL \ra \NIII \nu_e), \BR(\veL \ra \C e)$
    & $0.10$ , $0.86$ & $0.10$ , $0.62$ \\ \hline
\multicolumn{3}{l}{\underline{LEP161 cross sections:}} \\ 
$\sigma( \NI\NIII ), \sigma( \CI\CI )$  
    & $1500$ , -- & $2100$ , $2680$ \\
$\sigma( \NII\NII ), \sigma( \NII\NIII ), \sigma( \tI {\tI}^* )$
    & $120$ , -- , -- & $23$ , -- , $850$ \\
inclusive $\sigma( 2l + X ), \sigma( \ph\ph + X)$
    & $157$ , $100$ & $215$ , $17$ \\ \hline
\multicolumn{3}{l}{\underline{LEP190 cross sections:}} \\ 
$\sigma( \NI\NIII ), \sigma( \NI\NIIII )$
    & $1360$ , $24$ & $1500$ , $41$ \\
$\sigma( \NII\NII ), \sigma( \NII\NIII )$
    & $355$ , $227$ & $169$ , $150$ \\
$\sigma( \CI\CI ), \sigma( \CI\CII ), \sigma( \tI {\tI}^* )$
    & $880$ , -- , -- & $3110$ , -- , $760$ \\
inclusive $\sigma( 2l + X ), \sigma( \ph\ph + X)$
    & $302$ , $299$ & $254$ , $125$ \\
inclusive $\sigma( l\ph + X), \sigma( ll\ph + X)$
    & $56$ , $51$ & $78$ , $78$ \\ \hline
\multicolumn{3}{l}{\underline{Tevatron cross sections:}} \\ 
$\sigma( \eL\eL ), \sigma( \eR\eR )$   
    & $9.4$ , $4.0$ & -- , $7.5$ \\
$\sigma( \veL\veL ), \sigma( \eL\veL )$  
    & $10.5$ , $24.6$ & -- , -- \\
$\sigma( \NI\NIII ), \sigma( \CI\CI ), \sigma( \CII\CII )$  
    & $688$ , $681$ , $434$ & $1270$ , $1140$ , $298$ \\
$\sigma( \CI\NI ), \sigma( \CI\NII ), \sigma( \CI\NIII )$
    & $1590$ , $86$ , $575$ & $3430$ , $128$ , $974$ \\
$\sigma( \CII\NII ), \sigma( \CII\NIII ), \sigma( \CII\NIIII )$
    & $189$ , $29$ , $259$ & $218$ , $43$ , $283$ \\
inclusive $\sigma( 2l + X ), \sigma( \ph\ph + X )$ 
    & $1190$ , $43$ & $178$ , $16$ \\
inclusive $\sigma( l \ph + X ), \sigma( 2l \ph + X )$
    & $369$ , $279$ & $50$ , $48$ \\
inclusive $\sigma( l \ph\ph + X ), \sigma( 3l + X )$
    & $39$ , $654$ & $16$ , $7.5$ \\ \hline\hline
\end{tabular}
\end{center}
\end{small}
\caption{As in Fig.~\ref{sample-models12-table}, 
but for an $\eL + \eR$ model, and a model with 
a light stop.  Note that $\sigma \times \BR^2$ sums over
both $\eL$ and $\eR$ contributions for the $\eL + \eR$ model.}
\label{sample-models34-table}
\end{table}

\end{appendix}

\newpage




\begin{thebibliography}{99}
\bibitem{oldwork}
    The method we use here is very similar to what we used
    previously in Ref.~\cite{PRL}.  However, many aspects
    of the model building have been refined (and some 
    minor corrections made), so the results here should
    be taken to supersede and complete those of Ref.~\cite{PRL}.
\end{thebibliography}
\end{document}